\documentclass[11pt]{article}
\pdfoutput=1
\usepackage{jheppub}
\usepackage{float, extarrows, tikz-cd}
\usepackage{graphicx}
\usepackage{slashed}
\usepackage{tabularx,ragged2e}
\usepackage{subcaption}
\usepackage{amsmath,amssymb}
\usepackage{xcolor}
\usepackage{dsfont}
\usepackage{verbatim}
\usepackage{mathtools, xcolor,ytableau, amsfonts}
\usepackage[nobottomtitles*]{titlesec}

\usepackage{physics}
\newcolumntype{C}{>{\Centering\arraybackslash}X}

\numberwithin{equation}{section}

\newcommand{\mO}{\mathcal{O}}
\newcommand{\pd}{\partial}

\def\<{\langle}
\def\>{\rangle}

\author[a,b]{Gabriel Cuomo,} 
\author[c]{Yin-Chen He,}
\author[d]{Zohar Komargodski}

\affiliation[a]{Center for Cosmology and Particle Physics, Department of Physics, New York University, New York, NY 10003, USA}
\affiliation[b]{Department of Physics, Princeton University, Princeton, NJ  08544, USA}
\affiliation[c]{Perimeter Institute for Theoretical Physics, Waterloo, Ontario, Canada N2L 2Y5}
\affiliation[d]{Simons Center for Geometry and Physics, SUNY, Stony Brook, NY 11794, USA}

\emailAdd{gc6696@princeton.edu}		
\emailAdd{yhe@perimeterinstitute.ca}
\emailAdd{zkomargodski@scgp.stonybrook.edu}

\title{Impurities with a cusp: general theory and 3d Ising}

\abstract{
In CFTs, the partition function of a line defect with a cusp depends logarithmically on the size of the line with an angle-dependent coefficient: the cusp anomalous dimension. In the first part of this work, we study the general properties of the cusp anomalous dimension. We relate the small cusp angle limit to the effective field theory of defect fusion, making predictions for the first couple of terms in the expansion.  Using a concavity property of the cusp anomalous dimension we argue that the Casimir energy between a line defect and its orientation reversal is always negative (``opposites attract''). 
We use these results to determine the fusion algebra of Wilson lines in $\mathcal{N}=4$ SYM as well as pinning field defects in the Wilson-Fisher fixed points. 
In the second part of the paper we obtain nonperturbative numerical results for the cusp anomalous dimension of pinning field defects in the Ising model in $d=3$, using the recently developed fuzzy-sphere regularization. We also compute the pinning field cusp anomalous dimension in the $O(N)$ model at one-loop in the $\varepsilon$-expansion. Our results are in agreement with the general theory developed in the first part of the work, and we make several predictions for impurities in magnets.}

\begin{document}

\maketitle

\section{Introduction}

Conformal Field Theories (CFTs) appear often at either quantum or classical second order phase transitions. 
There has been a monumental effort to understand the properties of local operators, and in particular, their scaling exponents and correlation functions. For a review of this multifaceted subject see~\cite{Poland:2018epd}. 

In addition to local operators, CFTs also admit extended objects. Focusing on the case of conformal lines, in Lorentzian signature we can think about them as conformal impurities (point-like defects) if they are extended in time or as nonlocal operators if they act at a given time. In Euclidean signature, these are just conformal defects of dimension 1. Such objects are very natural in condensed matter systems as they describe point-like impurities in certain gapless theories and they are also natural in classical second order phase transitions, describing certain defects along a one-dimensional manifold.

Oftentimes, the parameters on line defects need to be tuned for the line defects to be conformal. This is clear in the impurity interpretation, since the impurity changes the Hamiltonian (it often also enlarges the Hilbert space by adding new degrees of freedom on the impurity)
\begin{equation} H\to H+H_{{\rm impurity}}\label{Hchange}\end{equation} and by doing so introduces some new local interaction parameters at the location of the impurity. The interaction parameters on the impurity typically flow to some conformal fixed point at long distances (though exceptions exist). 
We are very far from understanding the space of conformal defects in interacting CFTs. 

\begin{figure}[t]
\begin{center}
\includegraphics[scale=0.35]{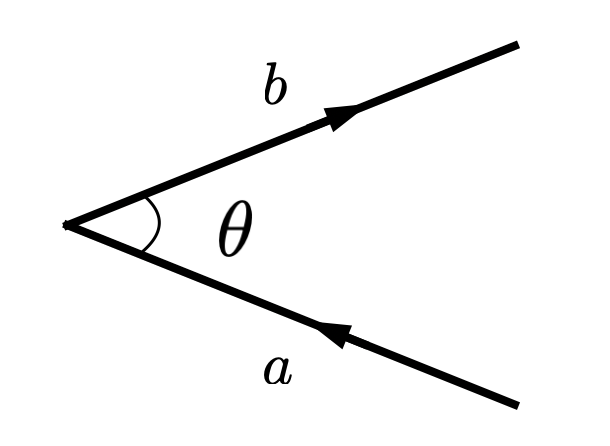}\caption{Defect lines $a$ and $b$ with a cusp at angle $\theta$.}\label{CuspRaw}
\end{center}
\end{figure}

Here we consider the problem of conformal line operators with a cusp. Most of our calculations will be in Euclidean signature with $d$ Euclidean dimensions. We will use our results to obtain observables concerning physical impurities in $d-1$ space dimensions.
The basic setup is that of two conformal lines $a,b$ with a cusp with angle $\theta$. This configuration preserves one conformal generator, namely, scale transformations around the cusp. In addition, it preserves transverse $SO(d-2)$ rotations. See figure~\ref{CuspRaw}.

There are various applications of this general setup. 
First, if we perform radial quantization around the cusp we find that the space of cusp operators is the Hilbert space of two impurities on $S^{d-1}$. In particular, the lowest dimension operator at the cusp corresponds to the ground state energy with two defects at angle $\theta$ on $S^{d-1}$. A lot of interesting data can be extracted from $\theta\simeq \pi$, such as the scaling dimension of the defect changing operator, and the two-point function of the displacement operator. In the opposite limit, when $\theta\ll1$ it is not very important that the defects are on $S^{d-1}$ and we can extract useful information from the cusp about flat space physics. For instance we can infer the  Casimir energy in the presence of two impurities, the quantum channel into which the impurities fuse, and other properties that we will discuss.

The contribution of the cusp in figure~\ref{CuspRaw} to the partition function is logarithmically divergent\footnote{It may be surprising to encounter a logarithm in conformal field theory, but here it has a simple interpretation as a trace anomaly at the cusp point itself since rescaling the ultraviolet scale $a\to \lambda a$ one produces a finite term that lives strictly on the cusp $e^{\Gamma_{ab}(\theta)\log \lambda}$. This is an allowed counter-term at the cusp since $\theta$ can be  defined locally. (It is a type B trace anomaly in 0 dimensions in the sense of~\cite{Deser:1993yx}.) (Alternatively, in more mundane terms,  it is wave-function renormalization.) This point of view also explains why one cannot obtain more complicated contributions, such as double logarithms -- those would not satisfy the Wess-Zumino consistency condition for trace anomalies. If one considers cusps in Minkowski signature and furthermore the cusp becomes null, double logarithms do appear, as has been shown in QCD~\cite{Korchemsky:1992xv}. This is permissible because the notion of locality on the light cone is more subtle as all the distances from the cusp to points along the defects vanish.} 
\begin{equation}\label{cuspdef}
\log Z_{ab}  = -\Gamma_{ab}(\theta) \log\left(L\over a\right)+\cdots ~, \end{equation} 
where $\Gamma(\theta)$ is the cusp anomalous dimension, $L$ is the size of the line operator (i.e. $L$ is some infrared cutoff), and $a$ some ultraviolet cutoff, which one can interpret as a microscopic scale on which the cusp is smoothed out. Via the state operator correspondence we can identify $\Gamma_{ab}$ as the ground state energy of two defects separated by an angle $\theta$ on $S^{d-1}$. (This has to be properly normalized, as we discuss in the text.)

In this work we only consider static (equilibrium) properties of defects, as can be read out from the cusp anomalous dimensions. 
It is also possible to consider dynamical properties such as quenches where an impurity switches from being at rest to (possibly a different impurity) moving at a constant velocity through the vacuum in flat space. This amounts to an analytic continuation of $\Gamma_{ab}(\theta)$ to complex $\theta$ and we do not discuss it here.  

The cusp function $\Gamma_{ab}(\theta)$ (with $a=b$) has been studied extensively in maximally supersymmetric Yang Mills theory,
where the conformal lines are various supersymmetric Wilson lines. At tree level in gauge theory, the computation was first done in~\cite{Polyakov:1980ca}. There are many perturbative as well as strong coupling results (see \cite{Drukker:1999zq,Makeenko:2006ds,Drukker:2007qr,Drukker:2011za,Correa:2012hh,Gromov:2015dfa,Grozin:2015kna} and references therein) in maximally supersymmetric Yang Mills theory.
However, very little is known about the cusp function for general critical points, such as the 3d Ising model. In fact, from the general theory we develop here we also find some new applications for the extensively studied case of maximally supersymmetric Yang Mills theory. 

The main contributions of this work are the following:
\begin{enumerate}
\item First, in Part \ref{Part1} we study the general properties of the cusp anomalous dimension, as well as its relation to other observables. We test our findings in several examples, ranging from free theories to exact results in $d=2$. We also speculate on some implications of our results for the fusion algebra of Wilson lines in maximally supersymmetric Yang Mills theory in four dimensions.
\item In Part \ref{Part2} we study the cusp anomalous dimension in a concrete experimentally relevant setup: the pinning field defect in the $O(N)$ model. Most notably, we obtain nonperturbative results in the $3d$ Ising model using the recently developed fuzzy-sphere technique. We also obtain perturbative results for the cusp anomalous dimension in the general $O(N)$ model within the $\varepsilon$-expansion.
Particularly important quantities that we focus on are the Casimir energies and fusion rules of the pinning field impurities in the $O(N)$ model.
\end{enumerate} 
In the next two sections we provide a detailed summary of the paper.

\subsection{Summary of General Properties of the Cusp}\label{GenPropInt}

\begin{figure}[t]
\begin{center}
\includegraphics[scale=0.35]{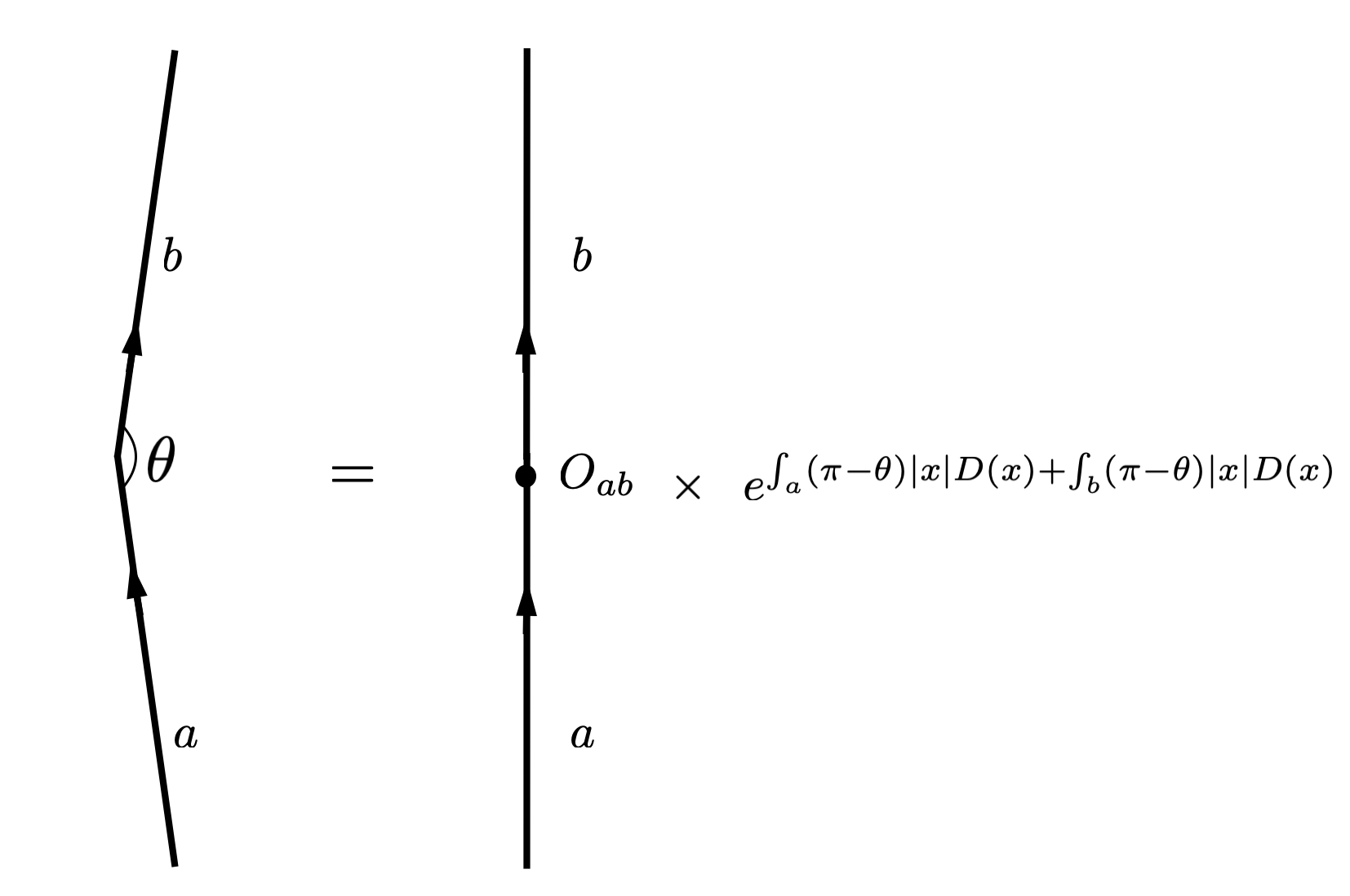}\caption{The cusp anomalous dimension near $\theta=\pi$ is computed perturbing the straight line with the displacement operator $D$.}\label{CuspExp}
\end{center}
\end{figure}

At $\theta\to\pi$ the cusp is very mild and one has an expansion around the perfectly straight infinite line using the displacement operator $D$, as in figure~\ref{CuspExp}. At zeroth order in $\pi-\theta$ the cusp is due to the scaling dimension of the defect changing operator $O_{ab}$ while at the next order there is a contribution from the connected four-point function $\langle O_{ab} D O_{ab} D\rangle$ (with the displacement operators integrated). We therefore obtain   
\begin{equation}\label{cuspexp}
\Gamma_{ab}(\theta)= \Delta_{ab}+\frac12\beta (\theta-\pi)^2+\cdots~,
\end{equation}
where $\Delta_{ab}>0$ the scaling dimension of the defect-changing operator $O_{ab}$ and $\beta$ must be computed from the integrated  four-point function.\footnote{For conformal defects invariant under transverse rotations and scalar $O_{ab}$ the three point function $\langle O_{ab} D O_{ab} \rangle=0$ and hence a linear term in~\eqref{cuspexp} is not allowed.}
In the special case where $b=a$ the defect changing operator is trivial $O_{aa}=1$ and then $\Delta_{aa} =0$ and the four-point function reduces to a two-point function and one has $\beta<0$ by reflection positivity. 
In the general case, $\beta$ is not sign definite as we will later see.

\begin{figure}[t]
\begin{center}
\includegraphics[scale=0.35]{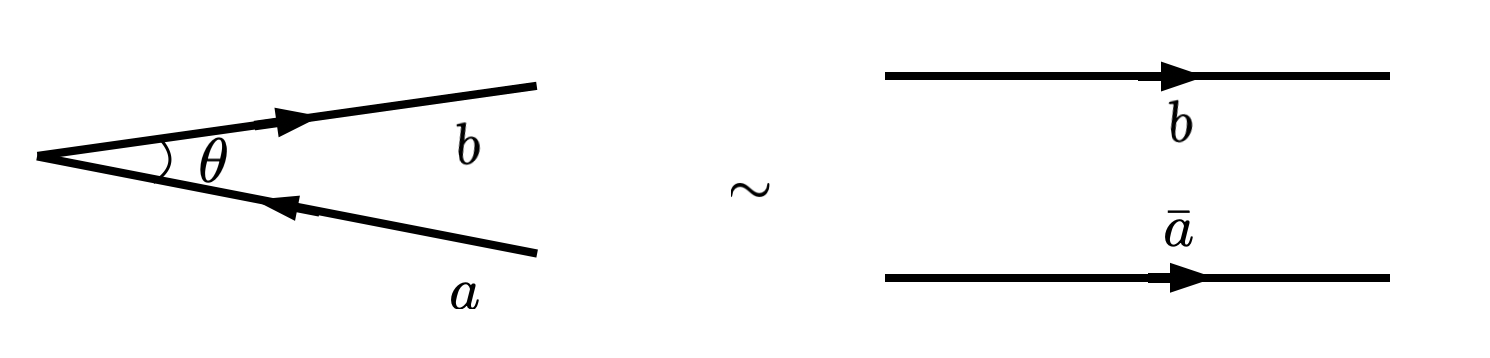}\caption{The cusp anomalous dimension for angle $\theta\ll 1$ admits an expansion fixed by the fusion of the half-lines $\bar{a}$ and $b$.}\label{CuspFuse}
\end{center}
\end{figure}

At $\theta\to0$ the cusp $\Gamma_{ab}$ is very sharp and we reinterpret the expansion as the fusion of the half-lines $a,\bar b$. This is schematically drawn in figure~\ref{CuspFuse}. 

We therefore consider the theory of fusion of conformal defects and find the following general form in the $\theta\to0 $ limit of the cusp anomalous dimension
\begin{equation}\label{cuspint}
\Gamma_{ab}(\theta) =  {C_{a\bar bc}\over \theta} +\Delta_{c1}+\alpha \theta^{\Delta_{irr}-1} +\cdots~,
\end{equation}
with $c$ the leading fused conformal line, $\Delta_{c1}>0$ the scaling dimension of the defect-creating operator, and $\Delta_{irr}>1$ the scaling dimension of an irrelevant operator on $c$. The coefficient ${C_{a\bar bc}\over \theta}$ is the energy one pays to fuse the conformal lines $a$ and $\bar b$ to $c$. If $a$ and $\bar b$ are attractive then $C_{a\bar bc}<0$ and otherwise it is positive. In flat space, the Casimir energy of the defects $a,\bar b$ separated by distance $r$ is given in terms of  $C_{a\bar bc}$ as 
\begin{equation}\label{CasimirFlat}
E(r)_{\mathbb{R}^{d-1}\times {\rm time}} = {C_{a\bar b c} \over r}~.
\end{equation}

In the special case $b=a$ we can further show that
\begin{equation}\label{convexity}
\Gamma_{aa}(\theta)<0~,\qquad \Gamma'_{aa}(\theta)>0~,\qquad  \Gamma''_{aa}(\theta)<0
\end{equation} 
for all $\theta\in(0,\pi)$ (for nontrivial conformal lines $a$)
which in particular implies that $C_{a\bar a c}<0 $ (and also implies $\beta<0$ in~\eqref{cuspexp} in the case $b=a$ as we already stated). 
Due to~\eqref{CasimirFlat} we can say that $C_{a\bar ac}<0$ implies that ``opposites attract'', as formerly proven in \cite{Bachas:2006ti}. (A theorem in the same spirit is also known for dielectrics, although the method of proof is very different~\cite{kenneth2006opposites}.) Indeed for Wilson lines in gauge theories the line $c$ is typically trivial and the coefficient $C_{a\bar a1}$ is just the coefficient in front of $1/r$ in the Coulomb potential. In this sense~\eqref{convexity} generalizes the fact that opposite charges attract to general conformal defects.

Since the scaling dimension $\Delta_{c1}$ appears in~\eqref{cuspint}, the cusp anomalous dimension serves as an interesting probe of the fusion rules in the theory. Indeed $\Delta_{c1}=0$ if (and most likely only if) the fused defect is trivial. Hence, studying this $\theta^0$ term of the cusp can teach us about when the fused defect is transparent. 
We will see how we can learn about the fusion algebra from this $\theta^0$ term of the cusp in several concrete examples, including the maximally supersymmetric Yang-Mills theory in the planar limit, some simple conformal lines in free theory, and the $2d$ Ising model. 
In the case of supersymmetric Yang-Mills theory, we find some properties of the fusion of BPS Wilson lines that appears to be new, and hint at a discontinuity in the fusion algebra between Wilson lines between weak and strong coupling. We speculate on some possible explanations in the main text.

\subsection{Summary of Results for the Pinning Field Conformal Defect }

The pinning field defect is a localized external field that couples to the order parameter. Usually we consider the response of a system to a uniform external field coupling to the order parameter (for instance, in order to compute the susceptibility). By contrast, in the case of the pinning field line defect we consider the response to a spatially localized external field. Such a localized external field flows to a conformal defect in many examples. This type of defects were studied recently quite extensively, see~\cite{vojta2000quantum,Sachdev:2003yk,PhysRevLett.96.036601,Assaad:2013xua,2017PhRvB..95a4401P,2014arXiv1412.3449A,Cuomo:2021kfm,Gimenez-Grau:2022ebb,Barrat:2023ivo,Franchi:2022rkx,Rodriguez-Gomez:2022gbz,Popov:2022nfq,Gimenez-Grau:2022czc,Rodriguez-Gomez:2022gif,Giombi:2022vnz,Bianchi:2022sbz,Nishioka:2022qmj,Pannell:2023pwz,Hu:2023ghk,Zhou:2023fqu,Dey:2024ilw} for a partial list of references studying various aspects of the pinning field defect. 

The quintessential example of the pinning field defect is in the $O(N)$ models, where we consider the line operator 
\begin{equation}\label{eq_pinning_defect_def}
e^{-\vec h\cdot \int dt\vec\phi(t,\vec x=0)}~,
\end{equation}
where $\vec\phi$ is the vector of order parameters and $\vec h$ is an external localized magnetic field.
This becomes a nontrivial conformal defect for $2\leq d\leq 4$ space-time dimensions.\footnote{The case of $N>1$ in $d=2$ requires a separate discussion which we will not delve into here.} For the pinning field defect, changing the orientation of the contour does not alter the magnetic field; thus, $\overline{\vec{h}} = \vec{h}$.

First, we study the pinning field in the $N=1$ case, i.e. the $3d$ Ising model, with the recent fuzzy sphere technique \cite{Zhu:2022gjc,Hu:2023xak,Han:2023yyb,Zhou:2023qfi,Han:2023O3,Hu:2024pen}. This is a novel numerical method to study critical phenomena in $2+1$ dimensions. It involves realizing the $3d$ Ising model using electrons in the lowest Landau level on $S^2$, and exploits the state-operator correspondence to access the CFT data of the theory. The fuzzy sphere method was recently used to realize the pinning field DCFT and measure several interesting defect data in \cite{Hu:2023ghk,Zhou:2023fqu}.

In this work we generalize the setup of \cite{Hu:2023ghk,Zhou:2023fqu} to study the cusp anomalous dimension. In the Ising model, the magnetic field is a scalar and there are two nontrivial pinning field defects corresponding to the sign of the coupling in eq.~\eqref{eq_pinning_defect_def}. Hence, we have two possible cusp functions to consider, $\Gamma_{++}(\theta)$ and $\Gamma_{+-}(\theta)$. In figures~\ref{fig:cusp++},~\ref{fig:cusp++_casimir} and~\ref{fig:cusp+-} we summarize our nonperturbative fuzzy sphere results for these quantities. The result for $\Gamma_{++}$ is manifestly compatible with the properties~\eqref{convexity}. Additionally, fitting our results at small $\theta$ with the expansion~\eqref{cuspint} we obtain the following estimates for the Casimir energies $C_{+++}$ and $C_{+-1}$ between, respectively, aligned and anti-aligned pinning field defects:
\begin{equation}
   C_{+++}=-0.29(2)\,,\qquad
    C_{+-1}=1.4(2)\,.
\end{equation}

We also study the cusp anomalous dimension at one-loop in the $\varepsilon$-expansion in the general $O(N)$ model. For $N>1$ there is a manifold of fixed point for the defect~\eqref{eq_pinning_defect_def}, parametrized by a unit vector $\hat{m}$ as $\vec{h}=h^*\hat{m}$, where $h^*$ is given in eq.~\eqref{eq_h_fix} in the main text. The cusp anomalous dimension $\Gamma_{\vec{h}_1^*\vec{h}_2^*}(\theta)$ thus depends both on the angle $\theta$ and the relative orientation $\hat{m}_1\cdot\hat{m}_2$ in the internal $O(N)$ space.

We use our results to discuss the general properties of the cusp discussed before. In particular, our result provide a stringent test of the structure~\eqref{cuspint} of the small angle expansion, since the fusion between defects depends in a nontrivial way on $\hat{m}_1\cdot\hat{m}_2$. Additionally, the fusion algebra is discontinuous when $\varepsilon\to0$ since in $d=4$ the magnetic field is an exactly marginal parameter and it is thus additive in the fusion, while in $4-\varepsilon$ dimensions the magnetic field of the fused line flows to the fixed point value. Yet, we show that the cusp function is perfectly continuous thanks to the existence of a slightly irrelevant operator in $4-\varepsilon$ dimensions.

Finally, we combine exact results in $d=2$ with our $\varepsilon$-expansion calculation to construct an approximation for the cusp anomalous dimension $\Gamma_{++}(\theta)$ in the $3d$ Ising model; the comparison between this extrapolation and the fuzzy-sphere result is shown fig.~\ref{fig:Pade1}. We argue that the cusp between anti-aligned magnetic defects instead cannot vary continuously between $d=4$ and $d=2$, and thus we refrain from constructing a similar extrapolation for $\Gamma_{+-}(\theta)$ at arbitrary values of $\theta$. We nonetheless combine our results in $4-\varepsilon$ dimensions and $d=2$ to obtain analytical extrapolations to $d=3$ for the scaling dimension of the defect changing operator and the Casimir energy $C_{+-1}$. These extrapolations are summarized in table~\ref{tab:Pade}, and are in excellent agreement with the numerical fuzzy sphere results,

The rest of the paper is organized as follows. In section~\ref{sec_general_cusp} we study the fusion of conformal defects and derive the small angle expansion of the cusp anomalous dimension. We discuss the concavity property and the other positivity properties we reviewed above. In section~\ref{sec_examples} we discuss various examples where we test the results of section~\ref{sec_general_cusp}. In section~\ref{sec_fuzzy} we discuss the fuzzy sphere computation of the cusp anomalous dimension in the 3d Ising model, and in sec.~\ref{sec_epsilon} we discuss the $\varepsilon$-expansion calculation and compare the results. Some technical results are in collected in an appendix. 

\medskip
\noindent
\textbf{Note added:} As we were finalizing this work, we learned of two related works \cite{Diatlyk:2024qpr} and \cite{Kravchuk:2024qoh}. These works have some overlap with our discussion in part~\ref{Part1}. In particular, they provide independent proofs that the Casimir energy between defects $a$ and $\bar{a}$ is negative, and analyze the effective field theory of fusion; \cite{Kravchuk:2024qoh} further discusses the cusp anomalous dimension at small angle in general and in $\mathcal{N}=4$ SYM. We are grateful to the authors of \cite{Diatlyk:2024qpr} and \cite{Kravchuk:2024qoh} for sharing their manuscripts with us. We are also aware of an upcoming work~\cite{Max} whose results have some overlap with the ones in sec.~\ref{sec_epsilon} of this work.

\part{General properties of the cusp}\label{Part1}

\section{The cusp anomalous dimension: Definition and Properties}\label{sec_general_cusp}

\subsection{Proper Definition of the Cusp Anomalous Dimension on the cylinder}\label{subsecdef}

As we mentioned in the introduction, we can equivalently define the cusp anomalous dimension both in flat space and on the cylinder $\mathds{R}\times S^{d-1}$. In flat space, the cusp anomalous dimension is the coefficient of the logarithmically divergent term in the partition function with the defects as in fig.~\ref{CuspRaw}:
\begin{equation}\label{eq_cusp_flat}
    \log \frac{Z_{ab}}{Z_{CFT}}=-\Gamma_{ab}(\theta)\log\left(\frac{L}{a}\right)+\ldots\,,
\end{equation}
where we normalized by the CFT partition function and the dots stand for non logarithmically divergent terms (which generically include a cosmological constant contribution $\propto L$). We can think of the cusp as a local operator transforming in a linear representation of the $\mathds{R}$ group of dilations, hence justifying the characterization of $\Gamma_{ab}(\theta)$ as \emph{anomalous dimension} (more precisely, the dependence on the IR scale $L$ is fixed by the scaling dimension as usual, while the cutoff scale factor can be understood as a wave-function normalization factor for the operator).

As mentioned below eq.~\eqref{cuspdef}, by a Weyl rescaling 
$\Gamma_{ab}(\theta)$ is the ground state energy on $S^{d-1}$ in the presence of the defects $a$ and $b$ separated by angle $\theta$. However, a proper subtraction of the ambiguous worldline masses of $a$ and $b$ is required to define the ground state energy. The worldline mass is the only ambiguity in defining a conformal line defect in the continuum. This counter-term rescales the defect as $a_i\to e^{-m_i\int dt}a_i$. It is a c-number which affects very little of the actual infrared physics but it is important that physical observables should be independent of it. 

This subtraction is done by observing that a straight conformal line $a$, which has no cusp anomalous dimension, maps to two defects, $a,\bar a$ separated by $\theta=\pi$ on $S^{d-1}$ (whose radius we take to be $R=1$). Therefore we can divide by the corresponding partition function to subtract the $a$ worldline mass and similarly for $b$. 
\begin{equation}\label{eq_cusp_def_cyl}
-\log\left(\frac{Z_{ab}(\theta)}{\sqrt{Z_{aa}(\pi)Z_{bb}(\pi)}}\right)=\Gamma_{ab}(\theta)T~.
\end{equation}
where $T$ is the Euclidean time and we take $T\to\infty$ to project on the ground states, therefore, we can equivalently write  
\begin{equation}\label{eq_cusp_def_cyli}
E_{ab}(\theta)-\frac12E_{aa}(\pi)-\frac12E_{bb}(\pi)\equiv\Gamma_{ab}(\theta)~.
\end{equation}
Note that we work in conventions such that $Z_{ab}$ corresponds to the cylinder partition function with the insertion of defects $a$ and $\bar{b}$ as in fig.~\ref{CuspFuse}.

Also in the following it will be convenient to go back and forth between the flat space and the cylinder definition on the cusp, depending on the application we are interested in.

\subsection{\texorpdfstring{$\Gamma''_{aa}<0$}{Gamma''<0}}

Here we give a simple proof (that relies on unitarity and locality only) that in the special case $a=b$ the cusp function is concave.

\begin{figure}[t]
\begin{center}
\includegraphics[scale=0.35]{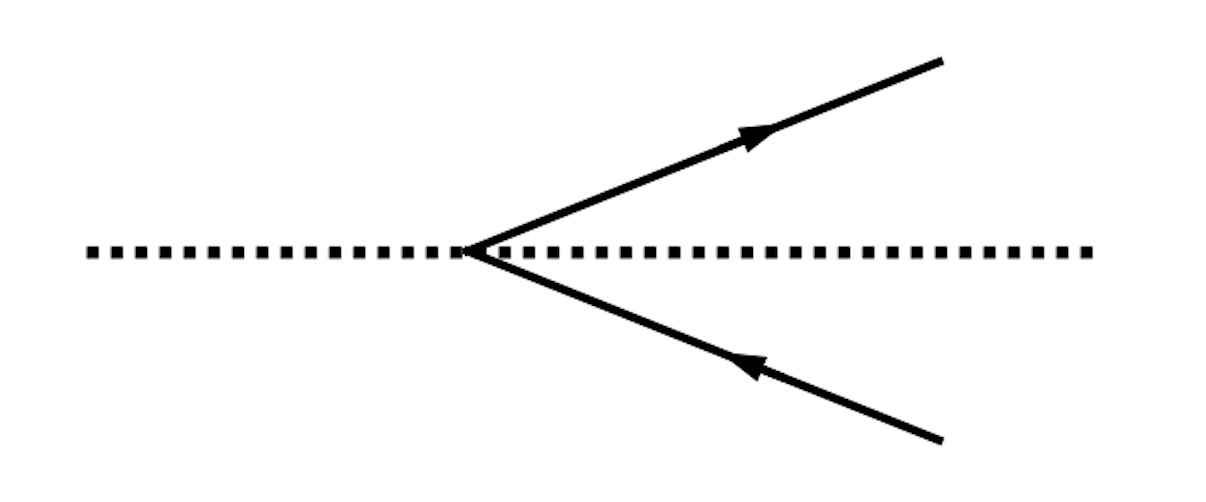}\caption{The cusp with the reflection plane used in the argument for concavity.}\label{cuspref}
\end{center}
\end{figure}

We start from the cusped flat space configuration in Euclidean signature in figure~\ref{cuspref}. We can change the angle of the cusp by inserting the integrated displacement operator on the bottom edge or the top edge. We can compute the second derivative with respect to the angle by 
\begin{equation}\label{Ward}
{d^2\over d\theta^2} \log Z_{ab} =\int d\gamma_bd\gamma_t \gamma_b\gamma_t \langle D_n(\gamma_b)D_n(\gamma_t)\rangle_c~, \end{equation} 
where $\gamma_{b,t}$ are proper positive parameters on the bottom (top) edges, respectively, and $D_n\equiv n^iD_i$ where $n^i$ is the normal to the contour.
In the above formula we used the fact that changing the angle by $\delta\theta$ changes the edge in the normal direction by $\gamma_{b,t}\delta\theta$. 
We could have placed both displacement operators on the same edge but then the answer is not manifestly convergent and subtractions are required. By placing the displacement operators on different edges we will now see that there are only mild divergences in accordance with what we expect. The subscript $c$ in~\eqref{Ward} stands for the fact that we need the connected two-point function. 

Since the cusp preserves scaling symmetry, and since the displacement operator has dimension 2, $\langle D_n(\gamma_b)D_n(\gamma_t)\rangle_c = {1\over \gamma_b^2\gamma_t^2}F(\gamma_b/\gamma_t)$. The function $F(\gamma_b/\gamma_t)$ cannot be further constrained by using the symmetries only.
Plugging the form of the two point function back into~\eqref{Ward} we find a logarithmically divergent integral over the product $\gamma_b\gamma_t$ and thus 
\begin{equation}\label{Wardi} {d^2\over d\theta^2} \log Z_{ab} =\log(L\over a) \int dx \,x^{-1} F(x)~, \end{equation} 
where $x=\gamma_b/\gamma_t$.

To show that~\eqref{Wardi} is convergent we need to discuss the limit of $x\to 0$ (which is similar to $x\to\infty$) of $F(x)$. This is best interpreted in radial quantization as the insertion of two very far apart (separated in time) displacement operators on the two defects. Then we can replace the configuration created by the displacement operator in the past as a sum over states. This is just the OPE of the displacement operator with the cusp.
Since we are interested in the connected two-point function we must project out the ground state from the displacement-cusp OPE.
Therefore we find \begin{equation}\label{smallx} x\to 0:\qquad
F(x)\sim x^{\Delta_{min}}~, \end{equation} with $\Delta_{min}>0$ the energy gap to the first excited state of the cusp. 
Similarly, as $x\to\infty$, $F(x)\sim x^{-\Delta_{min}}$.
This is sufficient for the convergence of~\eqref{Wardi}.

So far our discussion was for general defects $a,b$.  In the special case $a=b$ we can use reflection positivity, which is evident in figure~\ref{cuspref}.\footnote{For this, it is important that the configuration with a cusped defect can be prepared evolving the system with a (Euclidean) local Hamiltonian (more precisely, the cusp should be replaced by a smooth configuration over a short distance scale $a$). Alternatively, reflection positivity becomes manifest if we perform a conformal transformation that leaves the plane in fig.~\ref{cuspref} fixed and maps the point at infinity at finite distance from the origin, so that the defect is now given by two arcs that join the two cusps.} 
Reflection positivity is usually quoted as a property of the full two-point function, but it is also a property of the connected two-point function, which is what we need here. 
Therefore we find that ${d^2\over d\theta^2} \log Z_{aa}=-\log\left(\frac{L}{a}\right){d^2\over d\theta^2} \Gamma_{aa} >0$, as claimed.

Note that, since $\Gamma_{aa}(\theta=\pi)=\Gamma_{aa}'(\theta=\pi)=0$ (as we will see in the next section), it follows from the concavity condition that 
\begin{equation}
\Gamma_{aa}(\theta) < 0~,\qquad \Gamma'_{aa}(\theta)> 0~
,\end{equation} 
for all $\theta\in (0,\pi)$ and for all nontrivial conformal lines $a$. 

\subsection{The \texorpdfstring{$\theta\to\pi$ and $\theta\to0$}{theta to Pi and theta to 0} Limits of the Cusp Anomalous Dimension}\label{subsec_limits}

Let us begin with the $\theta\to\pi$ limit in which the cusp is almost straight. This limit is well studied, e.g. it was discussed in~\cite{Correa:2012at} for $1/2$ BPS Wilson lines and in \cite{Bianchi:2015liz,Wang:2021lmb} for monodromy defects. We review the results below for completeness.

We work in flat space. The expansion around that point proceeds in accordance with eq.~\eqref{Wardi}.  As we argued in~\eqref{cuspexp} the linear term is absent and hence the first term arises from the two point function of the displacement operator in the presence of the defect changing operators $\mO_{ab}$, as in figure~\ref{CuspExp}. In general this four point function is complicated and the result is not sign-definite. However for $a=b$ the defect changing-operator is trivial $\mO_{aa}=1$ and the two-point function of the displacement operator is fixed by conformal symmetry to be
\begin{equation}\label{eq_DD_2pt}
\langle D_i(\tau)D_j(0)\rangle_c=\frac{C_D\delta_{ij}}{\tau^4}~,
\end{equation}
which we can plug into~\eqref{Ward} to obtain
\begin{equation}
-\Gamma_{a a}''(\theta)\log\left(\frac{L}{a}\right)
=\int_0^\infty d\tau_1\int_0^\infty d\tau_2\frac{C_D\tau_1\tau_2}{(\tau_1+\tau_2)^4}=\frac{C_D}{6}\int_0^\infty \frac{d\tau}{\tau}\,.
\end{equation}
Therefore,
\begin{equation}\label{eq_cusp_CD}
\Gamma_{a a}''(\theta)\vert_{\theta=\pi}=-\frac{C_D}{6}\,~,
\end{equation}
in accordance with the expectation for a concave cusp anomalous dimension. Note that here we have $F(x) \sim \frac{1}{(x^{1/2}+x^{-1/2})^4} $
and the behavior at small $x$ is $F(x)\sim x^2$, which simply corresponds to the state associated with the displacement operator in accordance with~\eqref{smallx}.

Now let us discuss the small $\theta$ limit. As suggested in figure~\ref{CuspFuse}, we have to understand in detail the fusion of conformal defects (see~\cite{Bachas:2007td,Bachas:2013ora,Konechny:2015qla} for previous works of defect fusion in $d=2$ and~\cite{Diatlyk:2024zkk} for a recent discussion in $d>2$). So let us consider two conformal impurities on $\mathbb{R}^d$, $a_i,a_j$. 
We bring the defects distance $r$ apart and perform the renormalization group.
At long distances (much larger than $r$) we ought to get a sum over conformal impurities $a_k$: 
\begin{equation}\label{fusionimp} 
a_ia_j=\bigoplus_k e^{-\int dt {C_{ijk}\over r}}a_k~.
\end{equation} 
As explained in subsection~\ref{subsecdef} the conformal lines have an ambiguity $a_i\to e^{-m_i\int dt}a_i$. The meaning of the exponential term $e^{-\int dt {C_{ijk}\over r}}$ is that it leads to a physical (scheme independent) shift of the worldline mass of the fused line. Therefore, if we fuse lines with wordline masses $m_i,m_j$, the worldline mass of the fused line is $m_i+m_j+{C_{ijk}\over r}$. We can think of the $C_{ijk}$ as generalized Casimir energies: $C_{ijk}$ is the energy cost for the two impurities $a_i$, $a_j$ to become $a_k$. Of course, in the limit of small $r$ only the line(s) with the smallest $C_{ijk}$ should be retained, while the other conformal lines' contributions are exponentially small. (We keep the exponentially small contributions for completeness, and we will discuss power suppressed contributions below.)

Some comments are in order: 
\begin{itemize} 
\item Unlike the OPE of local operators, we see that in~\eqref{fusionimp} we find exponential singularities. 
\item In general, conformal defects $a$ and $\bar a$ do not necessarily contain the unit operator (transparent defect) in their fusion. This is unlike for topological defects, and unlike many cases where we fuse Wilson lines.
\item By genericity arguments, we expect that all possible operators will be generated on the conformal lines appearing in the fusion. These include off-diagonal operators too (off diagonal operators are also called defect-changing, or domain wall, operators - see \cite{Gabai:2022vri,Gabai:2022mya,Gabai:2023lax,Nagar:2024mjz} for some recent applications). Therefore we expect that the lines appearing in the sum~\eqref{fusionimp} are stable under defect RG. Of course there may be special situations where some multi-critical lines appear on the right hand side of~\eqref{fusionimp}, but we do not expect it in general. If we additionally assume that there are only finitely many conformal line defects with no relevant operators (as it is believed to be the case in many models), we are led to speculate that the right hand side of~\eqref{fusionimp} is a finite sum, unlike the usual OPE of local operators. 
\item In the infrared only the line with the smallest Casimir energy should be retained. Therefore,  we expect that the result of the fusion is a simple line, i.e. it cannot be written as a sum of multiple defects, once we drop sub-exponential contributions.\footnote{In the degenerate case where multiple lines have the same Casimir energy, the infrared line is determined by the defect changing operators.} 
\item There are power corrections in $r$ in addition to the exponentially small corrections due to conformal lines with non-minimal $C_{ijk}$. These arise from irrelevant operators that perturb the stable lines appearing in eq.~\eqref{fusionimp} via terms of the form $\sim r^{\Delta_{irr}-1}\int dt \ O_{irr}dt$.  In~\eqref{fusionimp} we have taken the infrared limit so that all the irrelevant operators on the lines on the right hand side of~\eqref{fusionimp} have decayed away.
\end{itemize}

To use the relation~\eqref{eq_cusp_def_cyli} to compute the cusp anomalous dimension at small $\theta$, we briefly discuss the generalization of fusion to impurities on the cylinder.
In general we expect that eq.~\eqref{fusionimp} holds in the same form with the distance $r$ now identified with the geodesic distance between the defects, so that $r\simeq \theta+O(\theta^3)$.\footnote{Note that different notions of distance, such as the chordal distance, differ in the form of the subleading terms; we expect that it is possible to switch from one notion of distance to another upon modifying the coefficients of the couplings to the curvature operators $\sim\mathcal{R}^n$ on the worldline.} Additionally, on the cylinder we generically expect that curvature terms of the form $r^{n-1}\int dt \mathcal{R}^n$ will be generated among the irrelevant operators.

From this discussion it is clear that the ground state energy~\eqref{eq_cusp_def_cyl} is dominated by the conformal line $c$ with the smallest $C_{a\bar bc}$ coefficient. Power-law corrections arise from the least irrelevant operator on $c$ with scaling dimension $\Delta_{irr}$. An important fact is that since we are now carrying out the fusion on $S^{d-1}$, the partition function of $c$ itself is given by the scaling dimension of the end-point operator $\Delta_{c1}$. Therefore we obtain the general ansatz
\begin{equation}\label{cusp}
\Gamma_{ab} =  {C_{a\bar bc}\over \theta} +\Delta_{c1}+\alpha \theta^{\Delta_{irr}-1} +\cdots~,
\end{equation}
where $\alpha$ depends on the product of coefficient of the least irrelevant opertor and its one-point function. The relation between the Casimir energy and the cusp anomalous dimension was previously discussed for Wilson lines in the context of $\mathcal{N}=4$ SYM, see e.g. \cite{Correa:2012hh}.
Unitarity requires $\Delta_{c1}\geq 0$. For $b=a$ from concavity it follows that 
\begin{equation}
C_{a\bar a c}\leq  0 ~,
\end{equation} 
which is a generalization of the concept that ``opposites attract.''\footnote{The method we use applies more generally for the Casimir force of reflection-symmetric geometries in a large class of QFTs.} 

Corrections to eq.~\eqref{cusp} arise from further irrelevant operators and from higher order in conformal perturbation theory. In general, we always expect a $\sim\theta$ correction to~\eqref{cusp} from the expansion of the geodesic distance $r= \theta+O(\theta^3)$ and from the curvature operator $\mathcal{R}$ on the line.

\section{Examples}\label{sec_examples}

\subsection{Free fields}\label{subsec_free}

Let us consider the cusp between two pinning field defects~\eqref{eq_pinning_defect_def} in the theory of $N$ free scalars in four dimensions:
\begin{equation}
S=\int d^4x\frac12(\pd\vec{\phi})^2\,.
\end{equation}
As explained in the introduction, the pinning field defect is defined by the expression,
\begin{equation}
    \mathcal{D}_{\vec{h}}=\exp\left[-\int d\tau\,\vec{h}\cdot\vec{\phi}(x(\tau))\right]
\end{equation}
where $x(\tau)$ is the defect contour and $\vec{h}$ is an exact marginal parameter. The insertion of a pinning field defect breaks the $O(N)$ symmetry group to $O(N-1)$ (hence it fully breaks the internal $\mathds{Z}_2$ symmetry for $N=1$).

\begin{figure}[t]
   \centering
		\subcaptionbox{  \label{fig:tree1}}
		{\includegraphics[width=0.2\textwidth]{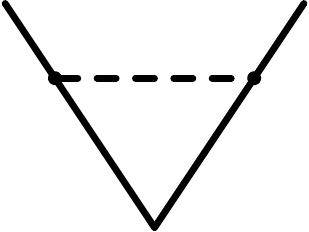}}
	\hspace*{3cm}
		\subcaptionbox{ \label{fig:tree2}}
		{\includegraphics[width=0.2\textwidth]{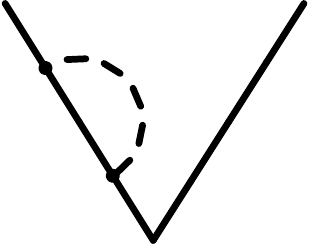}}
        \caption{Diagrams contributing the cusp anomalous dimension at tree-level. The plain black lines denote the defects, the dashed line is the scalar propagator and  the dots represent insertions of the defect coupling $\vec{h}$. }
\label{fig:Diagramstree}
\end{figure}

Let us now consider the cusp between two pinning field defects $\vec{h}_1$ and $\vec{h}_2$ at angle $\theta$. 
It is simplest to work in flat space, where the free theory propagator reads
\begin{equation}
  \langle \phi_i(x)\phi_j(0)\rangle
  =
  \frac{\delta_{ij}}{4\pi^2x^2}\,.
\end{equation}
The diagrams contributing to the result are shown in fig.~\ref{fig:Diagramstree}.
Summing them, we obtain:
\begin{equation}
\begin{split}
\log Z_{\vec{h}_1\vec{h}_2} &=
\frac{\vec{h}_1\cdot\vec{h}_2}{4\pi^2}\int_0^\infty d\tau_1\int_0^\infty d\tau_2\frac{1}{\left(\tau_1-\tau_2\cos\theta\right)^2+\tau_2^2\sin^2\theta}\\[0.5em]
&+\frac{\vec{h}_1^2+\vec{h}_2^2}{8\pi^2}
\int_0^\infty d\tau_1\int_0^\infty d\tau_2\frac{1}{|\tau_1-\tau_2|^2}\,.
\end{split} 
\end{equation} 
This expression contains a cutoff divergence from the region where $\tau_1\rightarrow\tau_2$ in the second integral, corresponding to the self-energy diagram in fig.~\ref{fig:tree1}. This divergence can be reabsorbed in a cosmological constant counterterm on the line $\sim a^{-1}\int d\tau$. Neglecting this term, the result takes the form
\begin{equation}
\log Z_{\vec{h}_1\vec{h}_2} =
\left[\frac{\vec{h}_1\cdot\vec{h}_2(\pi-\theta)}{4\pi^2\sin\theta}-\frac{\vec{h}_1^2+\vec{h}_2^2}{8\pi^2}\right]\int_0^\infty \frac{d\tau}{\tau}+\text{cosmological constant}\,.
\end{equation}
Cutting off the integral with a short distance regulator $a$ and an IR cutoff $L$, we see that the result takes the expected form~\eqref{cusp} and we obtain the cusp anomalous dimension
\begin{equation}\label{eq_tree_level}
\Gamma_{\vec h_1\vec h_2}(\theta)=-\frac{\vec{h}_1\cdot\vec{h}_2}{4\pi^2}\frac{\pi-\theta}{\sin\theta}+\frac{\vec{h}_1^2+\vec{h}_2^2}{8\pi^2}\,.
\end{equation}

Let us now analyze in detail the result~\eqref{eq_tree_level}. Consider first the case $\vec{h}_1=\vec{h}_2$. The result~\eqref{eq_tree_level} is manifestly concave.  It is also simple to check the validity of eq.~\eqref{eq_cusp_CD}. The displacement operator is given by $D_i=\vec{h}\cdot\pd_i\vec{\phi}$, and therefore the tree-level two-point function gives
\begin{equation}
C_D=\frac{\vec{h}^2}{2\pi^2}\,.
\end{equation}
Then the second derivative of eq.~\eqref{eq_tree_level} gives
\begin{equation}
\Gamma_{\vec h_1\vec h_1}''(\theta)\vert_{\theta=\pi}=-\frac{\vec{h}^2}{12 \pi ^2}=-\frac{C_D}{6}\,,
\end{equation}
in agreement with eq.~\eqref{eq_cusp_CD}.

For $\theta=\pi$ and $\vec{h}_1\neq\vec{h}_2$, the result~\eqref{eq_tree_level} yields the scaling dimension of the defect changing operator between the impurities with field $\vec{h}_1$ and $\vec{h}_2$:
\begin{equation}\label{eq_tree_level_Delta_creation}
\Delta_{\vec h_1\vec h_2}=
\Gamma_{\vec h_1\vec h_2}(\pi)
=\frac{(\vec{h}_1-\vec{h}_2)^2}{8\pi^2}\,.
\end{equation}
In particular for $\vec{h}_2=0$, eq.~\eqref{eq_tree_level_Delta_creation} gives the scaling dimension of the defect creation operator.

Finally we discuss the small angle limit of eq.~\eqref{eq_tree_level}:
\begin{equation}\label{eq_tree_level_small}
\Gamma_{\vec h_1\vec h_2}(\theta)=-\frac{\vec{h}_1\cdot\vec{h}_2}{4\pi \theta}+
\frac{(\vec{h}_1+\vec{h}_2)^2}{8\pi^2}-\frac{\vec{h}_1\cdot\vec{h}_2}{24\pi}\theta+\ldots\,\,.
\end{equation}
This expression is consistent with the prediction from fusion, that in this case results in a defect with pinning field $\vec{h}_1+\vec{h}_2$. The first term is the Casimir energy. The second term is the scaling dimension of the creation operator of the defect with pinning field $\vec{h}_1+\vec{h}_2$, in agreement with eq.~\eqref{eq_tree_level_Delta_creation}. 
In this example we clearly see how the $O(\theta^0)$ term illuminates the fusion algebra through the scaling dimension of the fused defect creation operator. 

Finally the $O(\theta)$ term arises both from the expansion of the geodesic distance in the Casimir energy and from the leading irrelevant deformations.\footnote{The latter include the Ricci scalar, which is nonzero on $\mathds{R}\times S^{d-1}$, and the operator $\vec{h}_1\pd_\theta\phi-\vec{h}_2\cdot\pd_\theta\vec{\phi}$, both of dimension $\Delta_{irr}=2$ in agreement with eq.~\eqref{cusp}.}

Let us briefly consider another trivial example: the cusped Wilson line in free Maxwell theory. We work in the normalization where the Euclidean action reads
\begin{equation}
 S=\frac{1}{4 e^2}\int d^4x F_{\mu\nu}^2 \,,
\end{equation}
so that a charge $q$ Wilson line is defined by
\begin{equation}
W_q=e^{i q \int d x^\mu A_\mu(x(\tau))}\,.
\end{equation}
Proceeding as before we find that the cusp anomalous dimension reads
\begin{equation}\label{eq_tree_level_gauge}
\Gamma_{qq}(\theta)=-\frac{e^2q^2}{4\pi^2}\left(\frac{ \pi-\theta}{\tan\theta}+1\right)\,.
\end{equation}
Again, the result is manifestly concave.
In the limit $\theta\rightarrow\pi$ eq.~\eqref{eq_tree_level_gauge} reads
\begin{equation}
    \Gamma_{qq}(\theta)=-\frac{e^2q^2}{12\pi^2}(\pi-\theta)^2+O\left((\pi-\theta)^4\right)\,.
\end{equation}
This result is consistent with eq.~\eqref{eq_cusp_CD}. Indeed the displacement operator is $D_{i }=q F_{i\tau}$ and thus we get
\begin{equation}
    C_D=\frac{e^2q^2}{\pi^2}\,.
\end{equation}
In the limit $\theta\rightarrow 0 $ we obtain instead
\begin{equation}\label{eq_tree_level_gauge_small}
    \Gamma_{qq}(\theta)=-\frac{e^2q^2}{4\pi^2\theta}+\frac{e^2q^2\theta}{12\pi}+O\left(\theta^2\right)\,.
\end{equation}
The first term of eq.~\eqref{eq_tree_level_gauge_small} is the well known Coulomb potential between two charges. The absence of a $\theta^0$ term is also consistent with the fusion. Indeed, since the orientation reversal of the Wilson line is a line of opposite charge $\overline{W}_q=W_{-q}$, the fusion results in a trivial line. Therefore the scaling dimension of the defect creation operator of the fused line vanishes. Finally, the subleading term is again due to the expansion of the geodesic distance in the first term and the first irrelevant deformations (that include the Ricci and the electromagnetic tensor).

We have not discussed a more general cusp anomalous dimension $\Gamma_{q_1q_2}$ for the following reason: The free Maxwell theory has an electric $U(1)$ one-form symmetry 
and hence defect changing operators of finite scaling dimension do not exist. However, in theories like 3d QED it would be interesting to consider $\Gamma_{q_1q_2}$ indeed.

\subsection{Exact results in 2d}\label{subsec_2d}

Line defects in $2d$ are interfaces. In general it is complicated to compute the cusp anomalous dimension of conformal 2d interfaces. But a special case is that of interfaces which are a tensor product of two boundaries, so there is no transmission. It was already noted long ago by Cardy and Peschel that the partition function of a cusped boundary in two dimensions can be computed exactly \cite{Cardy:1988tk}.
The result can be easily generalized to interfaces which are a product of two boundary conditions on opposite sides. We review these results below, and use them them to check the general properties of the cusp anomalous dimension that we discussed before.

Let us first review the case of two boundaries $a$ and $b$ in half space $\text{Im}z>0$ (in the usual complex coordinates $z=x+iy $) separated by a boundary changing operator at the origin.  The expectation value of the stress tensor in this geometry reads \cite{Cardy:2004hm}\footnote{Here we work in the standard $2d$ conventions where the stress tensor is normalized as
\begin{equation}
T_{\mu\nu}=2\pi\frac{\delta \log Z}{\delta g^{\mu\nu}}\,.
\end{equation}
}
\begin{equation}\label{eq_2d_T}
\langle T(z)\rangle=\frac{\Delta_{|a\rangle |b\rangle }}{z^2}\,,
\end{equation}
where $\Delta_{|a\rangle |b\rangle }$ is the scaling dimension of the boundary changing operator.  The partition function in the presence of a smooth cusp is then extracted using the conformal anomaly equation,
\begin{equation}
L\frac{\pd}{\pd L} \log Z_{BCFT}=-\frac{1}{2\pi}\int d^2x\langle T_\mu^\mu\rangle\quad\implies\quad
 \log Z_{BCFT}\vert_{\theta=\pi}=-\Delta_{|a\rangle |b\rangle }\log\left(\frac{L}{a}\right)\,,
\end{equation}
which is the obvious result. Here we used that from eq.~\eqref{eq_2d_T} it follows that
\begin{equation}\label{eq_int_2d}
\begin{split}
\int d^2 x\langle T_\mu^\mu\rangle
=
- i\oint dw \,w \langle T(w)\rangle+i\oint d\bar{w}\, \bar{w} \langle \bar{T}(\bar{w})\rangle
=2\pi \Delta_{|a\rangle |b\rangle }\,,
\end{split}
\end{equation}
which is equivalent to the standard Ward-identity $\langle T_\mu^\mu(x)\rangle=2\pi\Delta_{|a\rangle |b\rangle }\delta^2(x)$.

\begin{figure}[t]
   \centering
		\includegraphics[width=0.8\textwidth]{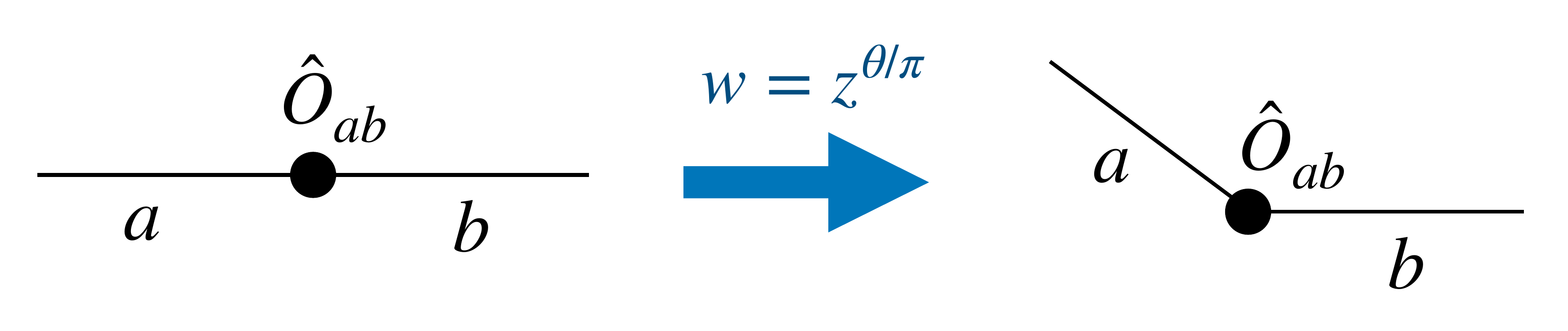}
        \caption{A conformal map relates a flat boundary and a cusped configuration. }
\label{fig:conformal_map2d}
\end{figure}

To obtain the partition function in a wedge with angle $\theta$, we perform the following conformal transformation
\begin{equation}
w=z^{\theta/\pi}\,,
\end{equation}
which is singular at the origin, see fig.~\ref{fig:conformal_map2d}. The expectation value of the stress tensor in the cusped geometry follows from the Schwartzian transformation
\begin{equation}
T(w)dw^2=T(z)dz^2-\frac{c}{12}\{w,z\}dz^2\,,\qquad
\{w,z\}=\frac{w'''w'-\frac32(w'')^2}{(w')^2}\,,
\end{equation}
where $c$ is the central charge of the CFT. We find
\begin{equation}
\langle T(w)\rangle=\frac{1}{w^2}
\left[\frac{c}{24}  \left(1-\frac{\pi ^2}{\theta ^2}\right)+\frac{\pi ^2 \Delta_{|a\rangle |b\rangle } }{\theta ^2}\right]\,.
\end{equation}
Using the conformal anomaly equation as before, we obtain \cite{Cardy:1988tk}:
\begin{equation}\label{eq_BCFT_2d}
\log Z_{BCFT}=-\frac{\theta}{\pi}\left[
\frac{c}{24}  \left(1-\frac{\pi ^2}{\theta ^2}\right)+\frac{\pi ^2 \Delta_{|a\rangle |b\rangle } }{\theta ^2}\right]
\log\frac{L}{a}\,.
\end{equation}
The overall factor of $\theta/\pi$ arises because the integration in eq.~\eqref{eq_int_2d} is now restricted to a wedge.

It is finally straightforward to generalize this result to interfaces which are the product of two boundary conditions on opposite sides. Therefore the interface $a$ is essentially $|a\rangle \langle a|$ and similarly for $b$. The interface changing operator obeys $\Delta_{ab} = 2\Delta_{|a\rangle |b\rangle }$.
To obtain the cusp anomalous dimension for interfaces we simply need to sum the result~\eqref{eq_BCFT_2d} with the contribution from the opposite side with angle $2\pi-\theta$. Explicitly, we find the cusp anomalous dimension
\begin{equation}\label{eq_cusp_2d}
\Gamma_{ab}(\theta)=\frac{2 \pi ^2 \Delta_{|a\rangle |b\rangle }}{(2 \pi -\theta ) \theta }-\frac{c (\pi -\theta )^2}{12 (2 \pi -\theta ) \theta }\,.
\end{equation}

It is now straightforward to see that the result~\eqref{eq_cusp_2d} obeys all the general properties that we discussed.
First, for $\Delta_{ab}=0$ eq.~\eqref{eq_cusp_2d} is concave as predicted. Additionally, from the $\theta\rightarrow 0$ expansion
\begin{equation}\label{eq_cusp2d_small}
\Gamma_{ab}(\theta)=\frac{\pi  \Delta_{|a\rangle |b\rangle } -\frac{\pi  c}{24}}{\theta }+\frac{c+8 \Delta_{|a\rangle |b\rangle }}{16} +O\left(\theta\right)\,,
\end{equation}
we extract the Casimir energy and the scaling dimension of the defect creation operator of the fused interface: $\frac{1}{16} (c+8 \Delta_{|a\rangle |b\rangle } )$. Note that the latter is always positive for $c>0$. 

As a nontrivial application,  let us discuss the result~\eqref{eq_cusp2d_small} for the pinning field in the Ising model.  The results below will be relevant in part~\ref{Part2} of this paper as well.

In the Ising model we have $c=1/2$ and the pinning field defect corresponds to setting equal Dirichlet boundary conditions on both sides of the interface.   We then obtain the following results for the cusp anomalous dimension $\Gamma_{++}(\theta)$ between two equal pinning fields:
\begin{equation}\label{eq_Gamma++_2d}
\Gamma_{++}(\theta)=-\frac{(\pi -\theta )^2}{24 (2 \pi -\theta ) \theta }\,.
\end{equation}
Similarly,  since the scaling dimension of the boundary changing operator between the two opposite Dirichlet boundary condition is $\Delta_{|+\rangle |-\rangle}=\frac{1}{2}$~\cite{Cardy:1989ir} (and hence $\Delta_{+-}=1$ for the interface),  the cusp anomalous dimension for pinning fields of opposite sign is
\begin{equation}\label{eq_Gamma+-_2d}
\Gamma_{+-}(\theta)=\frac{\pi ^2}{(2 \pi -\theta ) \theta }-\frac{(\pi -\theta )^2}{24 (2 \pi -\theta ) \theta }\,.
\end{equation}

Below we extract the scaling dimensions of the defect creation operator of the fused defect from the small angle limit of these results, and check that the results agree with the well known classification and solution of line defects in the $2d$ Ising model~\cite{Oshikawa:1996dj}.

First, note that the fusion between two pinning fields with the same orientation obviously results in the same defect. Therefore from the $\theta^0$ term in the expansion of eq.~\eqref{eq_Gamma++_2d} we obtain that the scaling dimension of the defect creation operator for the pinning field is
\begin{equation}\label{eq_2d_DW++}
\Delta_{+0}=\frac{1}{32}\,.
\end{equation}
For the cusp between opposite pinning fields, we note that in $2d$ the fusion between two magnetic line defects of opposite orientation results in an interface with opposite Dirichlet boundary conditions at the two sides.  The expansion of the cusp anomalous dimension~\eqref{eq_Gamma+-_2d} then predicts that the defect creation operator dimension for this interface is
\begin{equation}\label{eq_2d_DW+-}
\Delta_{(+-)0}=\frac{9}{32}\,.
\end{equation}
Both eq.s~\eqref{eq_2d_DW++} and~\eqref{eq_2d_DW+-} are in agreement with the existing results, see~\cite{Oshikawa:1996dj}.\footnote{By the folding trick, an interface in the Ising model is equivalent to a boundary in the $\mathds{Z}_2$ orbifold of the free boson at radius $R=1$. Using this fact, the authors of \cite{Oshikawa:1996dj} show that all the $\mathds{Z}_2$ preserving interfaces of the Ising model have $g=1$ and are labeled by an angle $\phi\in (0,\pi)$. This family of lines includes as limiting cases the non-simple line given by the sum of two pinning fields of opposite orientation at $\phi=0$, and the non-simple line consisting of the two possible interfaces with opposite Dirichlet boundary conditions at the two sides at $\phi=\pi$ (recall that a Dirichlet boundary has $g=1/\sqrt{2}$, hence these non-simple lines have $g=1$). The scaling dimension of the defect creation operator for any of the these lines is found to be
\begin{equation}
\Delta(\phi)=\frac{1}{2}\left[\frac{1}{\pi}\left(\phi-\frac{\pi}{4}\right)\right]^2\,,
\end{equation}
which agrees with the pinning field result~\eqref{eq_2d_DW++} for $\phi=0$ and with eq.~\eqref{eq_2d_DW+-} at $\phi=\pi$.}

The fusion of two opposite pinning fields in $2d$ is to be contrasted with the higher dimensional case. In $d>2$ we expect that the fusion between pinning fields with opposite orientation lands on a trivial line defect, as we verified for a free scalar in $d=4$ in sec.~\ref{subsec_free}.  Correspondingly, the cusp anomalous dimension $\Gamma_{+-}$ cannot contain a $\theta^0$ term for small angle in $d>2$ - we will verify that this is indeed the case in the $\varepsilon$-expansion at one-loop in sec.~\ref{sec_epsilon}.

\subsection{Planar \texorpdfstring{$\mathcal{N}=4$}{N=4} SYM at weak and strong coupling}

As a final example, we take $\mathcal{N}=4$ super Yang-Mills (SYM) with gauge group $SU(N)$ and consider the cusp anomalous dimension for a $1/2$ BPS Wilson line in the fundamental representation in the planar limit. We will be particularly interested in the small angle limit and its intepretation in terms of the fusion.

We remind the reader that the $1/2$ BPS Wilson line is defined by (in Euclidean signature)
\begin{equation}\label{eq_N4_WL}
W_F=\text{Tr}\left[P
\exp\left(i\int  dx^\mu A_\mu+\int\ |dx|\zeta_I\Phi^I\right)\right]\,,
\end{equation}
where the trace is taken in the fundamental representation, $\zeta^I$ is a unit $SO(6)_R$ vector and $\Phi^I$ the bottom component of the vector multiplet. The insertion of a Wilson line in a straight or circular contour preserves $16$ of the $32$ original supercharges and breaks the $SO(6)$ $R$-symmetry group down to $SO(5)$. 

We are interested in two $1/2$ BPS lines in the fundamental representation connected by a cusp angle $\theta$ and different orientations for the scalar coupling $\zeta_I$. We denote the corresponding cusp anomalous dimension $\Gamma_{F}(\theta,\phi)$, where $\phi$ is the angle between the two scalar couplings, such that $\Gamma_{F}(\theta,0)$ is the result for the cusp between two identical lines. The configuration with $\phi=\pm(\pi-\theta)$ is supersymmetric and therefore $\Gamma_{F}(\theta,\pi-\theta)=0$

The quantity $\Gamma_{F}(\theta,\phi)$ has been the subject of several previous works and many results are available in the literature \cite{Drukker:1999zq,Kruczenski:2002fb,Drukker:2011za,Correa:2012nk,Correa:2012at,Correa:2012hh,Drukker:2012de,Grozin:2015kna}. In particular, the result for $\pi-\theta\ll 1 $ and $\phi\ll 1$ can be computed exactly from localization and reads \cite{Correa:2012at}
\begin{equation}\label{eq_N4_small}
    \Gamma_{F}(\theta,\phi)=-\left[(\pi-\theta)^2-\phi^2\right]B(\lambda)+\ldots\,,\qquad
    B(\lambda)=\frac{\sqrt{\lambda}\,I_2(\sqrt{\lambda})}{4\pi^2 I_1(\sqrt{\lambda})}\,,
\end{equation}
where $\lambda =g_{YM}^2 N$ is the 't Hooft coupling. Note that $B(\lambda)>0$ in agreement with concavity at $\phi=0$. The value of $\Gamma_{F}(\theta,\phi)$ for arbitrary angles and 't Hooft coupling is formally given by the solution of a system of TBA equations \cite{Correa:2012hh,Drukker:2012de}. Below we will content ourselves with analyzing the leading order results at small and large 't Hooft coupling as given, e.g., in \cite{Drukker:2011za}. 

The weak coupling result is given by
\begin{equation}
\Gamma_{F}(\theta,\phi)=-\frac{\lambda}{8\pi^2}\frac{\pi-\theta}{\sin\theta}\left(\cos\theta+\cos\phi\right)+O\left(\frac{\lambda^2}{(4\pi)^4}\right)
\end{equation}
For $\phi=0$ the result is concave. For $\theta\rightarrow 0$ the result reads
\begin{equation}\label{eq_N4_weak_small_angle}
\Gamma_F(\theta,\phi)=-\frac{\lambda  (1+\cos \phi )}{8 \pi  \theta }+\frac{\lambda  (1+\cos \phi)}{8 \pi ^2}+O\left(\theta\right)    \,.
\end{equation}
The existence of a nontrivial $\theta^0$ term signals that the fusion is nontrivial. Indeed, the orientation reversal of the line~\eqref{eq_N4_WL} is a $1/2$ BPS line in the antifundamental. Therefore the fusion yields both a trivial line and a Wilson line in the adjoint, with some coupling to the scalar $\Phi^I$. In the planar limit at weak coupling we expect that the adjoint line dominates the result since its expectation value grows as $N^2$.  We did not investigate in detail the quantitative interpretation of the $\theta^0$ term, but we note that at tree-level both the line operator $\Phi^I$ and the defect creation operator (which is also given by $\Phi^I$ at the end of the line) are marginal - and thus both are expected to contribute to the $\theta^0$ term in the expansion~\eqref{eq_N4_weak_small_angle}.\footnote{The line operator $\Phi^I$ becomes slightly irrelevant once we account for interactions \cite{Polchinski:2011im,Beccaria:2022bcr}, while the fate of the defect creation operator is not known.}

Unfortunately, the analysis of the small angle limit at subleading orders in $\lambda$ requires a resummation of infinitely many diagrams; for instance, it is known that the quark-antiquark potential does not admit an analytic expansion at small coupling \cite{Pineda:2007kz,Gromov:2016rrp}.
It would be interesting to analyze further the small angle limit of the cusp at small coupling.

Let us now consider the cusp anomalous dimension at strong coupling. This is given by~\cite{Drukker:2011za}
\begin{equation}\label{eq_N4_strong_Gamma}
\Gamma_F(\theta,\phi)=\frac{\sqrt{\lambda}}{\pi\,b}\sqrt{\frac{1+b^2}{1-k^2}}
    \left[\left(1-k^2\right)K\left(k^2\right)-
    E\left(k^2\right)\right]
    +O\left(\lambda^0\right)\,,
\end{equation}
where the parameters $b$ and $k$ are related to the angles $\theta$ and $\phi$ as 
\begin{align}\label{eq_N4_strong_theta}
    \theta &=\frac{2 b \left(1-k^2\right)}{\sqrt{\left(1+b^2\right) \left(b^2+k^2\right)}}\left[
    \Pi \left(\frac{b^2+k^2}{1+b^2}\Bigg|k^2\right)-K\left(k^2\right)
    \right]\,,\\ \label{eq_N4_strong_phi}
    \phi&=2 \sqrt{\frac{1-\left(2+b^2\right) k^2}{1+b^2}}K\left(k^2\right)\,.
\end{align}
Here $K(k^2)$ and $E(k^2)$ are, respectively, complete elliptic integrals of the first and second kind, while $\Pi \left(x^2|k^2\right)$ is the incomplete elliptic integral of the third kind.\footnote{Explicitly, these functions are defined by
\begin{equation}
\begin{aligned}
 &K(k^2)=\int_0^{\frac{\pi}{2}} d\theta   \frac{1}{\sqrt{1-k^2\sin^2\theta}}\,,\qquad
  E(k^2)=\int_0^{\frac{\pi}{2}} d\theta   \sqrt{1-k^2\sin^2\theta}\,,\\
&\Pi \left(x^2|k^2\right)=\int_0^{\frac{\pi}{2}} d\theta
\frac{1}{\left(1-x^2 \sin ^2\theta \right)\sqrt{1-k^2 \sin ^2\theta } }\,.
  \end{aligned}
\end{equation}
}

\begin{figure}[t!]
    \centering
    \includegraphics[width=0.6\textwidth]{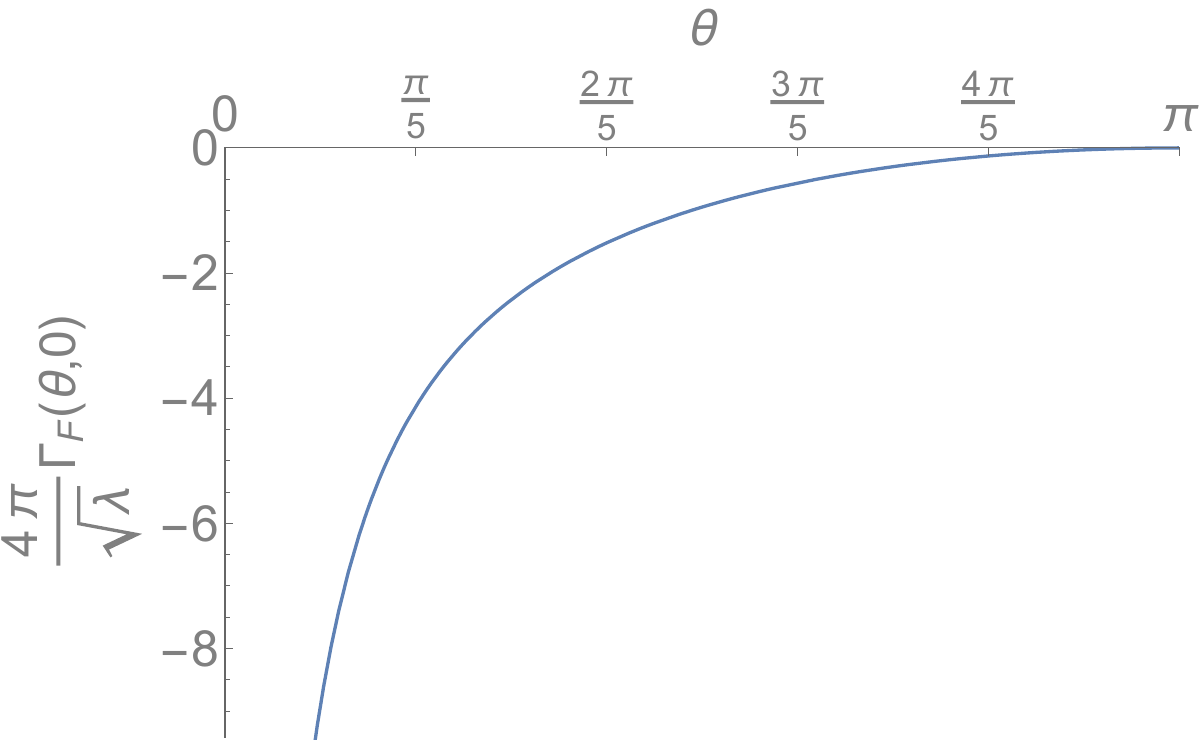}
    \caption{Plot of the cusp anomalous dimension $\frac{4\pi}{\sqrt{\lambda}}\Gamma_F(\theta,0)$ in planar $\mathcal{N}=4$ SYM at strong coupling. }
    \label{fig:N4}
\end{figure}

Let us first study the function at $\phi=0$ and arbitrary $\theta$. To this aim we simply need to solve eq.~\eqref{eq_N4_strong_phi} setting $k^2=1/(2+b^2)$ and plot the curve $(\theta(b),\Gamma_F(\theta(b),0))$ in fig.~\ref{fig:N4}. The result is manifestly concave.

Let us now consider the small angle limit at fixed $\phi$. This is equivalent to the limit $b\rightarrow 0$ with fixed $k$ in eq.s~\eqref{eq_N4_strong_Gamma},~\eqref{eq_N4_strong_theta} and~\eqref{eq_N4_strong_phi}. Solving for $\theta$ in terms of $b$ perturbatively, we find
\begin{equation}\label{eq_N4_strong_small}
\frac{4\pi}{\sqrt{\lambda}} \Gamma_F(\theta,\phi)=-
\frac{E_{F}(\phi)}{\theta}
+c_1(\phi)\theta+O\left(\theta^3\right)\,,   
\end{equation}
where $\phi$ is implicitly expressed through a parameter $k_0=k+O(b^2)$ as
\begin{equation}
    \phi=\sqrt{4-8 k_0^2} \,K\left(k_0^2\right)\,,
\end{equation}
and in eq.~\eqref{eq_N4_strong_small} we defined the following functions
\begin{align}
   E_{F}\left(\phi(k_0)\right)&= \frac{8 }{k_0 \sqrt{1-k_0^2}  }\left[E\left(k_0^2\right)-\left(1-k_0^2\right) K\left(k_0^2\right)\right]^2
   \,,\\[0.5em]
c_1\left(\phi(k_0)\right)&=   \frac{\left(1-2 k_0^2\right) E\left(k_0^2\right)-
\left(1-4k_0^2+3 k_0^4\right) K\left(k_0^2\right)
}{3 k_0 \sqrt{1-k_0^2} \left[E\left(k_0^2\right)-\left(1-k_0^2\right) K\left(k_0^2\right)\right]}\,.
\end{align}
The result simplifies for $\phi=0$, i.e. $k_0=1/\sqrt{2}$:
\begin{equation}
\frac{4\pi}{\sqrt{\lambda}} \Gamma_F(\theta,0)=-
\frac{4\, \Gamma \left(\frac{3}{4}\right)^4}{\pi  \,\theta }
+\frac{32\,\Gamma \left(\frac{5}{4}\right)^4}{3 \pi ^2}\theta+O\left(\theta^3\right)\,,   
\end{equation}
Note also that the expansion is singular for $\phi\rightarrow \pi$, i.e. $k_0\rightarrow 0$ - this is because supersymmetry ensures $\Gamma_F(\theta,\pi-\theta)=0$ as mentioned earlier.

The most notable feature of the result~\eqref{eq_N4_strong_small} is the absence of a $\theta^0$ term. This is in sharp contrast with the weak coupling result~\eqref{eq_N4_WL}. 
The simplest interpretation is that the fusion at strong coupling results in a trivial line. This could be due to the Casimir energy difference between the singlet and the adjoint, and the effect of the domain wall operator. (This kind of instability is different than the one discussed in \cite{Aharony:2022ntz,Aharony:2023amq}, which is due to a relevant operator on a simple line. Perhaps, this instability might be related to the discussion in \cite{Klebanov:2006jj}.)
We leave further investigations of the fusion algebra of supersymmetric lines in $\mathcal{N}=4$ SYM for future work.

We finally comment that the cusp anomalous dimension in $\mathcal{N}=4$ SYM is also related to the Regge limit of certain scattering amplitudes in the Coulomb branch \cite{Caron-Huot:2014gia,Bruser:2018jnc}. It would be interesting to investigate potential implications of our discussion for that setup.

\part{Results in the \texorpdfstring{$O(N)$}{O(N)} model}\label{Part2}

\section{Ising model: non-perturbative fuzzy sphere results}\label{sec_fuzzy}

\subsection{The method and previous results}

We are interested in the Ising model with two pinning field defects at cusp angle $\theta$. As explained in the introduction, the pinning field defect is the infrared fixed-point of eq.~\eqref{eq_pinning_defect_def}. For $N=1$ 
the coupling $h$ is a scalar dimensionful parameter that can be either positive or negative, and the fixed point is schematically associated with the limit $h\rightarrow\infty$.

To study the pinning field directly in $d=3$ we use the recently developed fuzzy-sphere regularization \cite{Zhu:2022gjc} (see also \cite{Hu:2023xak,Han:2023yyb,Zhou:2023qfi,Han:2023O3,Hu:2024pen}). In short, the fuzzy sphere regularization realizes 3D CFTs on the cylinder $S^2\times \mathbb R$ by using quantum mechanical systems defined on a fuzzy (noncommutative) sphere. Specifically, one considers a system of fermions on a spatial two-dimensional sphere in the background of a monopole flux $4\pi s$, restricted to the lowest Landau level~\cite{Haldane1983Fractional}. By appropriately choosing the Hamiltonian and filling factor of fermions, one could expect to realize various CFTs in the continuum limit $s\rightarrow \infty$. Importantly, this scheme has a surprisingly small finite size effect, hence one can conveniently extract CFT data using the state-operator correspondence by Hamiltonian diagonalization at small system.

In practice, the fuzzy sphere model is defined on $N_{\textrm{orb}}=2s+1$ sites labeled by $m=-s,-s+1,\cdots, s$, where these sites correspond to states (i.e. Landau orbitals) of the lowest Landau level and they form the spin-$s$ representation of the $SO(3)$ symmetry group of the sphere. In comparison to the familiar lattice model, $N_{\textrm{orb}}^{1/2}$ corresponds to the system size, or equivalently, $N_{\textrm{orb}}^{-1/2}$ corresponds to the lattice spacing.  To realize the 3D Ising CFT, one considers two-component fermions $\mathbf c_m^\dag = (c_{m,\uparrow}^\dag, c_{m,\downarrow}^\dag) $ at half filling (i.e. one particle per-site), where the Ising $\mathbb Z_2$ symmetry exchanges the fermion flavor $\uparrow$ and $\downarrow$. The fuzzy sphere Ising Hamiltonian is conceptually similar to the transverse Ising Hamiltonian on the lattice, but it takes a more complicated form due to the requirement of \( SO(3) \) sphere rotation symmetry. We refer to \cite{Zhu:2022gjc} for the precise form of the fuzzy sphere Ising Hamiltonian.
To study the cusp between pinning field defects, we add two localized pinning fields with strengths $h_{1,2}$ to the original fuzzy sphere Ising Hamiltonian, 
\begin{equation}\label{eq_Himp_fuzzy}
H(h_1,h_2, \theta) = H_{\textrm{bulk}} + h_1 n^z(\theta=\varphi=0) + h_2 n^z(\theta, \varphi=0).
\end{equation}
Here the first pinning field is located at the north pole, and we vary the position of the second pinning field. The pinning field breaks both the Ising symmetry and the sphere rotation symmetry, and it takes the form,
\begin{equation}
n^z(\theta,\varphi)=\frac{1}{2s+1}\sum_{m_1,m_2=-s}^s \mathbf c^\dag_{m_1} \sigma^z \mathbf c_{m_2} \bar Y^{(s)}_{s,m_1}(\theta,\varphi) Y^{(s)}_{s,m_2}(\theta,\varphi),  
\end{equation}  
where the monopole harmonics $Y^{(s)}_{s,m}$ are the wave functions of the states on the lowest Landau level,
\begin{equation}
    Y^{(s)}_{s,m} (\theta,\varphi) = \sqrt{\frac{(2s+1)!}{(s+m)!(s-m)!}}\, e^{im\varphi} \cos^{s+m}(\theta/2) \sin^{s-m}(\theta/2) \,.
\end{equation}

When we set $h_1=h_2$ and $\theta=\pi$ in eq.~\eqref{eq_Himp_fuzzy}, the setup describes the standard pinning field defect. In \cite{Hu:2023ghk} the fuzzy sphere was used to demonstrate that the system flows to a DCFT in the infrared and several interesting DCFT data were extracted numerically, including the scaling dimension of the leading deformation $\Delta_{\hat \phi}$ and the normalization of the two-point function of the displacement $C_D$; in particular, it was found that $\Delta_{\hat \phi}>1$ and thus the DCFT is infrared stable. Additionally in \cite{Zhou:2023fqu} the Hamiltonian~\eqref{eq_Himp_fuzzy} with different values for the two couplings (and still $\theta=\pi$) was used to compute the dimension of the defect creation operator $\Delta_{+0}$, and the dimension of the domain wall operator between opposite pinning fields $\Delta_{+-}$, as well as the $g$-function of the defect. We summarize some of these results in table~\ref{tab:Fuzzy_pinning} for reference.

\begin{table}[t]
    \centering
    \begin{tabular}{|c|c|c|c|c|}
    \hline
         $\Delta_{\hat \phi}$ &  $\Delta_{+0}$ & $\Delta_{+-}$ & $C_D$ & $g$-function \\
         \hline
        1.63(6) &  0.108(5) & 0.84(5) & 0.27(1) & 0.602(2)
        \\ \hline
    \end{tabular}
    \caption{Fuzzy sphere results from \cite{Hu:2023ghk,Zhou:2023fqu} for the pinning field defect. $\Delta_{\hat \phi}$, $\Delta_{+0}$ and $\Delta_{+-}$ are, respectively, the dimensions of the lowest dimensional defect operator, the defect creation operator, and the domain wall operator. $C_D$ is the normalization of the displacement operator two-point function~\eqref{eq_DD_2pt}. Notice that $g<1$ in agreement with the $g$-theorem~\cite{Cuomo:2021rkm}.}
    \label{tab:Fuzzy_pinning}
\end{table}

\subsection{Numerical results for the cusp}

According to the definition~\eqref{eq_cusp_def_cyli}, we compute the cusp anomalous dimension via the fuzzy sphere regularization as
\begin{align}
\Gamma_{h_1,h_2} (\theta) &= \lim_{N_{\textrm{orb}}\rightarrow\infty} \alpha_{N_{\textrm{orb}}} \left(E_{h_1,h_2}(\theta) - \frac{E_{h_1,h_1}(\theta=\pi)+E_{h_2,h_2}(\theta=\pi)}{2}\right) \,,
\end{align}
where $E_{h_1,h_2}(\theta)$ is the ground state energy of $H(h_1,h_2, \theta)$ \eqref{eq_Himp_fuzzy}, and $\alpha_{N_{\textrm{orb}}}$ is a size dependent numerical factor which rescales the bulk energy gaps to the scaling dimensions. In particular, we choose $\alpha_{N_{\textrm{orb}}} = \Delta_\sigma/\delta E_{\sigma}$, such that the rescaled gap of the bulk $\sigma$ state equals to the bootstrap data $\Delta_\sigma \approx 0.518149$.\footnote{One can also choose different ways of define $\alpha_{N_{\textrm{orb}}}$, e.g., requiring the rescaled stress tensor to have $\Delta=3$. Different choices will yield the same results in the thermodynamic limit, but would have small differences at finite $N_\textrm{orb}$.} 
Furthermore, the first-order derivative of $\Gamma(\theta)$ can be computed exactly using
\begin{equation}\label{eq_fuzzy_first_der}
\frac{d\, \Gamma_{h_1,h_2}(\theta)}{d\theta} = \lim_{N_{\textrm{orb}}\rightarrow \infty} \alpha_{N_{\textrm{orb}}} \langle \Psi_{h_1,h_2}(\theta) | \frac{d\, H(h_1,h_2,\theta)}{d\theta} | \Psi_{h_1,h_2}(\theta) \rangle ,
\end{equation}
where $| \Psi_{h_1,h_2}(\theta) \rangle$ is the ground state of $H(h_1,h_2,\theta)$, and 
\begin{equation}
    \frac{d \,H(h_1,h_2,\theta)}{d\theta} = h_2 \frac{d \,n^z(\theta,\varphi=0)}{d\theta}\,.
\end{equation}

\begin{figure}[t!]
    \centering
\subcaptionbox{\label{fig:cusp++1}}{\includegraphics[width=0.49\textwidth]{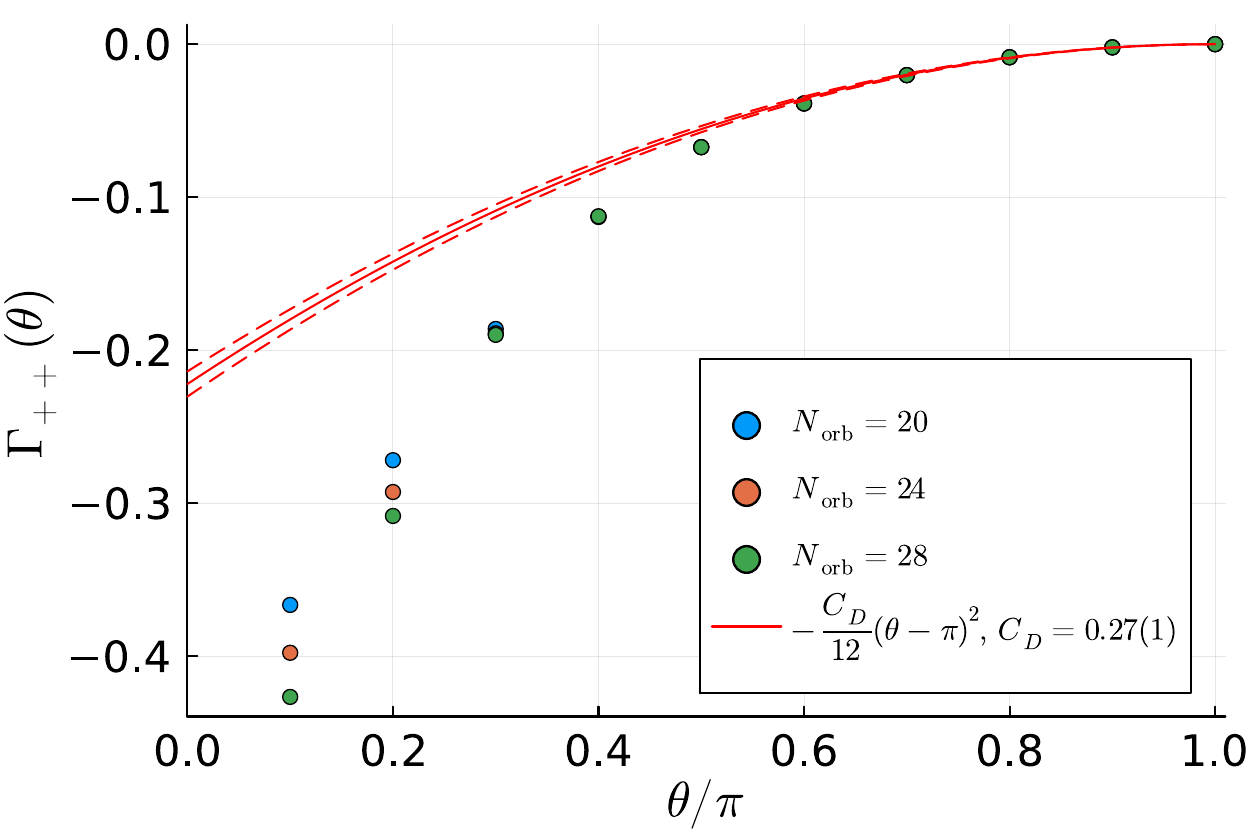}}
\subcaptionbox{\label{fig:cusp++2}}{
\includegraphics[width=0.49\textwidth]{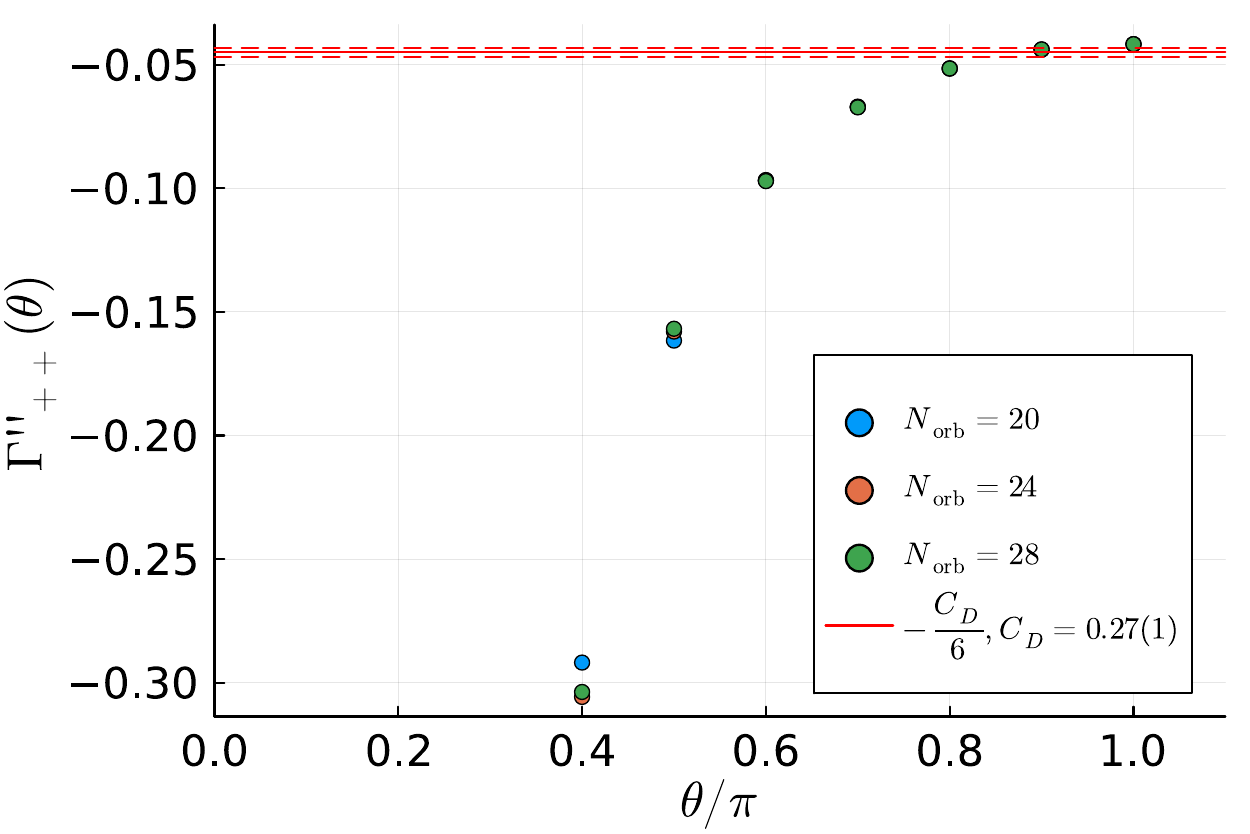}}
    \caption{\ref{fig:cusp++1} Cusp anomalous dimensions $\Gamma_{++}(\theta)$. \ref{fig:cusp++2} The second order derivatives $\Gamma''_{++}(\theta)$. The data points for the cusp reliably converged for $\theta\geq 0.4\pi$.}
    \label{fig:cusp++}
\end{figure}

\begin{figure}[t!]
    \centering
    \subcaptionbox{\label{fig:cusp++_casimir1}}
{\includegraphics[width=0.49\textwidth]{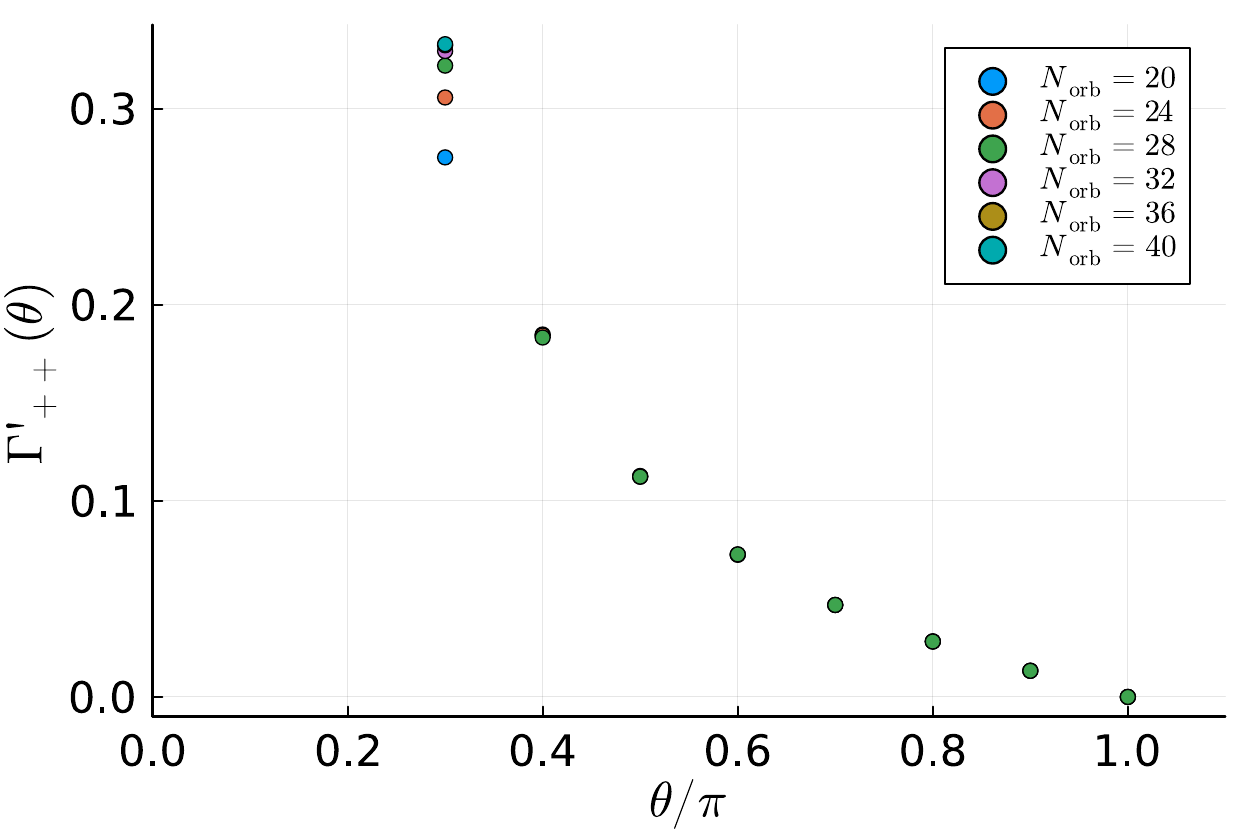}}
\subcaptionbox{\label{fig:cusp++_casimir2}}{
\includegraphics[width=0.49\textwidth]{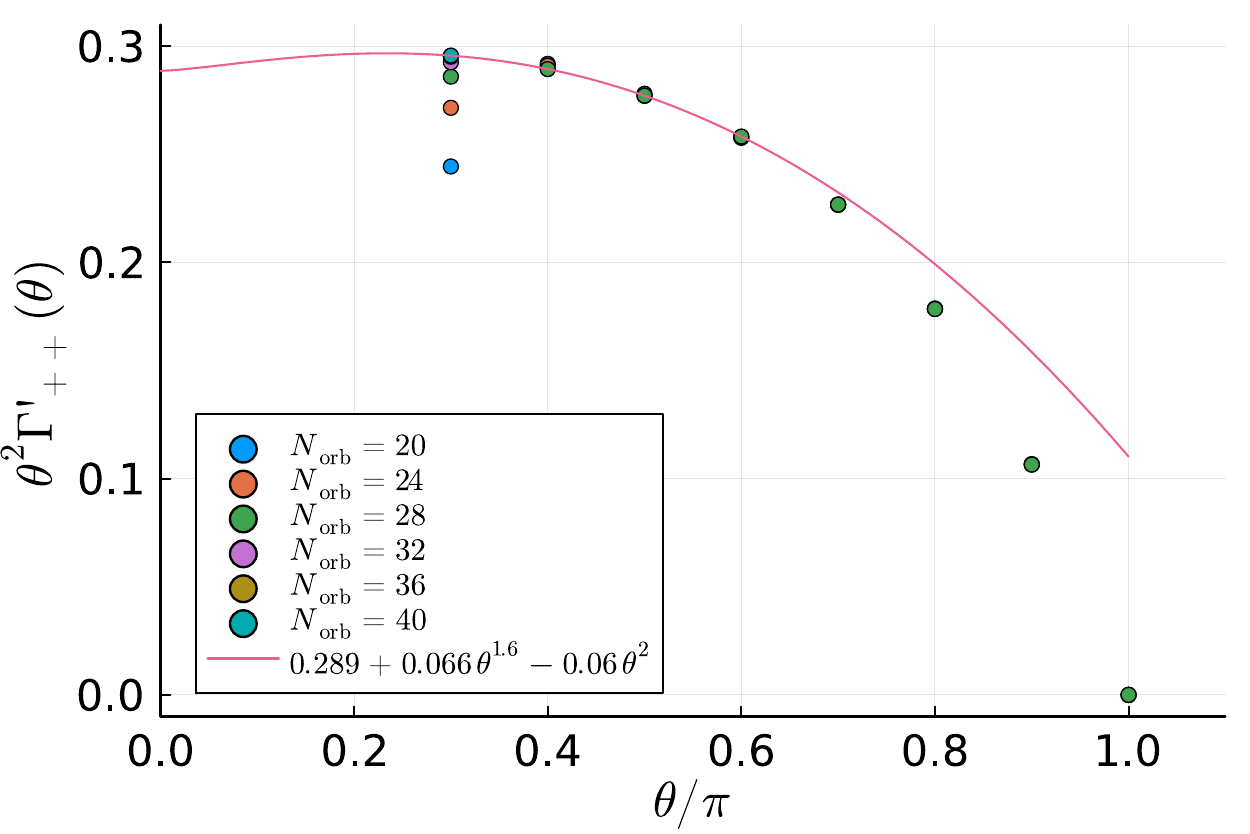}}
    \caption{\ref{fig:cusp++_casimir1} First order derivative $\Gamma'_{++}(\theta)$ - the results are numerically stable for $\theta\geq 0.3\pi$. \ref{fig:cusp++_casimir2} Fitting the Casimir energy through $\theta^2\Gamma'_{++}(\theta)=-C_{+++} + (\Delta_{\hat\phi}-1)a\,\theta^{\Delta_{\hat\phi}}+b\,\theta^2+\cdots$, yielding $C_{+++}=-0.29(2)$.}
    \label{fig:cusp++_casimir}
\end{figure}

\begin{figure}[t!]
    \centering
     \subcaptionbox{\label{fig:cusp+-1}}
{\includegraphics[width=0.49\textwidth]{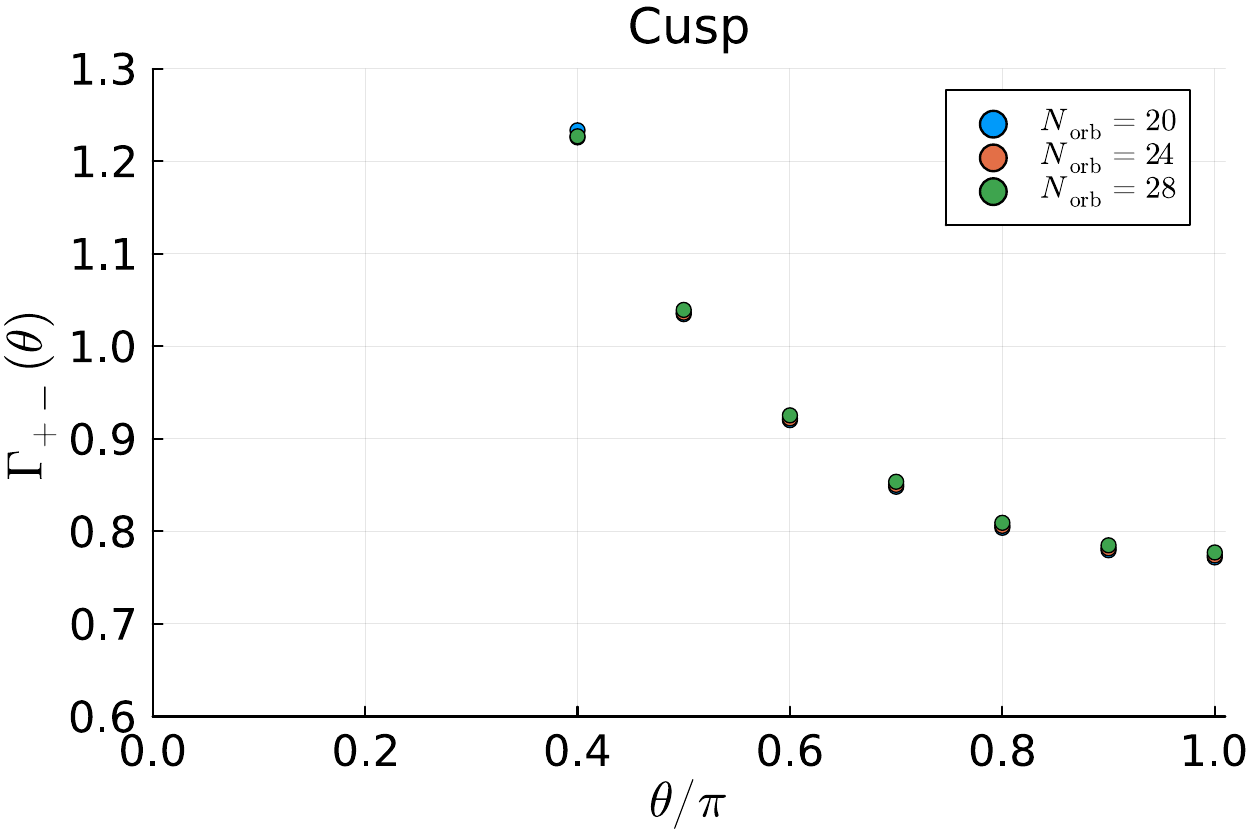}}
\subcaptionbox{\label{fig:cusp+-2}}{
\includegraphics[width=0.49\textwidth]{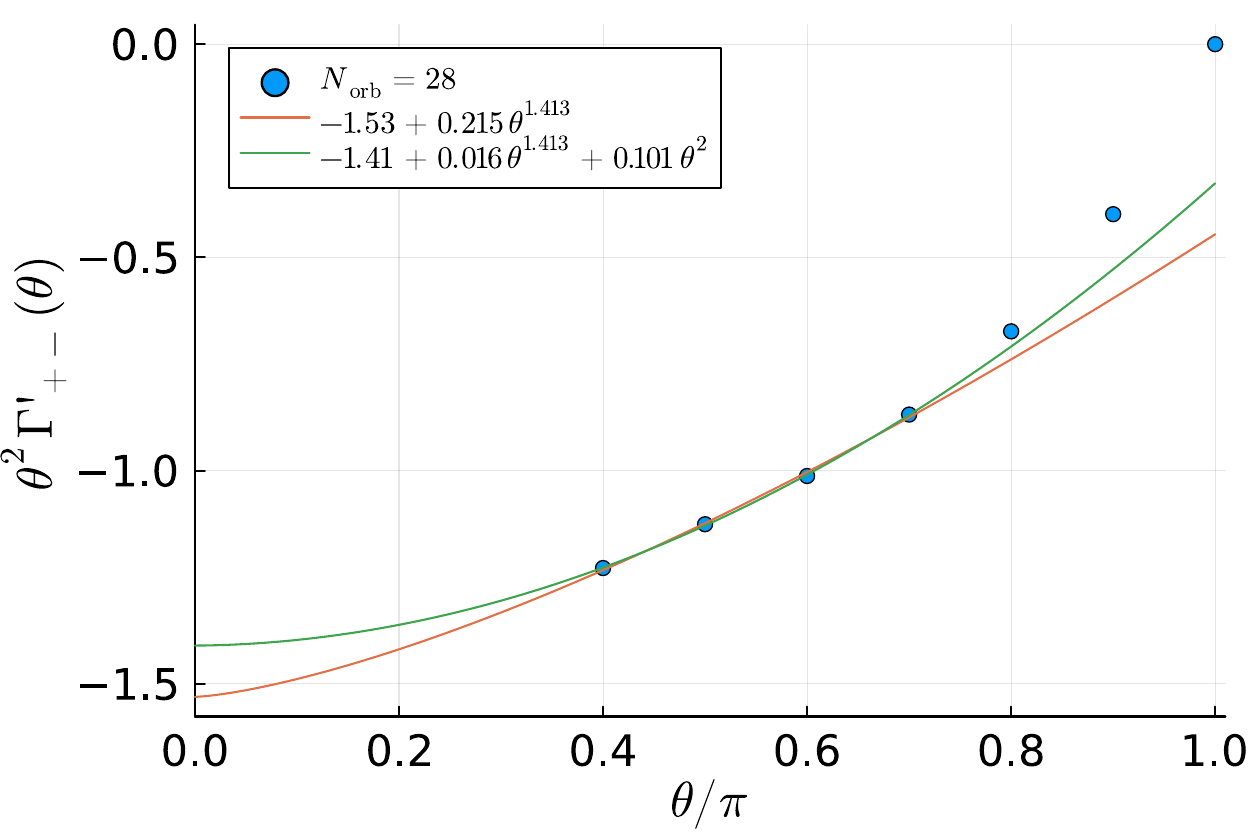}}
    \caption{\ref{fig:cusp+-1} Cusp anomalous dimensions $\Gamma_{+-}(\theta)$. \ref{fig:cusp+-2} Fitting the Casimir energy through $\theta^2\Gamma'_{+-}(\theta)=-C_{+-0} + (\Delta_{\epsilon}-1)a\,\theta^{\Delta_{\epsilon}}+b\,\theta^2$, yielding $C_{+-0}=1.4(2)$. The results are numerically stable for $\theta\geq 0.4\pi$.}
    \label{fig:cusp+-}
\end{figure}

We consider two different types of cusps: i) $\Gamma_{++}$ which can be achieved by taking $h_1>0$ and $h_2>0)$, and ii) $\Gamma_{+-}$ which is obtained setting $h_1>0$ and $h_2<0$.
Theoretically, different choices of the magnetic fields $h_{1,2}$ give the same $\Gamma$ in the continuum limit $N_\textrm{orb}\rightarrow \infty$. However, in practice, different field strengths result in different finite size effects at small $N_\textrm{orb}$. We find that large values for $h_{1,2}$ have smaller finite size effects; therefore, for most computations we choose $h_{1,2}=\pm 2000$. Our results are shown in fig.s~\ref{fig:cusp++}, \ref{fig:cusp++_casimir}, and \ref{fig:cusp+-}. Different colors for the data points correspond to different system sizes $N_\textrm{orb}$. All the data are computed using the density matrix renormalization group (DMRG) method~\cite{SWhite1992,Feiguin2008}, and the maximal bond dimension is $\chi=3000$ which gives truncation errors around $10^{-9}$ or smaller. The DMRG simulation was performed using the ITensor package~\cite{ITensor1,ITensor2}. The results of $\Gamma_{++}(\theta)$ and $\Gamma_{+-}(\theta)$ at different angles are listed in Table \ref{tab:cusp}.

\begin{table}[t]
    \centering
    \begin{tabular}{c|cccccccc}
    \hline
         &$\theta=0.9\pi$ &  $\theta=0.8\pi$ & $\theta=0.7\pi$ & $\theta=0.6\pi$ & $\theta=0.5\pi$ & $\theta=0.4\pi$ & $\theta=0.3\pi$  \\
     $\Gamma_{++}(\theta)$  & -0.0021(2) & -0.0085(5) & -0.020(1) & -0.039(3) & -0.067(3) & -0.112(7) & -0.19(1) \\ 
    $\bar{\Gamma}_{+-}(\theta)$ & 0.0077(4) & 0.032(2) & 0.076(4) & 0.148(8) & 0.262(15) & 0.45(3)  \\ \hline
    \end{tabular}
    \caption{\label{tab:cusp} Table of cusp anomalous dimensions  $\Gamma_{++}(\theta)$ and $\bar{\Gamma}_{+-}(\theta)=\Gamma_{+-}(\theta)-\Gamma_{+-}(\pi)$ at different angles. Numerically, we find $\Gamma_{+-}(\pi) = 0.84(5)$~\cite{Zhou:2023fqu} converges slowly with the system size, but $\Gamma_{+-}(\theta)-\Gamma_{+-}(\pi)$ converges quickly.  The error bars are estimated based on the data at different sizes and different pinning field strengths. }
\end{table}

Fig.~\ref{fig:cusp++1} shows the cusp anomalous dimensions $\Gamma_{++}(\theta)$ at different $\theta$ angles; from the plots we see that the data points for $\theta\ge 0.4\pi$ converged numerically within the system size that we considered ($N_\textrm{orb}=28$), while the results for smaller angles are numerically unstable.\footnote{$\theta=0.3\pi$ converges at a larger system size $N_{\textrm{orb}}=36\sim 40$, and is shown in other plots.} We plot the second-order derivative $\Gamma''_{++}(\theta)$ in fig.~\ref{fig:cusp++2},\footnote{In contrast to the first-order derivative, that can be evaluated using~\eqref{eq_fuzzy_first_der}, $\Gamma''_{++}(\theta)$ is computed by discretizing the angle $\theta$ and using $\Gamma''_{++}(\theta) = (\Gamma'_{++}(\theta+\delta \theta)-\Gamma'_{++}(\theta))/\delta \theta$, where we take $\delta\theta=0.01\pi$.} while in fig.~\ref{fig:cusp++_casimir1} we  show the value of $\Gamma'_{++}(\theta)$. These results are consistent with the theoretical expectations that, 1) $\Gamma_{++}(\theta)\sim -\frac{C_D}{12}(\theta-\pi)^2$ near $\theta=\pi$ with $C_D=0.27(1)$;   2) $\Gamma_{++}(\theta)$ is concave. 

We also estimate the Casimir energy in fig.~\ref{fig:cusp++_casimir2} fitting the result for $\Gamma'_{++}(\theta)$ with
\begin{equation}\label{eq_fuzzy_Casimir}
\Gamma'_{++}(\theta) = -\frac{C_{+++}}{\theta^2} + (\Delta_{\hat \phi}-1)\, a \,\theta^{\Delta_{\hat \phi}-2} + b+ \cdots\,,
\end{equation}
where $a$ and $b$ are fitting parameters. Eq~\eqref{eq_fuzzy_Casimir} follows from the derivative of eq.~\eqref{cusp}, which conveniently removes the constant $\theta^0$ term. Extrapolating $\Gamma'_{++}(\theta)$ for $\theta=0.3\pi\sim 0.6\pi$ we get the following estimate for the Casimir energy
\begin{equation}\label{eq_fuzzy_Casimir++}
   C_{+++}=-0.29(2)\,. 
\end{equation}

Fig.~\ref{fig:cusp+-1} shows the results of the cusp between opposite pinning fields $\Gamma_{+-}(\theta)$. The size dependence of $\Gamma_{+-}(\theta)$ is stronger than $\Gamma_{++}(\theta)$, and it mainly comes from the size dependence of the scaling dimension of the defect changing operator $\Gamma_{+-}(\theta=\pi)$. So the results for the difference $\Gamma_{+-}(\theta)-\Gamma_{+-}(\theta=\pi)$ converged numerically within the system sizes that we used. In Fig.~\ref{fig:cusp+-2} we estimate the Casimir energy $C_{+-0}$ similarly to what we did for $\Gamma_{++}$ fitting the result with 
\begin{equation}
\Gamma'_{+-}(\theta) = -\frac{C_{+-0}}{\theta^2} + (\Delta_{\epsilon}-1)\, a \,\theta^{\Delta_{\epsilon}-2} + b + \cdots\,.
\end{equation}
This yields
\begin{equation}\label{eq_fuzzy_Casimir+-}
    C_{+-0}=1.4(2)\,.
\end{equation}

\begin{figure}[t!]
    \centering
     \subcaptionbox{\label{fig:cusp++const}}
{\includegraphics[width=0.49\textwidth]{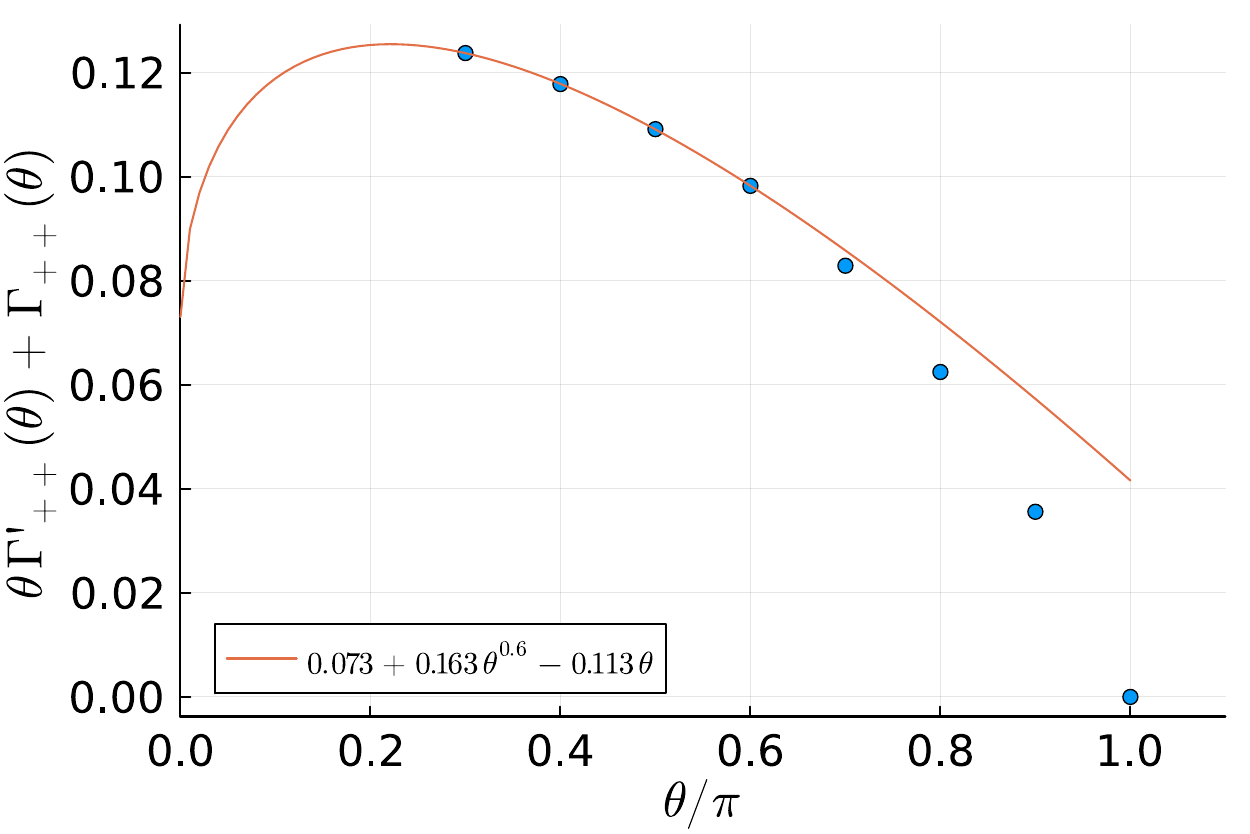}}
\subcaptionbox{\label{fig:cusp+-const}}{
\includegraphics[width=0.49\textwidth]{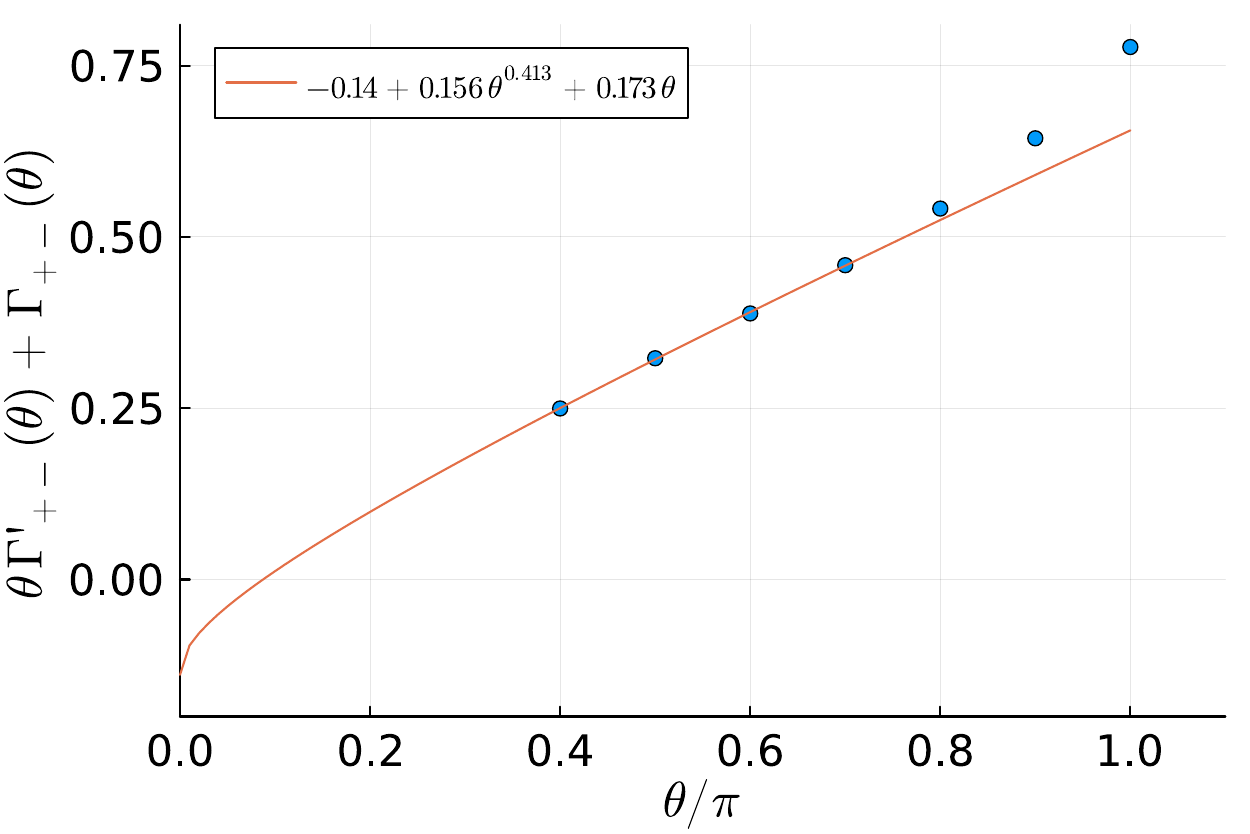}}
    \caption{Fitting the $\theta^0$ term of \ref{fig:cusp++const} $\Gamma_{++}(\theta)$ and \ref{fig:cusp+-const} $\Gamma_{+-}(\theta)$ in the $\theta\rightarrow0$ limit \eqref{cuspint}.}
    \label{fig:const} 
\end{figure}

At last, we check if the $\theta^0$ term of $\Gamma_{++}(\theta)$  and $\Gamma_{+-}(\theta)$ in the $\theta\rightarrow 0$ limit can be correctly reproduced by our numerical data. Fig.~\ref{fig:const} shows the results for $\theta \Gamma'_{++}(\theta) + \Gamma_{++}(\theta)$ and $\theta \Gamma'_{+-}(\theta) + \Gamma_{+-}(\theta)$, where the singular Casimir energy terms cancel. We further fit the data at small $\theta$ with $\Delta_{c0} + \Delta_{irr}\, a \, \theta^{\Delta_{irr}-1} + 2b\,\theta + \cdots $, where $\Delta_{irr}\approx 1.6$ for the $++$ cusp and $\Delta_{irr}\approx 1.413$ for the $+-$ cusp. The fitting results have a considerable discrepancy from the theoretical expectations $\Delta_{+0}=0.108(5)$ and $\Delta_{(+-)0}=0$. The discrepancy might be caused by the contributions of the higher order terms.

\section{\texorpdfstring{$\varepsilon$}{Epsilon}-expansion results in the \texorpdfstring{$O(N)$}{O(N)} model}\label{sec_epsilon}

\subsection{Pinning field defect in the \texorpdfstring{$\varepsilon$}{epsilon}-expansion}\label{subsec_eps_review}

We consider the critical $O(N)$ Wilson-Fisher model in $d=4-\varepsilon$ dimensions. As a reminder, the renormalized action in flat space reads
\begin{equation}\label{eq_BulkAction_epsilon}
S=\int d^dx\left[\frac{1}{2}(\pd\phi_a)^2+
\mu^{\varepsilon}
\frac{\lambda}{4!}\left(\phi_a^2\right)^2\right]\,,
\end{equation}
where $\mu$ is the sliding scale. In dimensional regularization within the minimal subtraction scheme the beta function of the coupling reads \cite{Kleinert:2001ax}
\begin{equation}\label{eq_beta_BULK}
\beta_{\lambda}=
-\varepsilon\lambda+\frac{N+8}{3}\frac{\lambda^2}{(4\pi)^2}
-\frac{3N+14}{3}\frac{\lambda^3}{(4\pi)^4}
+O\left(\frac{\lambda^4}{(4\pi)^6}\right)\,.
\end{equation}
The beta function \eqref{eq_beta_BULK} admits a zero $\beta(\lambda^*)=0$ at the Wilson-Fisher fixed point, for which:
\begin{equation}\label{eq_lambda_fix}
\frac{\lambda^*}{(4\pi)^2}=\frac{3 \varepsilon }{N+8}+
\frac{9 (3 N+14) \varepsilon ^2}{(N+8)^3}
+O\left(\varepsilon^3\right)\,.
\end{equation}
This fixed point describes the long distance behavior of correlation functions and is $O(N)$ invariant and weakly coupled for $\varepsilon \ll 1$.

The pinning field defect is obtained by perturbing the action as 
\begin{equation}\label{eq_DefectAction_epsilon}
S\rightarrow S+\mu^{\varepsilon/2}\int d\tau\, \vec{h}\cdot \vec{\phi}\left(x(\tau)\right)\,,
\end{equation}
where the magnetic field $h$ explicitly breaks the $O(N)$ symmetry to $O(N-1)$ for $N>1$ and fully breaks the $\mathbb{Z}_2$ symmetry for $N=1$.  The beta-function of the defect coupling $h=|\vec{h}|$ to order $O\left(\lambda^2\right)$ is given by \cite{Allais:2014fqa,Cuomo:2021kfm}:
\begin{equation}\label{eq_beta_h}
\beta_{h^2} =-\varepsilon h^2+\frac{\lambda }{(4\pi)^2}\frac{h^4}{3}+
\frac{\lambda^2}{(4\pi)^4}\left(
\frac{N+2}{18} h^2-\frac{N+8}{18}h^4-\frac{h^6}{6}
\right)
+O\left(\frac{\lambda^3}{(4\pi)^6}\right)\,.
\end{equation}
This beta function admits an infrared fixed point that describes the pinning field DCFT at
\begin{equation}\label{eq_h_fix}
(h^*)^2=(N+8)+
\varepsilon\frac{4 N^2+45 N+170}{2 N+16}+
O\left(\varepsilon^2\right)\,,
\end{equation}
where we used eq.~\eqref{eq_lambda_fix}. Note that, unlike the bulk coupling $\lambda^*$, the defect coupling at the fixed point is not small. It is nonenetheless straightforward to study digrammatically the fixed point for small bulk coupling and arbitrary values of the magnetic field.

Several perturbative results for the DCFT data of the theory were obtained in \cite{Allais:2014fqa,Cuomo:2021kfm} within the $\varepsilon$-expansion (and also at large $N$ in \cite{Cuomo:2021kfm}). In the following we will only need the scaling dimension of the lightest $O(N-1)$ singlet. This can be identified with the operator $\hat{\phi}\equiv\vec{h}\cdot\vec{\phi}/|\vec{h}|$ on the line, and from eq.~\eqref{eq_beta_h} it follows that its scaling dimension reads 
\begin{equation}\label{eq_gamma1_epsilon}
\Delta_{\hat{\phi}}=1+\varepsilon-\varepsilon^2
\frac{3 N^2+49 N+194}{2 (N+8)^2}
+O\left(\varepsilon^3\right)\,.
\end{equation}
We also note that for $N>1$ the defect admits $N-1$ exactly marginal deformations, the \emph{tilt} operators, whose effect is equivalent to a change in the orientiation of the magnetic field $\vec{h}$. This implies that there exists a defect conformal manifold isomorphic to the coset $O(N)/O(N-1)$ (see e.g. \cite{Drukker:2022pxk,Herzog:2023dop} for a general discussion of similar defect conformal manifolds).

\subsection{One-loop results}

\subsubsection{Sketch of the calculation}

Let us now compute the cusp anomalous dimension between two pinning field defects. To this aim, it is simplest to work directly on the cylinder Weyl frame. This is because on the cylinder dilations, i.e.  translation in the (Euclidean) time coordinate $\tau$ of the cylinder,  are a symmetry of the model also away from the fixed point, and the partition function factorizes in the expected form $\log Z_{\vec{h}_1\vec{h}_2}\propto T$ for arbitrary values of the couplings also at intermediate steps in the calculation. In flat space instead dilations become a symmetry only when we set the couplings to their fixed point value~\eqref{eq_lambda_fix}, and we recover the form~\eqref{cuspexp} for the defect expectation value only after we plug in the result the fixed point value for the couplings.

\begin{figure}[t]
   \centering
		\subcaptionbox{  \label{fig:treeCyl1}}
		{\includegraphics[width=0.15\textwidth]{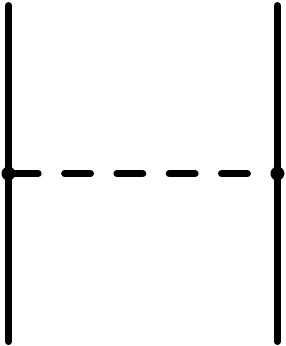}}
 \hspace*{3cm}
		\subcaptionbox{ \label{fig:treeCyl2}}
		{\includegraphics[width=0.15\textwidth]{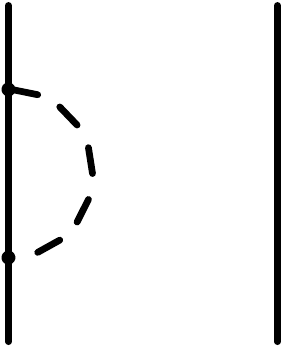}}
  \\[1em]
  \subcaptionbox{ \label{fig:OneLoopTriangle}}
		{\includegraphics[width=0.15\textwidth]{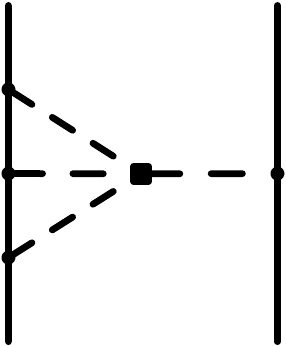}}
  \hspace*{1.5cm}
    \subcaptionbox{ \label{fig:OneLoopCross}}
		{\includegraphics[width=0.15\textwidth]{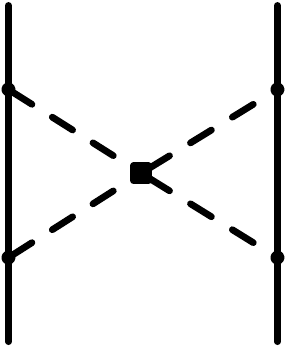}}
 \hspace*{1.5cm}
    \subcaptionbox{ \label{fig:OneLoopSelf}}
		{\includegraphics[width=0.15\textwidth]{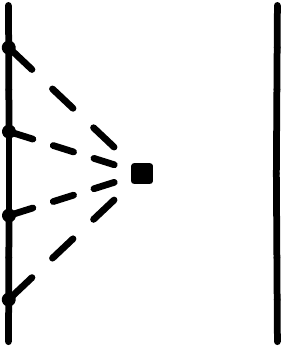}}
        \caption{Diagrams contributing the cusp anomalous dimension at one-loop on the cylinder. The notation is as in figure~\ref{fig:Diagramstree}, with the addition of square vertices denoting insertions of the bulk coupling $\lambda$. We neglected a diagram proportional to a massless tadpole, since it vanishes in dimensional regularization.}
\label{fig:DiagramsLoop}
\end{figure}

To extract the cusp anomalous dimensions, we compute first the partition function with two defect insertions at angular distance $\theta$ and arbitrary values of the renormalized coupling $\vec{h}_1$ and $\vec{h}_2$ (not necessarily at the fixed point). The diagrams contributing to the result are shown in fig.~\ref{fig:DiagramsLoop}. We provide some details on the calculation in appendix~\ref{app_eps_exp_calc}. Eventually we find that the partition function on the cylinder reads
\begin{align} \nonumber
&\frac{\log Z_{\vec{h}_1\vec{h}_2}(\theta)}{T}=\frac{\vec{h}_1\cdot\vec{h}_2}{4\pi^2} f_d(\cos\theta)\left[1+\frac{\varepsilon}{2}   \log (\pi e^{\gamma}\mu^2)-
\frac{\lambda  (\vec{h}_1^2+\vec{h}_2^2) }{192 \pi ^2}\log \left(64 \pi  e^{\gamma-1}\mu^2\right)
\right]\\[0.7em] \nonumber
&
-\frac{\lambda}{6(4\pi^2)^4}\left[\vec{h}_1^2\,\vec{h}_1\cdot\vec{h}_2+\vec{h}_2^2\,\vec{h}_1\cdot\vec{h}_2\right]I_{11}(\cos\theta)-\frac{\lambda}{12(4\pi^2)^4}\left[\vec{h}_1^2\,\vec{h}_2^2+2(\vec{h}_1\cdot\vec{h}_2)^2\right]I_{12}(\cos\theta)
\\[0.7em]
&+\text{self-energies}+O(\varepsilon^2,\varepsilon\lambda,\lambda^2)\,,
\label{eq_partition_function_finite}
\end{align}
where we work in units such that the cylinder radius is $R=1$.
In eq.~\eqref{eq_partition_function_finite} ``self-energies" stand for diagrams that do not connect the two defects (fig.~\ref{fig:treeCyl2} and~\ref{fig:OneLoopSelf}) and thus cancel-out when normalizing the partition function to obtain the physical result according to eq.~\eqref{eq_cusp_def_cyl}; we thus neglect the self-energy contributions in the main text. In  the first line of eq.~\eqref{eq_partition_function_finite} we defined the following $d$-dimensional convergent integral
\begin{equation}\label{eq_fd_def}
f_d(\cos\theta)=\int d\tau 
\frac{e^{-\frac{d-2}{2}|\tau|}}{(1-2\cos\theta e^{-|\tau|}+e^{-2|\tau|})^{\frac{d-2}{2}}}
\stackrel{d=4}{=}\frac{\pi -\theta }{ \sin (\theta )}\,.
\end{equation}
We list some properties of $f_d(x)$ for $d\neq 4$ in the appendix.
In the second and third line of eq.s~\eqref{eq_partition_function_finite} $I_{11}$ and $I_{12}$ stand for the following integrals:
\begin{align}
\label{eq_I11}
I_{11}(\hat{n}_1\cdot\hat{n}_2)&=\int d^3\hat{n}\left[
f_4^3(\hat{n}\cdot\hat{n}_1)f_4(\hat{n}\cdot\hat{n}_2)-
\frac{\pi^3 f_4(\hat{n}_1\cdot\hat{n}_2)}{\left(2-2\hat{n}\cdot\hat{n}_1\right)^{\frac{3}{2}}}
\right]\,,\\
\label{eq_I12}
I_{12}(\hat{n}_1\cdot\hat{n}_2)&=\int d^3\hat{n}f_4^2(\hat{n}\cdot\hat{n}_1)f_4^2(\hat{n}\cdot\hat{n}_2)\,.
\end{align}
These integrals are manifestly finite and can be easily computed numerically for any value of $\hat{n}_1\cdot\hat{n}_2=\cos\theta$. Additionally, we obtained analytically the expansion of eq.s~\eqref{eq_I11} and~\eqref{eq_I12} for $\theta\rightarrow 0$ and $\theta\rightarrow \pi$ - see app~\ref{app_details_I11} and~\ref{app_details_I12} for details.
Finally, notice the dependence on the sliding scale $\mu$ in eq.~\eqref{eq_partition_function_finite}; as a nontrivial check of our result, it is simple to verify that the $\mu$-dependence cancels at the defect fixed point specified by the solution of eq.~\eqref{eq_beta_h}.\footnote{Note that to this order in perturbation theory the beta function of the bulk coupling~\eqref{eq_beta_BULK} is negligible, so we do not need to use the fixed-point value for the bulk coupling in eq.~\eqref{eq_lambda_fix} for the sliding scale dependence to cancel.}

From eq.~\eqref{eq_partition_function_finite} we obtain the cusp anomalous dimension setting the couplings to their fixed point value and using eq.~\eqref{eq_cusp_def_cyl}.  The nontrivial defect fixed points are parametrized by a $O(N)$ unit vector $\hat{m}$: $\vec{h}^*=h^*\hat{m}$. The cusp anomalous dimension $\Gamma_{\vec{h}_1^*\vec{h}_2^*}$ between two such fixed points thus depends on the angular distance $\theta$ and the product $\hat{m}_1\cdot\hat{m}_2$.  For $N=1$ obviously $\hat{m}_1\cdot\hat{m}_2=1$ or $\hat{m}_1\cdot\hat{m}_2=-1$. Additionally, setting $\vec{h}_2=0$ we obtain the scaling dimension of the defect creation operator $\Delta_{\vec{h}^*\,0}$.  We list and discuss these results below.

\subsubsection{Results and discussion}

\begin{figure}[t!]
    \centering
		\subcaptionbox{  \label{fig:1loop++}}
		{\includegraphics[width=0.475\textwidth]{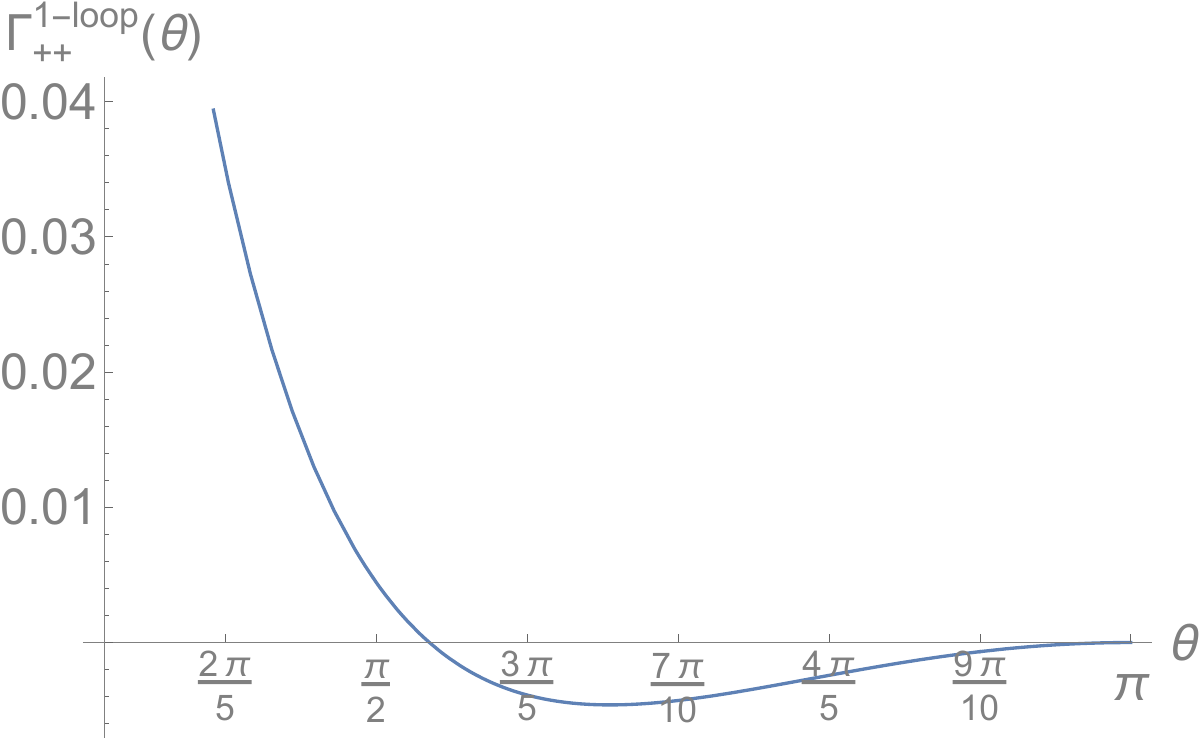}}
 \quad
		\subcaptionbox{ \label{fig:1loop+-}}
		{\includegraphics[width=0.475\textwidth]{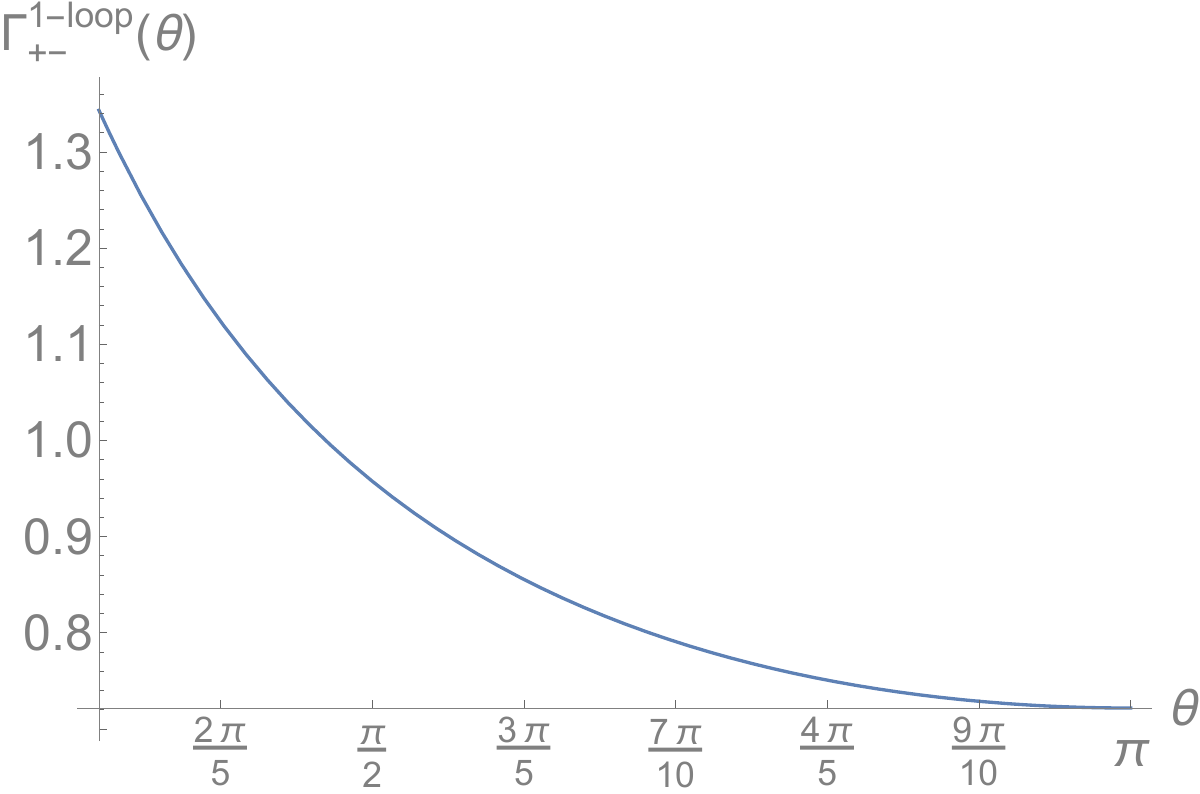}}
    \caption{Plot of the 1-loop correction to the cusp anomalous dimension, i.e. the term multiplied by $\varepsilon$ in eq.~\eqref{eq_Gamma12_res}, for $N=1$ and $\hat{m}_1\cdot\hat{m}_2=1$ (fig.~\ref{fig:1loop++}) and $\hat{m}_1\cdot\hat{m}_2=-1$ (fig.~\ref{fig:1loop+-}).}
    \label{fig:1loop}
\end{figure}

The cusp between pinning fields $\vec{h}_1=h_*\hat{m}_1$ and $\vec{h}_2=h_*\hat{m}_2$ is given by
\begin{equation}\label{eq_Gamma12_res}
\begin{split}
&\Gamma_{\vec{h}_1^*\vec{h}_2^*}(\theta)
=-\frac{N+8}{4\pi^2}\left(\hat{m}_1\cdot\hat{m}_2\frac{\pi-\theta}{\sin\theta}-1\right)\\
&+\varepsilon\frac{N+8}{4\pi^2}\left[
\left(\hat{m}_1\cdot\hat{m}_2\frac{\pi-\theta}{\sin\theta}-1\right)
\frac{N^2 (\log 64-5)+N (96 \log 2-61)+384 \log 2-234}{2 (N+8)^2}\right.
\\
&\left.\hspace*{4.5em}-
\left.\hat{m}_1\cdot\hat{m}_2\frac{d f_{4-\varepsilon}(\cos\theta)}{d
\varepsilon}\right\vert_{\varepsilon=0}
+\frac{\hat{m}_1\cdot\hat{m}_2}{4\pi^4}I_{11}(\cos\theta)
+\frac{1+2\left(\hat{m}_1\cdot\hat{m}_2\right)^2}{16\pi^4}I_{12}(\cos\theta)\right.\\
&
\left.\hspace*{4.5em}+
\frac{3 \zeta (3)}{2 \pi ^2}+\log 8-1
\right][+O\left(\varepsilon^2\right)\,.
\end{split}
\end{equation}
This lengthy expression can be evaluated numerically straightforwardly for any value of $\theta$, $N$ and $\hat{m}_1\cdot\hat{m}_2$. For illustration, we plot the term multiplied by $\varepsilon$ in eq.~\eqref{eq_Gamma12_res} in fig.~\ref{fig:1loop} for $N=1$. 

As formerly mentioned we can also use the partition function calculation~\eqref{eq_partition_function_finite} to compute the scaling dimension of the defect creation operator. We find:
\begin{equation}\label{eq_Delta_defect_creation}
\Delta_{\vec{h}^*\,0}=
\frac{N+8}{8 \pi ^2}+\varepsilon
\left[\frac{3 (N+8)^2 \zeta (3)+\pi ^2 (3 N^2+29N+106)}{16 \pi ^4 (N+8)}\right]+O\left(\varepsilon^2\right)\,.
\end{equation}
Note that the tree-level answer agrees with the free theory result~\eqref{eq_tree_level_Delta_creation} using eq.~\eqref{eq_h_fix}.

\begin{figure}[t!]
       \centering
		\subcaptionbox{  \label{fig:1loopLarge++}}
		{\includegraphics[width=0.475\textwidth]{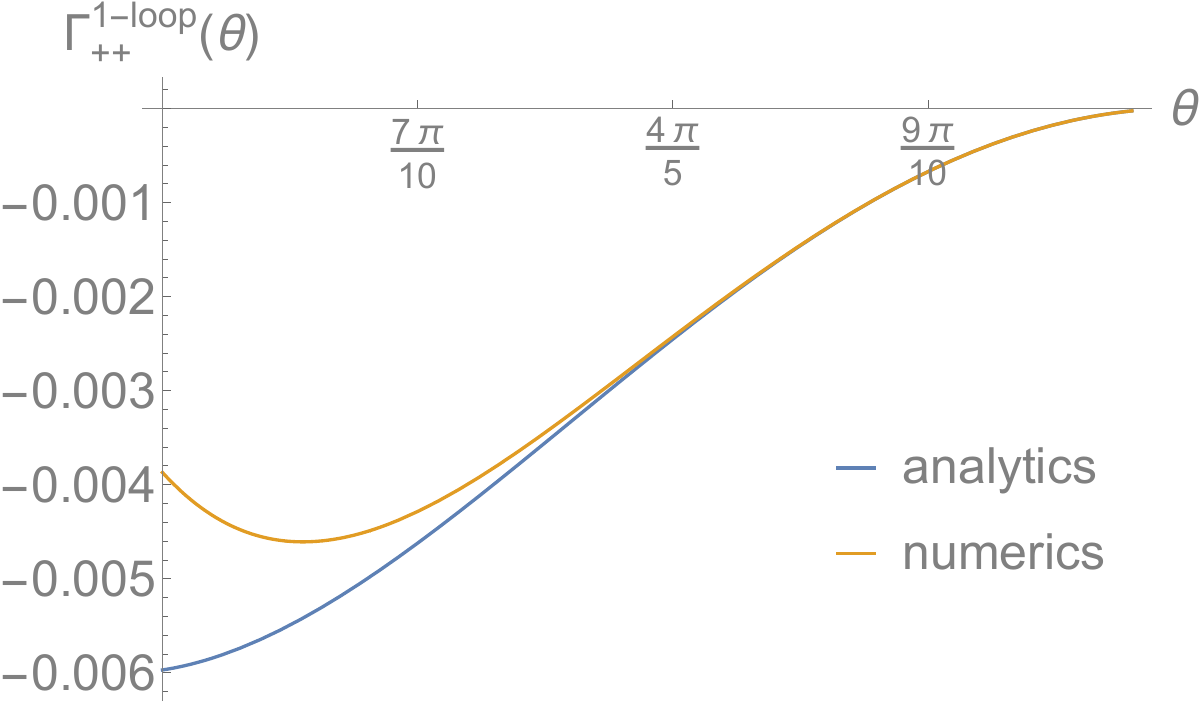}}
 \quad
		\subcaptionbox{ \label{fig:1loopLarge+-}}
		{\includegraphics[width=0.475\textwidth]{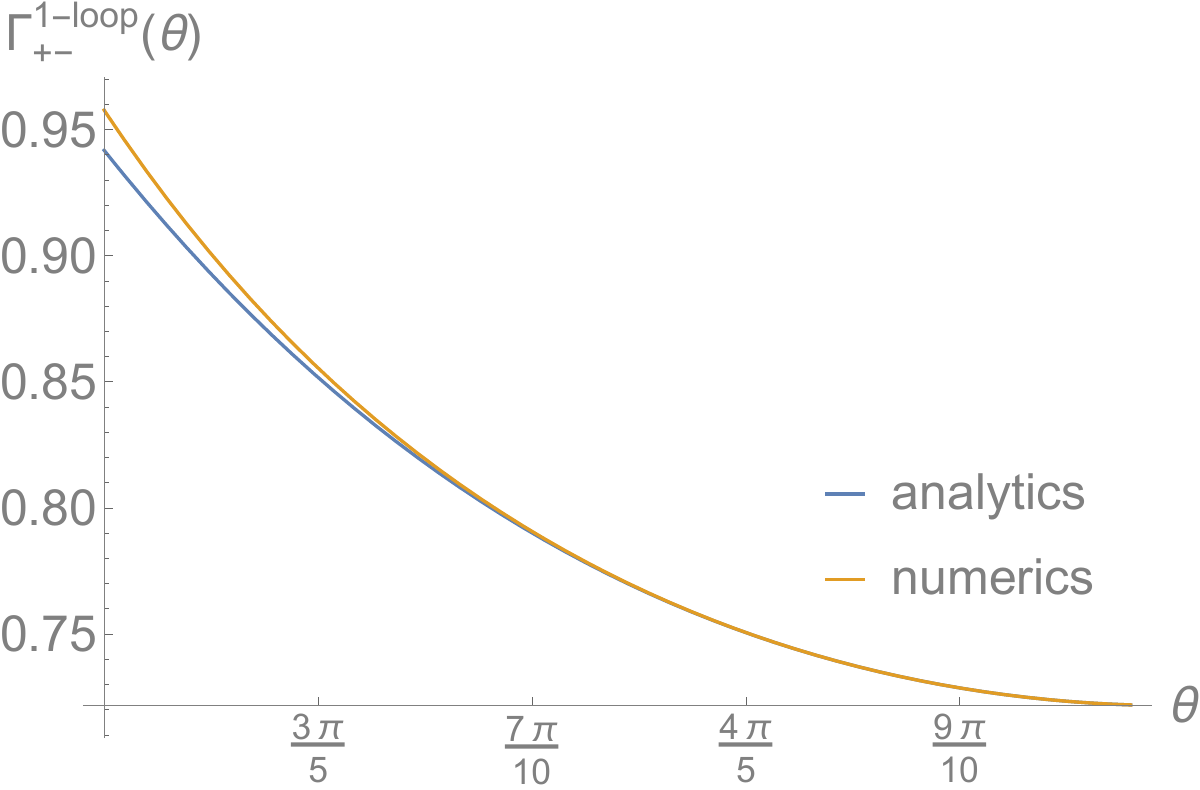}}
    \caption{Comparison of the analytical expansion around $\theta=\pi$ (including the fourth derivative term) and the numerical result~\eqref{eq_Gamma12_res} for the one-loop contribution to the cusp anomalous dimension at $N=1$ and $\hat{m}_1\cdot\hat{m}_2=1$ (fig.~\ref{fig:1loopLarge++}) and $\hat{m}_1\cdot\hat{m}_2=-1$ (fig.~\ref{fig:1loopLarge+-}).}
    \label{fig:1loopLarge}
\end{figure}

We now analyze analytically some limits of the result in what follows.
We discuss first the expansion for $\theta\simeq\pi$. This contains only even powers as predicted and reads
\begin{equation}
\Gamma_{\vec{h}_1^*\vec{h}_2^*}(\theta)\simeq \Gamma_{\vec{h}_1^*\vec{h}_2^*}(\pi)+\frac{1}{2}\Gamma_{\vec{h}_1^*\vec{h}_2^*}''(\pi)(\pi-\theta)^2+O\left((\pi-\theta)^4\right)\,,
\end{equation}
where
\begin{align}\nonumber
\Gamma_{\vec{h}_1^*\vec{h}_2^*}(\pi)&=\Delta_{\vec{h}_1^*\vec{h}_2^*}=
(1-\hat{m}_1\cdot\hat{m}_2)\frac{N+8}{4\pi^2}
\\
\nonumber
&+\varepsilon(1-\hat{m}_1\cdot\hat{m}_2)
\frac{\pi ^2 \left(3 N^2+29 N+106\right)+6 (N+8)^2 (1-\hat{m}_1\cdot\hat{m}_2) \zeta (3)}{8 \pi ^4 (N+8)}\\
&+O\left(\varepsilon^2\right)\,,
\label{eq_Gamma12_Pi}
\\ \nonumber
\frac{1}{2}\Gamma_{\vec{h}_1^*\vec{h}_2^*}''(\pi)&=
-\hat{m}_1\cdot\hat{m}_2\frac{N+8}{24\pi^2}
\\ \nonumber
&+\varepsilon\left\{\frac{(N+8) \left[18 \Sigma_2+3 \pi ^6+48 \pi ^2+4 \pi ^4 (16 \log 2-9)\right]}{1152 \pi ^6}\right. \\
\nonumber
&\quad\quad-\hat{m}_1\cdot\hat{m}_2\frac{\left[5 \pi ^2 N^2+6 N^2+36 (N+8)^2 \zeta (3)+23 \pi ^2 N+96 N+62 \pi ^2+384\right]}{144 \pi ^4 (N+8)}\\ \nonumber
&\quad\quad\left.+(\hat{m}_1\cdot\hat{m}_2)^2\frac{(N+8)  \left[18 \Sigma_2+3 \pi ^6+48 \pi ^2+4 \pi ^4 (16 \log 2-9)\right]}{576 \pi ^6}
\right\}\\
&+O\left(\varepsilon^2\right) \label{eq_Gamma_dd}
\,.
\end{align}
In eq.~\eqref{eq_Gamma_dd} $\Sigma_2\simeq -159.753$ can be expressed analytically in terms of a convergent sum, see eq.~\eqref{eq_app_Sigma2} in the appendix.   The result for $\Gamma_{\vec{h}_1^*\vec{h}_2^*}(\pi)$ is particularly relevant, since it corresponds to the dimension $\Delta_{\vec{h}_1^*\vec{h}_2^*}$ of the domain wall operator between two different pinning field defects.  Note that $\Gamma_{\vec{h}_1^*\vec{h}_2^*}(\pi)=0$ for $\hat{m}_1\cdot\hat{m}_2=1$. In fig.~\ref{fig:1loopLarge} we compare the analytical expansion near $\theta\simeq \pi$ with the numerical evaluation of eq.~\eqref{eq_Gamma12_res} for $N=1$

\begin{figure}[t!]
    \centering
 \subcaptionbox{  \label{fig:1loopSmall++}}
		{\includegraphics[width=0.475\textwidth]{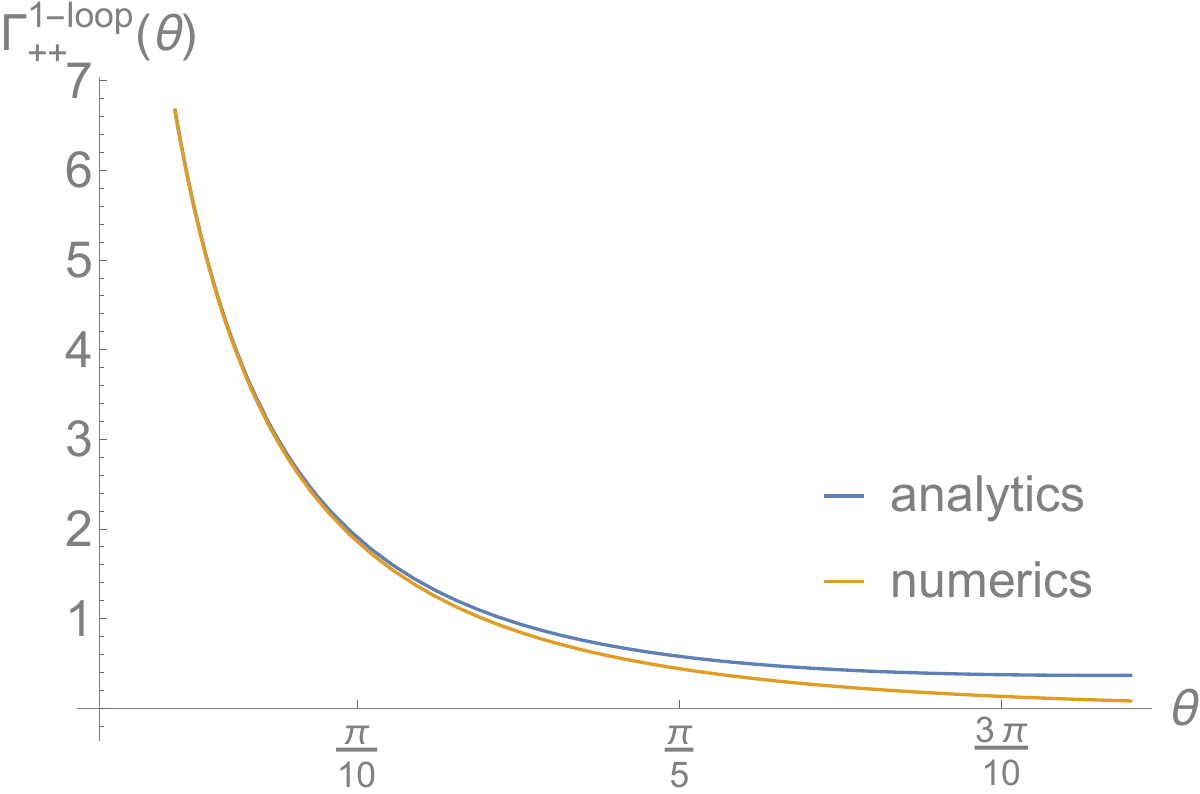}}
 \quad
		\subcaptionbox{ \label{fig:1loopSmall+-}}
		{\includegraphics[width=0.475\textwidth]{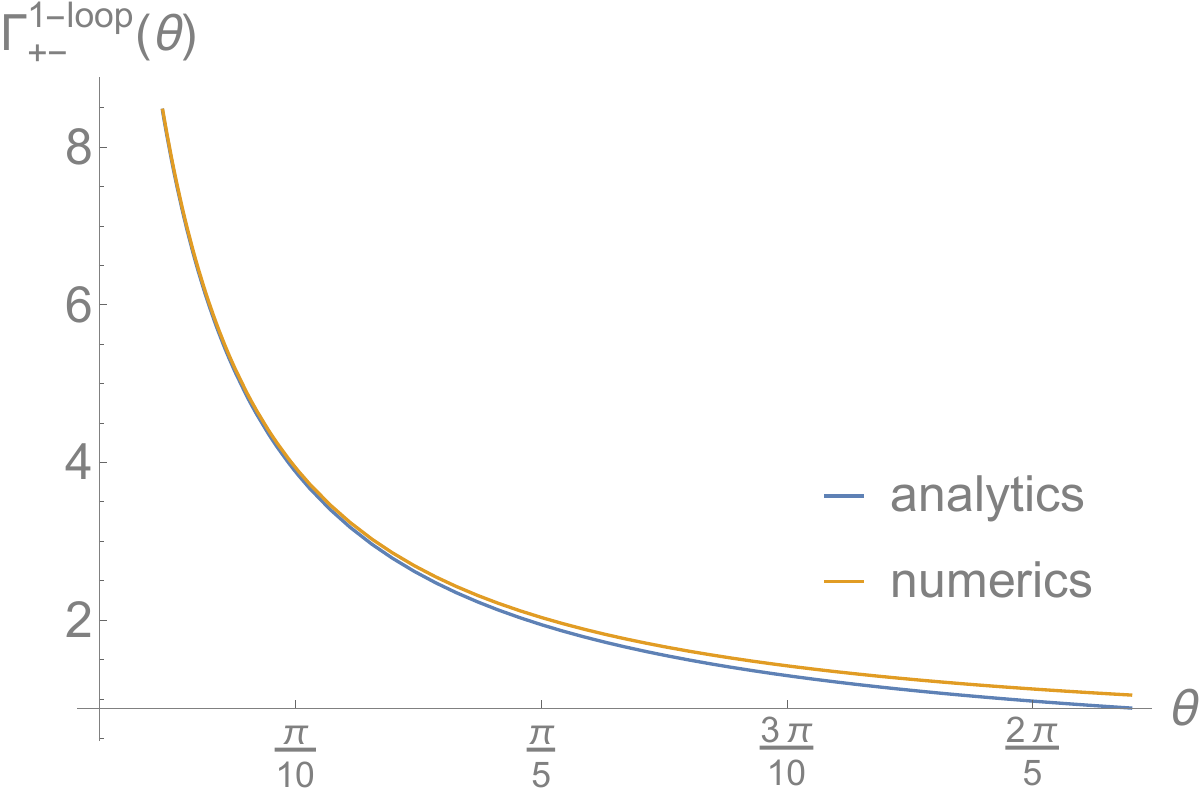}}
    \caption{Comparison of the analytical approximation around $\theta=0$ and the numerical result~\eqref{eq_Gamma12_res} for the one-loop contribution of the cusp at $N=1$ and $\hat{m}_1\cdot\hat{m}_2=1$ (fig.~\ref{fig:1loopSmall++}) and $\hat{m}_1\cdot\hat{m}_2=-1$ (fig.~\ref{fig:1loopSmall+-}).}
    \label{fig:1loopSmall}
\end{figure}

Let us now look at the expansion for $\theta\rightarrow 0$. 
To this order, the expansion for $\theta\rightarrow 0$ takes the form
\begin{equation}\label{eq_one_loop_small}
\Gamma_{\vec{h}_1^*\vec{h}_2^*}(\theta)=\frac{C_{\vec{h}_1^*,\vec{h}_2^*,\vec{h}_{fus}^*}}{\theta}+\delta\log\theta+\alpha+O\left(\theta\right)\,.
\end{equation}
The first term in eq.~\eqref{eq_one_loop_small} is the Casimir energy and reads
\begin{equation}
\begin{split}
C_{\vec{h}_1^*,\vec{h}_2^*,\vec{h}_{fus}^*}&=-\frac{N+8}{4\pi }\hat{m}_1\cdot\hat{m}_2 \\
&+
\varepsilon\left\{
\frac{\pi(N+8)}{64} \left[1+2 (\hat{m}_1\cdot\hat{m}_2)^2\right]-\hat{m}_1\cdot\hat{m}_2\frac{N^2-3N -22 }{8 \pi  (N+8)}\right\}
+O\left(\varepsilon^2\right)\,.
\end{split}
\end{equation}
This result agrees with the recent computation in~\cite{Diatlyk:2024zkk}. The coefficient of the logarithm in eq.~\eqref{eq_one_loop_small} is
\begin{equation}
\delta=\varepsilon\frac{(N+8)(1+\hat{m}_1\cdot\hat{m}_2)(1+2\hat{m}_1\cdot\hat{m}_2)}{4\pi^2}+O\left(\varepsilon^2\right)\,.
\end{equation}
Finally, the constant term in eq.~\eqref{eq_one_loop_small} is given by
\begin{equation}
\begin{split}
\alpha =&\frac{(N+8) (1+\hat{m}_1\cdot\hat{m}_2)}{4 \pi ^2}\\
+&\varepsilon\frac{1+\hat{m}_1\cdot\hat{m}_2}{(N+8)8\pi^2} \left[
\frac{6  \zeta (3)}{\pi ^2}(N+8)^2 (1+\hat{m}_1\cdot\hat{m}_2)
-N^2 (8 \hat{m}_1\cdot\hat{m}_2+1)\right.\\
-&N(35+128 \hat{m}_1\cdot\hat{m}_2)+2 (N+8)^2 (1+2 \hat{m}_1\cdot\hat{m}_2) \log 2-2 (75+256 \hat{m}_1\cdot\hat{m}_2)\Big]\\
+& O\left(\varepsilon^2\right)\,.
\end{split}
\end{equation}
In fig.~\ref{fig:1loopSmall} we compare the expansion~\eqref{eq_one_loop_small} with the numerical evaluation of eq.~\eqref{eq_Gamma12_res} for $N=1$.

It is interesting to compare the result~\eqref{eq_one_loop_small} with the predictions from fusion~\eqref{cusp}. As recently verified to one-loop order in \cite{Diatlyk:2024zkk}, the fusion of two pinning field defects results in another pinning field defect whose infrared coupling is
\begin{equation}\label{eq_h_fus}
\frac{\hat{m}_1+\hat{m}_2}{\sqrt{2+2\hat{m}_1\cdot\hat{m}_2}}h^*\,,
\end{equation}
as it was perhaps intuitively expected on geometric grounds. Eq.~\eqref{eq_h_fus} does not apply for $\hat{m}_1\cdot\hat{m}_2=-1$, in which case the fused defect is trivial.  The result~\eqref{eq_h_fus} can be justified at the nonperturbative level noting that the configuration with two defects at distance $D$ in flat space is invariant under the combined action of two symmetries. The first a is reflection in the $O(N)$ internal space across the line specified by $\hat{m}_1+\hat{m}_2$, such that it exchanges $\hat{m}_1$ and $\hat{m}_2$. The second symmetry action is simply a spatial rotation exchanging the location of the defects. Further using the invariance under the residual $O(N-2)$ symmetry group, this argument implies that the magnetic field of the fused defect must point in the $\hat{m}_1+\hat{m}_2$ direction (this argument does not specify the sign of the magnetic field), and in particular implies that when $\hat{m}_1+\hat{m}_2=0$ the fused defect must be trivial.

As mentioned in sec.~\ref{subsec_eps_review}, the first irrelevant deformation of the pinning field is an operator of dimension $1+\varepsilon+\ldots$, see eq.~\eqref{eq_gamma1_epsilon}. This is the origin of the $\log\theta$ in eq.~\eqref{eq_one_loop_small}.  We will show below that the the logarithm is the first term of an expansion, whose resummation yields a result in agreement with the theoretical expectation eq.~\eqref{cusp}. We note first however that both $\delta$ and $\alpha$ vanish for $\hat{m}_1\cdot\hat{m}_2=-1$, and thus there is no $\theta^0$ term in the expansion~\eqref{eq_one_loop_small}. This is in agreement with the expectations for a trivial fused defect,  providing a very nontrivial test of the structure~\eqref{cusp}.

To quantitatively understand the coefficient of the $\log\theta$ term, we note that, as discussed in sec.~\ref{subsec_free}, at tree-level the fused defect is a pinning field defect with coupling 
\begin{equation}
\vec{h}_{fus}=\vec{h}_1^*+\vec{h}_2^*\qquad\text{(tree-level)}\,.
\end{equation}
For generic relative orientiation, $|\vec{h}_{fus}|\neq |\vec{h}^*|$, and thus the tree-level coupling must flow to the fixed point under RG. Equivalently, the difference $|\vec{h}_{fus}|- |\vec{h}^*|$ can be thought as the coefficient of the leading irrelevant deformation on the pinning field defect.

As explained in sec.~\ref{subsec_limits}, we can think of the cusp anomalous dimension at small angle as the expectation value of the fused defect, with the Casimir energy identified with the cosmological constant term on the fused line. Therefore the constant and logarithmic term in eq.~\eqref{eq_one_loop_small} should be understood as due to the expectation value of a pinning field defect with coupling $\vec{h}_{fus}+O(\varepsilon)$ and UV cutoff $\theta$.  The renormalization group states that changes in the defect coupling are equivalent to rescaling of the UV scale,  and hence Callan-Symanzik equation implies
\begin{equation}\label{eq_Callan_Symanzik}
\frac{\pd \alpha}{\pd h_{fus}}\beta_{h_{fus}}-\theta\frac{\pd}{\pd\theta}\delta\log\theta=0\quad
\implies\quad
\delta=\frac{\pd \alpha}{\pd h_{fus}}\beta_{h_{fus}}\,,
\end{equation}
that is immediately checked at leading order using $\alpha=\frac{h^2_{fus}}{8\pi^2}+O(\varepsilon)$, and eq.s~\eqref{eq_lambda_fix} and~\eqref{eq_beta_h}.  From eq.~\eqref{eq_Callan_Symanzik} we also recognize that the $\log\theta$ term is the first of an infinite series that resums to
\begin{equation}\label{eq_one_loop_small_resum}
\alpha+\delta\log\theta\simeq
\frac{N+8+O\left(\varepsilon\right)}{8 \pi ^2 \left(1-\frac{1+2 \hat{m}_1\cdot\hat{m}_2}{2+2\hat{m}_1\cdot\hat{m}_2} \theta ^{\varepsilon }\right)}\simeq
\frac{h^2_{fus}(\theta)}{8\pi^2}\,,
\end{equation}
where
\begin{equation}
h_{fus}^2(\mu/\mu_0)\equiv\frac{N+8}{ 1-\frac{1+2 \hat{m}_1\cdot\hat{m}_2}{2+2\hat{m}_1\cdot\hat{m}_2} (\mu/\mu_0)^{\varepsilon }}\,,
\end{equation}
is the solution of the beta function equation~\eqref{eq_beta_h} to one-loop order.  Using eq.~\eqref{eq_one_loop_small_resum} in the result~\eqref{eq_one_loop_small}, we see that for finite $\varepsilon$, the constant term in the $\theta\rightarrow 0$ limit approaches the scaling dimension of the defect creation operator $\Delta_{\vec{h}^*\,0}$ to order $O(\varepsilon^0)$. We additionally obtain corrections $\sim\theta^{n\varepsilon}$ for any $n\in \mathds{N}^+$,  corresponding to $n$ insertions of the irrelevant operator $\hat{\phi}$ in conformal perturbation theory .

It is harder to verify that the coefficient of the $\theta^0$ term approaches $\Delta_{\vec{h}^*\,0}$ to order $O(\varepsilon)$, since for that we would need to compute the coefficient of the irrelevant operator $\hat{\phi}$ on the fused defect at subleading order.  We did not perform this check in full generality.  However, we note that for $\hat{m}_1\cdot\hat{m}_2=-\frac{1}{2}$ the coefficient of the logarithm vanishes and the result~\eqref{eq_one_loop_small} takes the form
\begin{equation}\label{eq_one_loop_small_coincidence}
\Gamma_{\vec{h}_1^*\vec{h}_2^*}(\theta)\vert_{\hat{m}_1\cdot\hat{m}_2=-\frac{1}{2}}=\frac{C_{\vec{h}_1^*,\vec{h}_2^*,\vec{h}_{fus}^*}}{\theta}+\Delta_{\vec{h}^*\,0}+O\left(\theta\right)\,,
\end{equation}
where $\Delta_{\vec{h}^*\,0}$ is the scaling dimension of the defect creation operator given in eq.~\eqref{eq_Delta_defect_creation}. Eq.~\eqref{eq_one_loop_small_coincidence} implies that the coefficient of the first irrelevant deformation $\hat{\phi}$ on the fused defect vanishes (to one-loop order at least). This is again geometrically intuitive, since $|\vec{h}^*_1+\vec{h}^*_2|=h^*$ for $\hat{m}_1\cdot\hat{m}_2=-\frac{1}{2}$.

\subsection{Pad\'e extrapolations and comparison with the fuzzy sphere results}

In this section we compare the $\varepsilon$-expansion calculations for the Ising model with the non-perturbative fuzzy sphere results discussed in sec.~\ref{sec_fuzzy}. In order to extrapolate our results in a more reliable way to $d=3$, we will use the exact result in $d=2$ to construct Pad\'e approximants. 
The same approach was used in~\cite{Cuomo:2021rkm} to predict the scaling dimension $\Delta_{\hat{\phi}}$ in $d=3$, and the result was found to be compatible with all the existing numerical calculations (including the fuzzy sphere one). Since we only use one-loop order results, below we always consider Pad\'e approximants of polynomial order $1$ (in $\varepsilon$) both in the numerator and the denominator.

\begin{figure}[t!]
    \centering
    \includegraphics[width=0.6\textwidth]{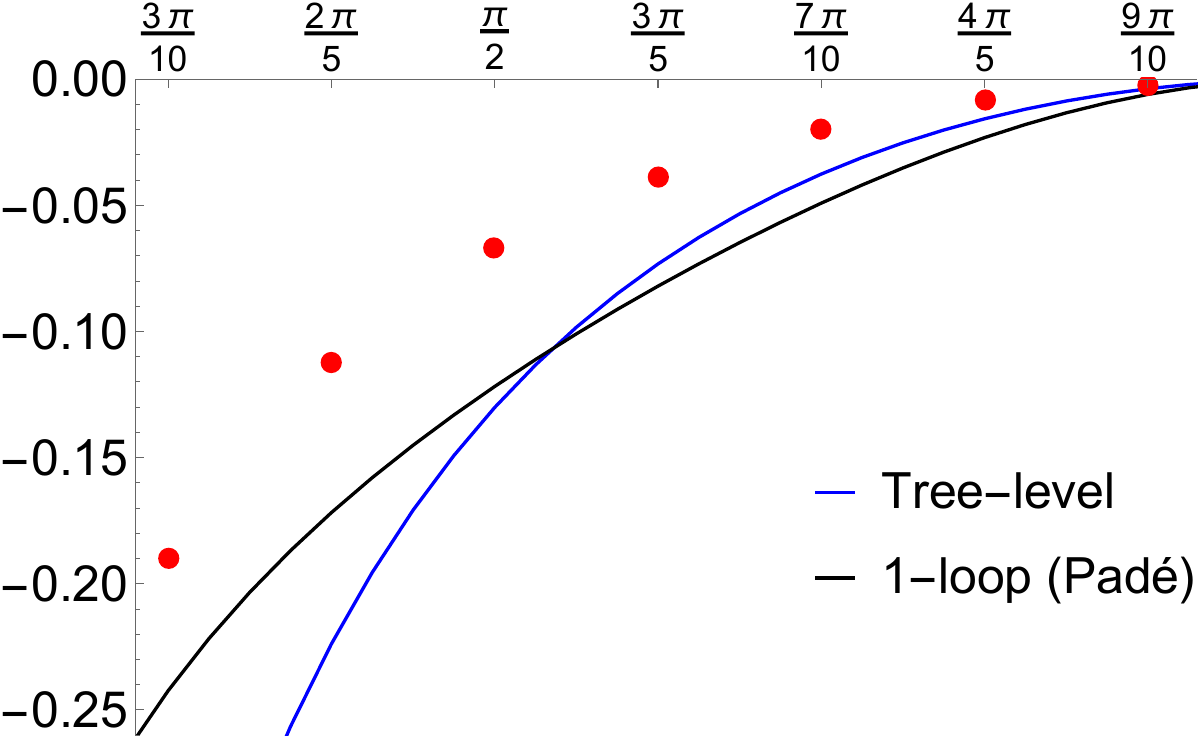}
    \caption{Plot of the 1-loop Pad\'e extrapolation for the cusp anomalous dimension $\Gamma_{++}(\theta)$ (black curve). The red dots are data points obtained with the fuzzy sphere regularization, and the blue curve is the tree-level result extrapolated to $\varepsilon=1$.}
    \label{fig:Pade1}
\end{figure}

Combining the $\varepsilon$-expansion result~\eqref{eq_Gamma12_res} and the $2d$ result~\eqref{eq_Gamma++_2d}, we may easily construct a Pad\'e approximant for the cusp anomalous dimension between equal pinning field defects. The extrapolated result is shown in fig.~\ref{fig:Pade1}, where we compare it both to the tree-level result and to the fuzzy sphere data-points. 

Clearly the one-loop corrected extrapolation seems to work better at small angles compared to the tree-level. However, the one-loop corrected result is further away than the tree-level from the fuzzy sphere data for $\theta\simeq \pi$. Quantitatively,  the extrapolated result close to $\theta=\pi$ is
\begin{equation}\label{eq_Gamma_++_Pade_large}
\Gamma_{++}(\theta)=\begin{cases}
-0.0379954(\pi-\theta)^2+\ldots &\text{tree-level}\\
-0.0621195(\pi-\theta)^2+\ldots&1-\text{loop (Pad\'e)}\,.
\end{cases}
\end{equation}
Comparing eq.~\eqref{eq_Gamma_++_Pade_large} with the prediction in eq.~\eqref{eq_cusp_CD}, we would estimate a value of $C_D\in  (0.45 , 0.75)$ for the displacement operator two-point function. This value is quite far from the fuzzy sphere measurement $C_D\simeq 0.27(2)$.

The small angle limit of the Pad\'e improved result is
\begin{equation}\label{eq_Gamma_++_Pade_small}
\Gamma_{++}(\theta)=\begin{cases}
-\frac{0.716197}{\theta }+0.455945+\ldots &\text{tree-level}\\
-\frac{0.185971}{\theta }+0.100866+0.0938353 \log \theta
+\ldots &1-\text{loop (Pad\'e)}\,,
\end{cases}
\end{equation}
which shows that the extrapolation predicts a smaller Casimir energy than the tree-level; therefore the one-loop result is closer than the tree-level answer to the fuzzy sphere data, but it is still not compatible with the result~\eqref{eq_fuzzy_Casimir} that we obtained in sec.~\ref{sec_fuzzy}.  Note also that the term proportional to $\log\theta$ in eq.~\eqref{eq_Gamma_++_Pade_small} is not compatible with fusion in $d=3$ and is just an artifact of the perturbative result. We interpret its coefficient as an estimate of the error on the $\theta^0$ term.  Therefore eq.~\eqref{eq_Gamma_++_Pade_small} suggest that the scaling dimension of the defect creation operator lies in the interval $\Delta_{+0}\in (0, 0.2)$, in agreement with the result of~\cite{Zhou:2023fqu} quoted in table~\ref{tab:Fuzzy_pinning}.

As mentioned in sec.~\ref{subsec_2d}, the fusion between pinning field defects of opposite strengths is discontinuous at $d=2$. This also implies that the cusp anomalous dimension between pinning field defects in opposite directions must be discontinuous\footnote{A similar discontinuity in the fusion occurs between a free scalar theory in $d=4$, where the strength of the pinning field is a marginal parameter,  and the interacting theory, in which there is a unique defect fixed point up the action of the symmetry. In that case, as we have seen around eq.~\eqref{eq_one_loop_small_resum}, the result for the cusp anomalous dimension is continuous thanks to the existence of an almost marginal operator near $d=4$. There is no analogous marginal operator on the interface of interest in $d=2$, and thus the result for the cusp cannot be continuous in this case.} as we approach $d=2$, and therefore we refrain from constructing Pad\'e approximants for the full result $\Gamma_{+-}(\theta)$. 

We nonetheless consider the extrapolation of our results for the scaling dimension $\Delta_{+-}$ of the domain wall operator, and for the Casimir energy $C_{+-0}$. We believe that this is justified since these two quantities can be defined independently of the cusp.  The result for the scaling dimension of the domain wall operator follows from the $2d$ result~\eqref{eq_Gamma+-_2d} and eq.~\eqref{eq_Gamma12_Pi}, while for the Casimir energy we interpolate between eq.s~\eqref{eq_Gamma+-_2d} and~\eqref{eq_one_loop_small}. We report the results in table~\ref{tab:Pade},  where we compare the extrapolations to the fuzzy sphere results. Both of the extrapolations are in excellent agreement with the non-perturbative fuzzy sphere results.

\begin{table}[t]
    \centering
    \begin{tabular}{|c|c|c|}
    \hline
          &  $\Delta_{+-}$ & $C_{+-0}$ \\
         \hline
        Pad\'e &  0.85 & 1.31 
        \\
Fuzzy sphere  &  0.84(4) & 1.4(2)  \\        
         \hline
    \end{tabular}
    \caption{In this table we compare the Pad\'e extrapolations of the $\varepsilon$-expansion results for the dimension of the domain wall operator $\Delta_{+-}$ and the Casimir energy $C_{+-0}$ to $d=3$ and the corresponding non-perturbative fuzzy sphere results.}
    \label{tab:Pade}
\end{table}

We conclude by mentioning that it should also be possible to obtain an estimate for the scaling dimension of the defect creation operator $\Delta_{+0}$ by directly extrapolating the $\varepsilon$-expansion result~\eqref{eq_Delta_defect_creation}, rather than comparing with the small angle expansion of the extrapolation as we did below eq.~\eqref{eq_Gamma_++_Pade_small}. However, the one-loop correction to the scaling dimension of the defect creation operator~\eqref{eq_Delta_defect_creation} has the wrong sign compared to the $2d$ result~\eqref{eq_2d_DW++}.  This implies that a first order Pad\'e approximant leads to a pole at finite $\varepsilon$ - we thus do not consider it.  We expect that this issue will be resolved upon constructing Pad\'e approximants of higher order with the two-loop $\varepsilon$-expansion result. Note also that the extrapolation of the naive tree-level $\varepsilon$-expansion result gives $\Delta_{+0}\simeq 0.11$, which is in good agreement with the fuzzy sphere result in table~\ref{tab:Fuzzy_pinning}.

\section*{Acknowledgements}

We thank H. Khanchandani, M. Metlitski, F. Popov, R. Lanzetta, A. Sever, Y. Wang, Z. Zhou, Y. Zou for useful discussions.  
GC is supported by the Simons Foundation grant 994296 (Simons Collaboration on Confinement and QCD Strings). ZK is supported in part by the Simons Foundation grant 488657 (Simons Collaboration on the Non-Perturbative Bootstrap), the BSF grant no. 2018204 and NSF award number 2310283. Research at Perimeter Institute is supported in part by the Government of Canada through the Department of Innovation, Science and Industry Canada and by the Province of Ontario through the Ministry of Colleges and Universities. 

\newpage

\appendix

\part*{Appendix}
\addcontentsline{toc}{part}{Appendix}

\section{Details on the \texorpdfstring{$\varepsilon$}{epsilon}-expansion}\label{app_eps_exp_calc}

\subsection{Bare perturbation theory in the minimal subtraction scheme}

In this section we provide some details on the scheme we used for calculations. We will work on the cylinder $\mathds{R}\times S^{d-1}$ with radius $R=1$ and metric
\begin{equation}
    ds^2=d\tau^2+d\hat{n}^2\,.
\end{equation}
The action reads
\begin{equation}\label{eq_app_BulkAction_epsilon}
S=\int d^dx\left[\frac{1}{2}(\pd\phi_a)^2+\frac{1}{2}m_d^2\phi_a^2+\frac{\lambda_0}{4!}\left(\phi_a^2\right)^2\right]\,,
\end{equation}
where 
\begin{equation}
m_d^2=\frac{d-2}{4(d-1)}\mathcal{R}=\frac{(d-2)^2}{4} \,.   
\end{equation}
is the conformal mass coupling. The model can be studied by  expanding perturbatively in the coupling $\lambda_0$ and using the propagator
\begin{equation}\label{eq_propagator_free_cyl}
\langle \phi_a(\tau,\hat{n}_x)\phi_b(0,\hat{n}_y)\rangle_{\lambda_0=0}=
\delta_{ab}G_{cyl}(\tau,\hat{n}_x\cdot\hat{n}_y)\,,
\end{equation}
where we defined
\begin{equation}\label{eq_cyl_prop}
G_{cyl}(\tau,\hat{n}_x\cdot\hat{n}_y)=\mathcal{N}_d\frac{e^{-\frac{d-2}{2}|\tau|}}{(1-2\hat{n}_x\cdot\hat{n}_ye^{-|\tau|}+e^{-2|\tau|})^{\frac{d-2}{2}}}\,,\qquad
\mathcal{N}_d=\frac{1}{(d-2)\Omega_{d-1}}\,.
\end{equation}
Here $\Omega_{d-1}=\frac{2\pi^{d/2}}{\Gamma(d/2)}$ is the volume of the $d-1$-dimensional sphere.

We work in dimensional regularization within the minimal subtraction scheme, for which the relation between the bare coupling and the renormalized (physical) one is expressed through an ascending series of poles at $\varepsilon= 0$:
\begin{equation}\label{eq_bare_renormalized_BULK}
\lambda_0=\mu^{\varepsilon}\left(\lambda+\frac{\delta\lambda}{\varepsilon}+\frac{\delta_2\lambda}{\varepsilon^2}+\ldots\right)\,,
\end{equation}
where $\mu$ is the sliding scale.  To two-loop order only $\delta\lambda$ and $\delta_2\lambda $ are nonzero and read \cite{Kleinert:2001ax}:
\begin{equation}
\begin{split}
\delta\lambda&=\frac{N+8}{3}\frac{\lambda^2}{(4\pi)^2}-
\frac{3N+14}{6}\frac{\lambda^3}{(4\pi)^4}
+O\left(\frac{\lambda^4}{(4\pi)^6}\right)\,,\\[0.7em]
\delta_2\lambda&=\frac{(N+8)^2}{9}\frac{\lambda^3}{(4\pi)^4}
+O\left(\frac{\lambda^4}{(4\pi)^6}\right)\,.
\end{split}
\end{equation}
From eq.~\eqref{eq_bare_renormalized_BULK} we easily extract the beta function of the coupling via
\begin{equation}\label{eq_app_beta_BULK}
\frac{\pd\lambda}{\pd\log \mu}\equiv\beta_{\lambda}=
-\varepsilon\lambda+
\lambda\frac{\pd\delta\lambda}{\pd\lambda}-\delta\lambda
\,,
\end{equation}
which gives eq.~\eqref{eq_beta_BULK} in the main text.

To model the pinning field defect we perturb the action as
\begin{equation}\label{eq_app_DefectAction_epsilon}
S\rightarrow S+\int_Dd\tau\, \vec{h}_0\cdot \vec{\phi}\left(x(\tau)\right)\,.
\end{equation}
We renormalize the defect coupling similarly to eq.\footnote{Note that here we use a different normalization for the counterterms compared to \cite{Cuomo:2021kfm}.}~\eqref{eq_bare_renormalized_BULK}:
\begin{equation}\label{eq_h0_renormalization}
\vec{h}_0=\mu^{\varepsilon/2}\vec{h}\left(1+\frac{\delta h}{\varepsilon}
+\frac{\delta_2 h}{\varepsilon^2}+\ldots\right)\,.
\end{equation}
We find \cite{Cuomo:2021kfm}
\begin{align}\label{eq_h0_renormalization2}
\delta h&=\frac{\lambda}{(4\pi)^2}\frac{h^2}{12}+
\frac{\lambda^2}{(4\pi)^4}\left(\frac{N+2}{72}-
\frac{N+8}{108} h^2-\frac{h^4}{48}\right)+
O\left(\frac{\lambda^3}{(4\pi)^6}\right)\,,\\[0.7em]
\delta_2h&=\frac{\lambda^2}{(4\pi)^4}
\left(\frac{N+8}{108} h^2 +\frac{h^4}{96}\right)
+O\left(\frac{\lambda^3}{(4\pi)^6}\right)\,.
\end{align}
Using the independence of the bare coupling $h_0$ from the sliding scale, from eq.~\eqref{eq_h0_renormalization} we find the beta-function of the defect coupling analogously to eq.~\eqref{eq_app_beta_BULK}; the result is given in eq.~\eqref{eq_beta_h} in the main text to order $O\left(\lambda^2\right)$.

\subsection{The cylinder partition function}\label{app_details_finite}

We consider the partition function for arbitrary pinning fields $\vec{h}_1$ and $\vec{h}_2$ (not necessarily at the fixed point):
\begin{equation}
\log Z_{\vec{h}_1\vec{h}_2}(\theta)=-TE_{12}(\theta)\,,\qquad E_{12}(\theta)=E_{12}^{(0)}(\theta)+\lambda E_{12}^{(1)}(\theta)+\ldots\,.
\end{equation}
We discuss below how to recast it in terms of finite quantities to one-loop order.  

The tree-level term is given by the sum of the diagrams~\ref{fig:treeCyl1} and~\ref{fig:treeCyl2}:
\begin{equation}
-E_{12}^{(0)}(\theta)=\vec{h}_{0,1}\cdot\vec{h}_{0,2}T_d(\cos\theta)+\frac{1}{2}\left(\vec{h}_{0,1}^2+\vec{h}_{0,2}^2\right)T_d(1)\,,
\end{equation}
where the subscript stress that all couplings are bare and
\begin{align}\label{eq_app_fd_def}
T_d(x)\equiv\int d\tau G_{cyl}(\tau,x)=\mathcal{N}_d f_d(x)\,,
\end{align}
where $f_d(x)$ is given in eq.~\eqref{eq_fd_def} in the main text. The terms proportional to $T_d(1)$ are self-energy diagrams and will thus drop from the normalized physical cusp anomalous dimension~\eqref{eq_cusp_def_cyl}. Below we list some properties of the function $f_d(x)$ that we will need in what follows. 

First, the result in $d=4$ is
\begin{equation}\label{eq_f4}
f_4(\cos\theta)=\frac{\pi -\theta }{ \sin (\theta )}\,,\qquad
\lim_{d\rightarrow 4}f_d(1)=-1\,,
\end{equation}
where the result at $\cos\theta=1$ is obtained via analytic continuation from $d<3$, i.e. renormalizing away a cutoff divergence in dimensional regularization.  
Expanding for $d=4-\varepsilon$, the next correction to eq.~\eqref{eq_f4} is given by the following integral:
\begin{equation}\label{eq_df4_deps}
\left.\frac{df_{4-\varepsilon}(x)}{d\varepsilon}\right\vert_{\varepsilon=0}=\int d\tau\frac{e^{-|\tau| } \left[| \tau| +\log \left(1-2 x e^{-| \tau| }+e^{-2 | \tau| }\right)\right]}{2 \left(1-2 x e^{-| \tau| }+e^{-2 | \tau| }\right)}\,.
\end{equation}
Eq.~\eqref{eq_df4_deps} admits a complicated closed form expression in terms of polylogarithms. In practice it is easier to evaluate numerically the integral for the desired value of the argument. Eq.~\eqref{eq_df4_deps} admits the following expansions for $\theta\rightarrow 0$ and $\theta\rightarrow\pi$: 
\begin{equation}
\left.\frac{df_{4-\varepsilon}(\cos\theta)}{d\varepsilon}\right\vert_{\varepsilon=0}=\begin{cases}
\displaystyle\frac{\pi  \log (2\theta )}{\theta }-1  +O\left(\theta\right) &\text{for }\theta\ll 1\\[0.5em]
\displaystyle 1+\frac{1}{18} (\pi-\theta  )^2+O\left((\pi-\theta )^4\right) &\text{for } (\pi-\theta)^2\ll 1\,.
\end{cases}
\end{equation}
Finally, we will also need the short distance behaviour $x\rightarrow 1$ of $f_d(x)$ for arbitrary number of dimensions $d$:
\begin{equation}\label{eq_app_fd_exp}
f_d(x)=\underbrace{\frac{\sqrt{\pi } 2^{\frac{3}{2}-\frac{d}{2}} \Gamma \left(\frac{d-3}{2}\right)}{(1-x)^{\frac{d-3}{2}} \Gamma \left(\frac{d}{2}-1\right)}}_{\equiv\tilde{f}_d(x)}+\frac{2^{3-d} \Gamma \left(\frac{3}{2}-\frac{d}{2}\right) \Gamma \left(\frac{d}{2}-1\right)}{\sqrt{\pi }}+O\left((1-x)^{\frac{5-d}{2}},(1-x)\right)\,.
\end{equation}

Let us now consider the one-loop correction to the parition function. This is the sum of three terms:
\begin{equation}\label{eq_1loop_part}
\begin{split}
-\lambda E_{12}^{(1)}(\theta)=&-\frac{\lambda}{3!}\left[\vec{h}_{0,1}^2\,\vec{h}_{0,1}\cdot\vec{h}_{0,2}+\vec{h}_{0,2}^2\,\vec{h}_{0,1}\cdot\vec{h}_{0,2}\right]L_{11}(\cos\theta)\\
&
-\frac{\lambda}{4}\left[\frac{1}{3}\vec{h}_{0,1}^2\,\vec{h}_{0,2}^2+
\frac{2}{3}\left(\vec{h}_{0,1}\cdot\vec{h}_{0,2}\right)^2\right]L_{12}(\cos\theta)
-\frac{\lambda}{12}\left[\left(\vec{h}_{0,1}^2\right)^2+\left(\vec{h}_{0,2}^2\right)^2\right]L_1
\end{split}
\end{equation}
where we defined
\begin{align}
L_{11}(\hat{n}_1\cdot\hat{n}_2)&=\int d^{d-1}\hat{n}\left[\int d\tau G_{cyl}(\tau,\hat{n}\cdot\hat{n}_1)\right]^3\left[\int d\tau G_{cyl}(\tau,\hat{n}\cdot\hat{n}_2)\right]\,,\\
L_{12}(\hat{n}_1\cdot\hat{n}_2)&=\int d^{d-1}\hat{n}\left[\int d\tau G_{cyl}(\tau,\hat{n}\cdot\hat{n}_1)\right]^2\left[\int d\tau G_{cyl}(\tau,\hat{n}\cdot\hat{n}_2)\right]^2\,,\\
L_1&=\int d^{d-1}\hat{n}\left[\int d\tau G_{cyl}(\tau,\hat{n}\cdot\hat{n}_1)\right]^4\,.
\end{align}
The last term in the last line of eq.~\eqref{eq_1loop_part} is the contribution from the self-energy diagrams~\ref{fig:OneLoopSelf},  that  do not connect the two lines, and thus will drop out from the normalized physical expression~\eqref{eq_cusp_def_cyl}. We will not disuss this term in detail, but we will nonetheless present its value at the end of this section. Note that at one-loop order the difference between bare and renormalized coupling for $\lambda$ is of higher order, and thus we do not distinguish between the two here. Below we discuss in detail the last two terms in eq.~\eqref{eq_1loop_part} individually.

The first term in eq.~\eqref{eq_1loop_part} arises from the diagram~\ref{fig:OneLoopTriangle} and requires some care, since it is not finite without regularization in $d=4$. Upon performing the $\tau$ integrals it becomes
\begin{equation}
L_{11}(\hat{n}_1\cdot\hat{n}_2)=\mathcal{N}_d^4\int d^{d-1}\hat{n}f_d^3(\hat{n}\cdot\hat{n}_1)f_d(\hat{n}\cdot\hat{n}_2)\,.
\end{equation}
Note now that $f_4^3(\hat{n}\cdot\hat{n}_1)\sim 1/\theta_1^3$ for $\theta_1\rightarrow 0$, while the measure only contributes as $\sqrt{g}\sim\theta_1^2$ in $d=4$.  Therefore there is a logarithmic divergence, that will be renormalized away by the defect counterterms. To isolate the divergent piece we perform a subtraction and recast the result as
\begin{equation}
L_{11}(\hat{n}_1\cdot\hat{n}_2)=\mathcal{N}_d^4\left[
f_d(\hat{n}_1\cdot\hat{n}_2)\int d^{d-1}\hat{n}\tilde{f}_d^3(\hat{n}\cdot\hat{n}_1)+I_{11}(\hat{n}_1\cdot\hat{n}_2)+O\left(\varepsilon\right)\right]\,,
\end{equation}
where $\tilde{f}_d(x)$ is the first term in eq.~\eqref{eq_app_fd_exp} and we defined the following finite integral
\begin{equation}
I_{11}(\hat{n}_1\cdot\hat{n}_2)=\int d^3\hat{n}\left[
f_4^3(\hat{n}\cdot\hat{n}_1)f_4(\hat{n}\cdot\hat{n}_2)-
f_4(\hat{n}_1\cdot\hat{n}_2)\tilde{f}_4^3(\hat{n}\cdot\hat{n}_1)
\right]\,.
\end{equation}
We will analyze this integral in detail in sec.~\ref{app_details_I11}.

The second term in eq.~\eqref{eq_1loop_part} arises from the diagram~\ref{fig:OneLoopCross} and is manifestly finite in $d=4$. Therefore we can simply set
\begin{equation}
L_{12}(\hat{n}_1\cdot\hat{n}_2)=\mathcal{N}_4^4I_{12}(\hat{n}_1\cdot\hat{n}_2)\,,
\end{equation}
where we defined the following finite integral
\begin{equation}
I_{12}(\hat{n}_1\cdot\hat{n}_2)=\int d^3\hat{n}f_4^2(\hat{n}\cdot\hat{n}_1)f_4^2(\hat{n}\cdot\hat{n}_2)\,.
\end{equation}
We will analyze this integral in detail in sec.~\ref{app_details_I12}.

Expanding the bare defect couplings using eq.~\eqref{eq_h0_renormalization}, we finally recast the partition function as
\begin{equation}
\begin{split}
&-E_{12}(\theta)=\mu^{2\varepsilon}\vec{h}_1\cdot\vec{h}_2 f_d(\cos\theta)\left[\frac{\mathcal{N}_d}{\mu^{\varepsilon}}\left(1+\frac{\delta h_1+\delta h_2}{\varepsilon}\right)-\frac{\lambda}{6 } \left(\vec{h}_1^2+\vec{h}_2^2\right)\mathcal{N}_d^4\int d^{d-1}\hat{n}\tilde{f}^3_d(\hat{n}\cdot\hat{n}_0)\right]\\
&
-\frac{\lambda}{6}\mathcal{N}_4^4\left[\vec{h}_1^2\,\vec{h}_1\cdot\vec{h}_2+\vec{h}_2^2\,\vec{h}_1\cdot\vec{h}_2\right]I_{11}(\cos\theta)-\frac{\lambda}{12}\mathcal{N}_4^4\left[\vec{h}_1^2\,\vec{h}_2^2+2(\vec{h}_1\cdot\vec{h}_2)^2\right]I_{12}(\cos\theta)\\
&+\frac{\vec{h}_1^2}{2}\left[T_d(1)\left(1+2\frac{\delta h_1}{\varepsilon}\right)-\frac{\lambda}{12}\vec{h}_1^2L_1\right]+
\frac{\vec{h}_2^2}{2}\left[T_d(1)\left(1+2\frac{\delta h_2}{\varepsilon}\right)-\frac{\lambda}{12}\vec{h}_2^2L_1\right]+O(\varepsilon^2,\varepsilon\lambda,\lambda^2)\,,
\end{split}
\end{equation}
where again the last line is just the self-energy and will drop from physical observables.  Evaluating the integral in the square parenthesis in the first line~\eqref{eq_cyl_prop}, we arrive at eq.~\eqref{eq_partition_function_finite} in the main text.

For completeness we report the result for the self-energy diagrams below
\begin{multline}
\frac{\vec{h}_1^2}{2}\left[T_d(1)\left(1+2\frac{\delta h_1}{\varepsilon}\right)-\frac{\lambda}{12}\vec{h}_1^2L_1\right]=
-\vec{h}_1^2\left\{\frac{1}{8 \pi ^2}
+\frac{\varepsilon  \left[2 \log \mu+\log\pi+\gamma +2\right]}{16 \pi ^2}
\right.\\
\left.-\frac{\lambda\vec{h}_1^2 \left[\pi ^2 \left(2 \log \mu+\log \pi +\gamma +3\right)-3 \zeta (3)\right]}{768 \pi ^6}+
O\left(\varepsilon^2,\varepsilon\lambda,\lambda^2\right)
\right\}\,,
\end{multline}
and identically for the term proportional to $\vec{h}_2$. Note that the result is finite. It can be checked that the dependence on the sliding scale drops out at the fixed point.

\subsection{The integral \texorpdfstring{$I_{11}$}{I11}}\label{app_details_I11}

The integral
\begin{equation}
I_{11}(\hat{n}_1\cdot\hat{n}_2)=\int d^3\hat{n}\left[
f_4^3(\hat{n}\cdot\hat{n}_1)f_4(\hat{n}\cdot\hat{n}_2)-
f_4(\hat{n}_1\cdot\hat{n}_2)\tilde{f}_4^3(\hat{n}\cdot\hat{n}_1)
\right]\,.
\end{equation}
can be easily computed numerically in hyperspherical coordinates setting
\begin{equation}\label{eq_app_n_def}
\begin{gathered}
\hat{n}_1=(1,0,0,0)\,,\qquad
\hat{n}_2=(\cos\theta,\sin\theta,0,0)\,,\\
\hat{n}=(\cos \theta_x,\sin\theta_x \cos\phi_x,\sin\theta_x \sin\phi_x \cos\xi_x,\sin\theta_x \sin\phi_x \sin\xi_x)\,.
\end{gathered}
\end{equation}
Therefore
\begin{equation}\label{eq_app_n_prod}
\hat{n}_1\cdot\hat{n}_2=\cos\theta\,,\quad
\hat{n}_1\cdot\hat{n}=\cos\theta_x\,,\quad
\hat{n}_2\cdot\hat{n}=\cos \theta  \cos \theta_x+
\sin \theta\sin\theta_x \cos\phi_x\,,
\end{equation}
and the measure is
\begin{equation}\label{eq_app_n_measure}
\int d^3\hat{n}=\int_0^\pi d\theta_x \sin^2\theta_x\int_0^\pi d\phi_x\sin\phi_x\int_0^{2\pi}d\xi_x\,.
\end{equation}
Note that the integrand is independent of $\xi_x$, while only $f_4(\hat{n}\cdot\hat{n}_2)$ depends on $\phi_x$. Therefore using
\begin{equation}
g_4(\theta,\theta_x)=2\pi\int_0^\pi d\phi_x \sin \phi_x f_4(\hat{n}\cdot\hat{n}_2)=
4 \pi\times\begin{cases}
\displaystyle \frac{\theta_x (\pi -\theta ) }{\sin\theta\sin\theta_x} &\text{for }\theta>\theta_x\\[0.9em]
\displaystyle \frac{\theta  (\pi -\theta_x)}{\sin\theta\sin\theta_x}  
&\text{for }\theta_x>\theta\,,
\end{cases}
\end{equation}
we arrive at
\begin{equation}
I_{11}(\cos\theta)=\int_0^\pi d\theta_x\sin^2\theta_x\left[
f_4^3(\cos\theta_x)g_4(\theta,\theta_x)
-4\pi\tilde{f}_4^3(\cos\theta_x)f_4(\cos\theta)\right]\,.
\end{equation}
This integral admits a closed form expression in terms of a sum of many polylogarithms, but in practice it is easier to evaluate it numerically upon by breaking the integration into two regions $\theta_x \in (0,\theta)\cup(\theta,\pi)$.

From the closed form result, we extract the limiting behaviours. For $\theta\ll 1$ we have
\begin{equation}\label{eq_II_small}
I_{11}(\cos\theta)=
\frac{4 \pi ^5 (\log \theta -\log 4+2)}{\theta }+
4 \left(3 \pi ^4 \log \theta+\pi ^4 \log 64 +6 \pi ^2 \zeta (3)-8 \pi ^4\right)+O\left(\theta\right)\,,
\end{equation}
while for $(\pi-\theta)^2\ll 1$ we obtain 
\begin{equation}\label{eq_I11_large}
\begin{split}
I_{11}(\cos\theta)&=8 \pi ^4-4 \pi ^4 \log 8-24 \pi ^2 \zeta (3)\\
&+\frac{4 \pi ^4-2\pi^2-2\pi^4\log 8-12 \pi ^2 \zeta (3)}{3}(\pi-\theta)^2+O\left((\pi-\theta)^4\right)\,.
\end{split}
\end{equation}

\subsection{The integral \texorpdfstring{$I_{12}$}{I12}}\label{app_details_I12}

The integral
\begin{equation}\label{eq_app_I12}
I_{12}(\hat{n}_1\cdot\hat{n}_2)=\int d^3\hat{n}f_4^2(\hat{n}\cdot\hat{n}_1)f_4^2(\hat{n}\cdot\hat{n}_2)\,.
\end{equation}
can computed numerically using \eqref{eq_app_n_def},~\eqref{eq_app_n_prod} and~\eqref{eq_app_n_measure}; explicitly:
\begin{equation}
I_{12}(\cos\theta)=2\pi
\int_0^\pi d\theta_x \int_0^\pi d\phi_x \sin^2\theta_x\sin\phi_x
f_4^2(\cos\theta_x)f_4^2(\cos \theta  \cos \theta_x+
\sin \theta\sin\theta_x \cos\phi_x)\,,
\end{equation}
where again in practice it is convenient to break the contour into two pieces $\theta_x \in (0,\theta)\cup(\theta,\pi)$ for the numerical evaluation.  At $\theta=\pi$ the integral over $\phi_x$ trivializes and we obtain the following exact result
\begin{equation}\label{eq_app_I12_Pi}
I_{12}(-1)=24 \pi ^2 \zeta (3)\,.
\end{equation}
Unfortunately the naive Taylor expansion around $\theta=\pi$ does not commute with the $\phi_x$ integration, and therefore we cannot naively expand the integrand to obtain the subleading correction as $\theta\rightarrow\pi$.

To obtain analytical approximations, we express the result as a sum over Gegenbauers. To this aim, we first obtained the decomposition of $f_4^2(\cos\theta)$ into Gegenbauer polynomials
\begin{equation}\label{eq_app_f4^2_geg}
f_4^2(\cos\theta)=\sum_{\ell=0}^\infty
g_{\ell}C^{(1)}_\ell(\cos\theta)\,,\qquad
g_{\ell}=\pi ^2-2 \psi ^{(1)}\left(\frac{\ell+2}{2}\right)\,,
\end{equation}
where we recall that the Gegenbauer polynomials are given by
\begin{equation}
C^{(1)}_{\ell}(\cos\theta)=\frac{\sin\left((\ell+1)\theta\right)}{\sin\theta}\quad\implies\quad
C^{(1)}_{\ell}(1)=1+\ell\,.
\end{equation} 
Eq.~\eqref{eq_app_f4^2_geg} was derived using the orthonormality relation
\begin{equation}
\int  d^3\hat{n}C^{(1)}_{\ell}(\hat{n}_x\cdot\hat{n})C^{(1)}_{\ell'}(\hat{n}\cdot\hat{n}_y)=\delta
_{\ell\ell'}\frac{2\pi^2}{\ell+1}
C^{(1)}_{\ell}(\hat{n}_x\cdot\hat{n}_y)\,,
\end{equation}
which implies
\begin{equation}
\frac{1}{2\pi^2}\int  d^3\hat{n}C^{(1)}_{\ell}(\hat{n}_x\cdot\hat{n})C^{(1)}_{\ell'}(\hat{n}\cdot\hat{n}_x)=\delta
_{\ell\ell'}\,.
\end{equation}

Using these formulas in eq.~\eqref{eq_app_I12}, we obtain
\begin{equation}\label{eq_app_I12_sum}
I_{12}(\cos\theta)=\sum_{\ell,\ell'}g_{\ell}g_{\ell'}\int d^3\hat{n}C^{(1)}_{\ell}(\hat{n}_1\cdot\hat{n})C^{(1)}_{\ell'}(\hat{n}\cdot\hat{n}_2)=
\sum_{\ell=0}^{\infty}\frac{2\pi^2g_{\ell}^2}{1+\ell}\frac{\sin\left((\ell+1)\theta\right)}{\sin\theta}\,.
\end{equation}
Below we use eq.~\eqref{eq_app_I12_sum} to expand~$I_{12}(\cos\theta)$ near $\theta=\pi$ and $\theta=0$.

The expansion near $\theta=\pi$ is straightforward, since $I_{12}(\cos\theta)$ is regular for zero or small cusp. Therefore all we have to do is to expand the argument of the summands near $\theta=\pi$:
\begin{equation}
\frac{\sin\left((\ell+1)\theta\right)}{\sin\theta}=(-1)^{\ell}\left[(\ell+1)-\frac{1}{6}  \ell(\ell+1)(\ell+2)(\pi-\theta )^2+O\left((\pi-\theta)^4\right)\right]\,.
\end{equation}
We then obtain
\begin{equation}
I_{12}(\cos\theta)=\sigma_0+\sigma_2(\pi-\theta)^2+\ldots\,,
\end{equation}
where
\begin{align}
\sigma_0&=2\pi^2\sum_{\ell=0}^{\infty}g_{\ell}^2(-1)^{\ell}\,, \\
\sigma_2&=-\frac{\pi^2}{3}\sum_{\ell=0}^{\infty}(-1)^{\ell}\ell(\ell+2)g_{\ell}^2 \,.
\end{align}
In practice, these sums are oscillatory, so it is best to subtract the first terms in the asymptotic expansion of $g_{\ell}^2$ to make the series absolutely convergent.  The sum over the asymptotic terms is then evaluated using
\begin{equation}
\sum_{\ell=1}^\infty\frac{(-1)^{\ell}}{\ell^k}=(2^{1-k}-1 ) \zeta(k)\,.
\end{equation}
We finally recast our result as 
\begin{align}
\sigma_0&=16 \pi ^4 \log (2)-\frac{\pi ^6}{9}+
\Sigma_0\simeq 284.732
\,,\\
\sigma_2&=\frac{1}{18} \pi ^2 \left[3 \pi ^4+48+4 \pi ^2 (16 \log 2-9)\right]+\Sigma_2\simeq 72.0467\,,
\end{align}
where we defined the following convergent sums
\begin{align}\nonumber
\Sigma_0&=2 \pi ^2\sum_{\ell=1}^{\infty}(-1)^{\ell}\left\{\left[\pi ^2-2 \psi ^{(1)}\left(\frac{\ell+2}{2}\right)\right]^2-\pi ^4+\frac{8 \pi ^2}{\ell}\right\} \\&
\simeq 
-688.749\,, \label{eq_app_Sigma0}
\\[0.6em]
\nonumber
\Sigma_2&=4 \pi ^2 \sum_{\ell=1}^{\infty}(-1)^{\ell}\frac{3 (\ell+2) \ell^2 \psi ^{(1)}\left(\frac{\ell+2}{2}\right) \left[\pi ^2-\psi ^{(1)}\left(\frac{\ell}{2}+1\right)\right]+12 \ell-2 \pi ^2 \left[3 \ell (\ell+1)-4\right]}{9 \ell}\\
&\simeq -159.753\,. \label{eq_app_Sigma2}
\end{align}
It can be verified numerically that $\sigma_0$ agrees with the former result~\eqref{eq_app_I12_Pi}. We also computed the coefficient of the $(\pi-\theta)^4$ term in the same way.

To analyze the short distance limit, we note that $I_{12}(\cos\theta)$ is singular at $\theta=0$, so we cannot just Taylor expand the summand like before.  This is because $\lim_{\theta\rightarrow 0}\frac{2\pi^2g_{\ell}^2}{1+\ell}\frac{\sin\left((\ell+1)\theta\right)}{\sin\theta}\sim 2\pi^6$ for $\ell\rightarrow\infty$.
Therefore we isolate the would be divergent contribution of the sum by expanding the coefficient multiplying the Gegenbauer polynomial for large $\ell$:
\begin{equation}
\begin{split}
I_{12}(\cos\theta)&\sim \sum_{\ell=1}^\infty\left[\frac{2 \pi ^6}{\ell}-\frac{2 \left(\pi ^2 \left(8 \pi ^2+\pi ^4\right)\right)}{\ell^2}+O\left(\frac{1}{\ell^3}\right)\right]\frac{\sin\left((\ell+1)\theta\right)}{\sin\theta} \\
&\simeq\frac{\pi ^7}{\theta }+\frac{\pi ^4}{3}  \left(48 \log \theta -\pi ^4-17 \pi ^2-48\right)\,.
\end{split}
\end{equation}
Restoring the finite term, we arrive at
\begin{equation}
I_{12}(\cos\theta)=\frac{\pi ^7}{\theta }+
16 \pi ^4 \log \theta -\frac{1}{9} \pi ^4 \left(9+\pi ^2\right) \left(16+3 \pi ^2\right)
+\tilde{\Sigma}_0
+O\left(\theta\right)
\end{equation}
where we defined the following convergent sum
\begin{equation} \label{eq_app_Sigmat0}
\tilde{\Sigma}_0=\sum_{\ell=1}^{\infty}\left\{\frac{2 \pi ^4 \left(8 \ell+\pi ^2+8\right)}{\ell^2}+8 \pi ^2 \psi ^{(1)}\left(1+\frac{\ell}{2}\right) \left[\psi ^{(1)}\left(\frac{\ell}{2}+1\right)-\pi ^2\right]\right\}\,.
\end{equation}
Below we show that this sum can be evaluated analytically to give
\begin{equation}\label{eq_app_conj1}
\tilde{\Sigma}_0=24 \pi ^2 \zeta (3)+\frac{\pi ^8}{3}+\frac{43 \pi ^6}{9}+16 \pi ^4 (\log 2-1)\simeq 7562.64\,.
\end{equation}
Therefore we arrive at
\begin{equation}
I_{12}(\cos\theta)=\frac{\pi ^7}{\theta }+
16 \pi ^4 \log \theta+ 16 \pi ^4 \log 2+24 \pi ^2 \zeta (3)-32 \pi ^4
+O\left(\theta\right)\,.
\end{equation}

To prove the relation~\eqref{eq_app_conj1}, we first write $\tilde{\Sigma}_0$ in terms of three convergent sums:
\begin{equation}
\tilde{\Sigma}_0=\sum_{\ell=1}^{\infty}
\frac{16 \pi ^4+2 \pi ^6}{\ell^2}+8\pi^4\sum_{\ell=1}^{\infty}\left[\frac{2}{\ell}-\psi ^{(1)}\left(1+\frac{\ell}{2}\right)\right]+8\pi^2\sum_{\ell=1}^\infty\left[ \psi ^{(1)}\left(1+\frac{\ell}{2}\right)\right]^2\,.
\end{equation}
The first sum is trivial:
\begin{equation}\label{eq_app_id1}
\sum_{\ell=1}^{\infty}
\frac{16 \pi ^4+2 \pi ^6}{\ell^2}=\frac{1}{3} \pi ^6 \left(8+\pi ^2\right)\,.
\end{equation}
The second is evaluated considering the following identity
\begin{equation}
\frac{2}{\ell}-\psi ^{(1)}\left(1+\frac{\ell}{2}\right)=
-\int_0^\infty dz\left[-\frac{2}{(\ell+z)^2}-\frac{1}{2} \psi ^{(2)}\left(1+\frac{\ell+z}{2}\right)\right]\,.
\end{equation}
We then exchange sum and integral and use the identity
\begin{equation}
\sum _{k=1}^{\infty } \frac{1}{\left(k+\frac{\ell+z}{2}\right)^3}=
-\frac{1}{2} \psi ^{(2)}\left(1+\frac{\ell+z}{2}\right)\,,
\end{equation}
to obtain
\begin{equation}
\sum_{\ell=1}^\infty\left[\frac{2}{\ell}-\psi ^{(1)}\left(1+\frac{\ell}{2}\right)\right]=-\int_0^\infty dz
\left[\sum_{\ell=1}^\infty\frac{-2}{(\ell+z)^2}+\sum_{\ell=1}^\infty\sum_{k=1}^\infty\frac{1}{\left(k+\frac{\ell+z}{2}\right)^3}\right]\,.
\end{equation}
Both sums can be performed in closed form for any $z$ and we obtain
\begin{align}\nonumber
\sum_{\ell=1}^\infty\left[\frac{2}{\ell}-\psi ^{(1)}\left(1+\frac{\ell}{2}\right)\right]=&-\frac14\int_0^\infty dz
\left[\zeta \left(3,\frac{z+3}{2}\right)-\zeta \left(3,\frac{z+4}{2}\right)-8 \psi ^{(1)}(z+1) \right.
\\&\hspace*{5em} \nonumber
\left.
+16 \psi ^{(1)}(z+3)+8 z \psi ^{(2)}(z+3)+12 \psi ^{(2)}(z+3)
\right]\\
=&\frac{5 \pi ^2}{12}-2\,. \label{eq_app_id2}
\end{align}

Using eq.~\eqref{eq_app_id1} and~\eqref{eq_app_id2} in eq.~\eqref{eq_app_Sigmat0} we recast $\tilde{\Sigma}_0$ in the form
\begin{equation}
\tilde{\Sigma}_0=\frac{1}{3} \pi ^4 \left(\pi ^4+18 \pi ^2-48\right)
+8\pi^2\sum_{\ell=1}^\infty\left[ \psi ^{(1)}\left(1+\frac{\ell}{2}\right)\right]^2\,,
\end{equation}
and thus eq.~\eqref{eq_app_conj1} is equivalent to
\begin{equation}\label{eq_app_conj2}
\sum_{\ell=1}^\infty\left[ \psi ^{(1)}\left(1+\frac{\ell}{2}\right)\right]^2=3 \zeta (3)-\frac{11 \pi ^4}{72}+\pi ^2 \log 4\,.
\end{equation}
To verify eq.~\eqref{eq_app_conj2} we use the following integral representation of the digamma function:
\begin{equation}
\psi ^{(1)}\left(z\right)=\int_0^{\infty} dt\frac{t e^{-t z}}{1-e^{-t}}\,.
\end{equation}
Using this in the sum and commuting with the integral we arrive at
\begin{equation}
\begin{split}
\sum_{\ell=1}^\infty\left[ \psi ^{(1)}\left(1+\frac{\ell}{2}\right)\right]^2 &=\int_0^\infty dt_1\int_0^\infty dt_2
\frac{e^{-t_1-t_2} t_1 t_2}{\left(1-e^{-t_1}\right) \left(e^{\frac{t_1}{2}+\frac{t_2}{2}}-1\right) \left(1-e^{-t_2}\right)} \\
&=\int_0^\infty dt_2\frac{t_2 \left[\pi ^2 \left(3 e^{\frac{t_2}{2}}+1\right)-24 e^{t_2} \text{Li}_2\left(e^{-\frac{t_2}{2}}\right)\right]}{6 \left(e^{t_2}-1\right)^2}\,.
\end{split}
\end{equation}
Thus upon changing variable as $e^{-t_2/2}=x$, eq.~\eqref{eq_app_conj2} becomes
\begin{equation}\label{eq_app_conj3}
\int_0^1 dx\frac{2 x \left[24 \text{Li}_2(x)-\pi ^2 x (x+3)\right] \log (x)}{3 \left(1-x^2\right)^2}=3 \zeta (3)-\frac{11 \pi ^4}{72}+\pi ^2 \log 4\,.
\end{equation}
We verified this identity through integration by parts using:
\begin{multline}
\frac{2 x \left[24 \text{Li}_2(x)-\pi ^2 x (x+3)\right] \log (x)}{3 \left(1-x^2\right)^2}=\\
\frac{d}{dx}\left\{
\frac{1}{6} \left[24 \text{Li}_2(x)-\pi ^2 x (x+3)\right] \left[\frac{2 x^2 \log (x)}{1-x^2}+\log (1-x^2)\right]\right\}\\
+\frac{1}{6} \left[\pi ^2 (2x+3)+\frac{24 \log (1-x)}{x}\right] \left[\frac{2 x^2 \log (x)}{1-x^2}+\log (1-x^2)\right]
\,.
\end{multline}

\bibliography{Biblio.bib}
	\bibliographystyle{JHEP.bst}

\end{document}